\documentclass[fleqn,usenatbib]{mnras}

\usepackage{newtxtext,newtxmath}
\usepackage{multirow}

\usepackage[T1]{fontenc}

\DeclareRobustCommand{\VAN}[3]{#2}
\let\VANthebibliography\thebibliography
\def\thebibliography{\DeclareRobustCommand{\VAN}[3]{##3}\VANthebibliography}

\usepackage{graphicx}	
\usepackage{amsmath}	
\newcommand*\standardrel{<}
\newcommand*\standardbin{+}
\makeatletter
\newcommand*\tabularrel[1]{%
  \mathrel{\mathpalette{\@tabularsym\standardrel}{#1}}%
}
\newcommand*\tabularbin[1]{%
  \mathbin{\mathpalette{\@tabularsym\standardbin}{#1}}%
}
\newcommand*\@tabularsym[3]{%
  \setbox\z@\hbox{$#2#1\m@th$}%
  \hbox to\wd\z@{\hss$#2#3\m@th$\hss}%
}

\newcommand{\kms}{\rm{km\ s}^{-1}} 

\title[\textit{JWST} FRESCO: $6.8<z<9.0$ {[\ion{O}{iii}]} emitters]{\textit{JWST} FRESCO: a comprehensive census of H$\beta$+[\ion{O}{iii}] emitters at $6.8<z<9.0$ in the GOODS fields}

\author[Meyer, R. A. et al.]{R. A. Meyer$^{1}$,\thanks{E-mail: romain.meyer@unige.ch},
P. A. Oesch$^{1,2}$,
E. Giovinazzo$^{1}$,
A. Weibel$^{1}$,
G. Brammer$^{2}$, 
J. Matthee$^{3}$, 
R. P. Naidu$^{4}$\thanks{Hubble Fellow}, \newauthor
R. J. Bouwens$^{5}$,
J. Chisholm$^{6}$, 
A. Covelo-Paz$^{1}$,
Y. Fudamoto$^{7}$,
M. Maseda$^{8}$,
E. Nelson$^{9}$,
I. Shivaei$^{10}$,\newauthor
M. Xiao$^{1}$,
T. Herard-Demanche$^{5}$,
G. D. Illingworth$^{11}$,
J. Kerutt$^{12}$,
I. Kramarenko$^{3}$,
I. Labbe$^{13}$,\newauthor
E. Leonova$^{14,15}$,
D. Magee$^{6}$,
J. Matharu$^{2}$,
G. Prieto Lyon$^{2}$,
N. Reddy$^{16}$,
D. Schaerer$^{1}$, 
A. Shapley$^{17}$,\newauthor
M. Stefanon$^{18,19}$,
M. A. Wozniak$^{16}$ and
S. Wuyts$^{29}$ \\
\emph{\normalsize Affiliations are listed at the end of the paper}
}

\graphicspath{{./}{plots/}}

\date{Accepted 2024 October 08. Received 2024 October 08; in original form 2024 May 16}

\pubyear{2023}

\begin{document}
\label{firstpage}
\pagerange{\pageref{firstpage}--\pageref{lastpage}}
\maketitle

\begin{abstract}
We present the census of H$\beta$+\rm{[\ion{O}{iii}]}\ $4960,5008\ \text{\AA}$ emitters at $6.8<z<9.0$ from the \textit{JWST} FRESCO survey over 124 arcmin$^2$ in the GOODS-North and GOODS-South fields. Our unbiased spectroscopic search results in 137 spectroscopically-confirmed galaxies at $6.8<z<9.0$ with observed [\ion{O}{iii}] fluxes $f_{\rm{ [\ion{O}{iii}]}}\gtrsim 1\times 10^{-18}\ \rm{ergs}\ \rm{s}^{-1} \ \rm{cm}^{-2}$. The rest-frame optical line ratios of the median stacked spectrum (median $M_{\rm{UV}}=-19.65^{\tabularrel+0.59}_{\tabularrel-1.05}$) indicate negligible dust attenuation, low metallicity ($12+\log(\rm{O/H})= 7.2-7.7$) and a high ionisation parameter $\log_{10}U \simeq -2.5$. We find a factor $\times 1.3$ difference in the number density of $6.8<z<9.0$ galaxies between GOODS-South and GOODS-North, which is caused by a single overdensity at $7.0<z<7.2$ in GOODS-North. The bright end of the UV luminosity function of spectroscopically-confirmed [\ion{O}{iii}] emitters is in good agreement with HST dropout-selected samples. Discrepancies between the observed [\ion{O}{iii}] LF, [\ion{O}{iii}]/UV ratio and [\ion{O}{iii}] equivalent widths, and that predicted by theoretical models, suggest burstier star-formation histories and/or more heterogeneous metallicity and ionising conditions in $z>7$ galaxies. We report a rapid decline of the [\ion{O}{iii}] luminosity density  at $z\gtrsim 6-7$ which cannot be explained by the evolution of the cosmic star-formation rate density. Finally we find that FRESCO detects in only $2$h galaxies likely accounting for $\sim 10-20\%$ of the ionising budget at $z=7-8$ (assuming an escape fraction of $10\%$), raising the prospect of directly detecting a significant fraction of the sources of reionisation with \textit{JWST}.
\end{abstract}




\section{Introduction}

Understanding the first billion years of cosmic history is of prime importance to astrophysics. This period saw the emergence of the first stars and galaxies \citep[][]{Bromm2013, Stark2016},  the early enrichment of the circumgalactic medium \citep[e.g.][for reviews]{Tumlinson2017,Maiolino2019}, the growth of the first supermassive black holes \citep[e.g.][]{Inayoshi2020,Volonteri2021}, and the reionisation of the intergalactic medium \citep[e.g.][]{Becker2015,Fan2022}. Understanding the complex processes behind the turbulent youth of the Universe thus sheds important light on astrophysics in general.

Pushing the redshift frontier to $z\gtrsim 6$ (e.g. $\lesssim 1\ \rm{Gyr}$ after the Big Bang) has been a challenging and decades-long endeavour. The Lyman-break technique \citep{Guhathakurta1990,Steidel1992}, originally developed for $z\sim 3$ galaxies, was applied to $z\sim 6-10$ galaxies with the first deep imaging fields and lensing clusters observed with large \textit{HST} programmes \citep[][]{Giavalisco2004, Beckwith2006, Postman2012, Bouwens2015,Lotz2017, Coe_2019,Steinhardt2020}. Whilst enabling a first census of the luminous galaxies in the Epoch of Reionisation \citep[e.g.][]{Ellis2013,Oesch2014, Bouwens2015,Livermore2017,Atek2018, Bouwens2021}, the \textit{HST} imaging data yielded limited information on these early objects as it only covered the rest-frame UV probing recent star-formation. Furthermore, spectroscopic confirmation suffered from the absence of near-infrared coverage. The reliance on the detection of the Lyman-$\alpha$ line from the ground, attenuated by the neutral IGM and redshifted to $1.0-1.5\ \mu\rm{m}$ observed range, made such work extremely time-consuming. Indeed, only $\sim 1\% $ of $z>7$ galaxies had spectroscopic redshifts in the pre-\textit{JWST} era \citep[e.g.][]{Pentericci2014,Zitrin2015,Oesch2016,Stark2017,Laporte2017,Jung2019}. In the years leading to the launch of \textit{JWST}, ALMA was thus the dominant observatory to obtain high-redshift spectroscopic confirmations via the detection of far-infrared fine structure lines, albeit for luminous galaxies \citep[e.g.][]{Hashimoto2018a,Hashimoto2019, Jarugula2021, Bouwens2022_Rebels}.

The advent of \textit{JWST} has transformed the field. Due to its unprecedented depth and near-IR imaging capabilities, many ultra-high-redshift galaxy candidates have been claimed at $z>10$, and some up to $z>15$  \citep{Finkelstein2022a,Finkelstein2023a,Harikane2023a,Naidu2022,Donnan2022,Adams2022,Yan2023,Bouwens2023, Robertson2023a, Robertson2023b, Hainline2024}. Disagreement between the different studies on the exact number of candidates in each field \citep{Bouwens2023a} could suggest that the contamination fraction of Lyman break-selected samples is still uncertain, especially at the highest redshifts \citep[e.g.,][]{Meyer2024}. Nonetheless, numerous objects have also been successfully confirmed with spectroscopy up to $z\sim 14$  \citep{CurtisLake2023,ArrabalHaro2023a, ArrabalHaro2023,Bunker2023_JADESrelease,Bunker2023_GNz11,Saxena2024, WangB2023,Harikane2024,Carniani2024}. Especially at $6<z<10$, where numerous rest-frame optical emission lines are detected, these observations have led to rapid advances in the characterisation of galaxies in the epoch of reionisation. 

Most of the studies listed above used the tried-and-tested method of photometric pre-selection (with ground-based, HST and/or \textit{JWST} photometry) with follow-up using the \textit{JWST}/NIRSpec instrument in multi-slit spectroscopy mode. Deep spectroscopy obtained in this way comes at the expense of (often) untractable selection functions, hindering inferences about the statistical properties of high-redshift galaxies and the nature of average sources. Such experiments are instead better suited to slitless spectroscopic observations, which provide unbiased flux-limited samples in a given field, at the cost of surveying only the brightest objects. Early results with the Wide Field Slitless Spectroscopy (WFSS) modes on NIRCam \citep[][]{nircam_performance}{}{} and NIRISS \citep[][]{Willott2022}{}{} have demonstrated their power for blind searches for high-redshift galaxies \citep[e.g.][]{Sun2022_WFSS,Wang2023,Matthee2023_EIGER, Roberts-Borsani2022a}. Already, the number of spectroscopic redshifts from WFSS data is ahead of multi-slit ones even in deep legacy fields \citep[e.g.][]{Rieke2023_JADESrelease,Helton2023}. Such blind surveys are necessary to have a complete view of galaxies from Cosmic Noon to Cosmic Dawn, enabling the study of large-scale structures, unbiased population properties, as well as finding rare objects and constituting legacy samples for future follow-up. The community interest in this approach is reflected in the large increase in direct and pure-parallel programmes using NIRCam/NIRISS WFSS in \textit{JWST} Cycles 2 \& 3. 

In this paper, we present the results of an unbiased spectroscopic search for emission line selected galaxies at $6.8<z<9.0$ in the two Great Observatories Origins Deep Survey (GOODS) fields (GOODS-North and GOODS-South) from the First Reionisation Epoch Spectroscopically Complete Observations \citep[FRESCO,][]{Oesch2023} \textit{JWST} Cycle 1 programme \#1895. FRESCO surveys a contiguous area of $2\times 62\ \rm{arcmin}^2$ in the GOODS-North and GOODS-South fields using the NIRCam WFSS mode. With the use of the longest wavelength NIRCam filter (F444W), FRESCO delivers hundreds of spectroscopic redshifts from Cosmic Noon to Cosmic Dawn, leveraging for example the Paschen-$\alpha$ line at $1.0<z<1.7$ \citep[e.g.][]{Neufeld2024,Shivaei2024}{}{}, the H$\alpha$ line at $4.85<z<6.75$ \citep[e.g.][]{Helton2023, Herard-Demanche2023}{}{}, and the  [\ion{O}{iii}] $5008, 4960\ \text{\AA}$ doublet at $6.8\lesssim z\lesssim 9.0$. FRESCO has also led to the discovery of faint obscured AGN at high-redshift \citep["Little Red Dots"][]{Matthee2024_LRD}{}{}, the first measurement of galaxy rotation using \textit{JWST} WFSS data \citep[][]{Nelson2023}{}{} and constraining the efficient formation of massive optically-dark galaxies at $z\sim 5.5$ \citep[][]{Xiao2023}{}{}. The aim of this work is to provide an unbiased flux-limited sample of [\ion{O}{iii}] emitters as well as characterise the key statistical properties of these high-redshift line emitters: their UV and  [\ion{O}{iii}] luminosity function, median physical properties, number densities, and contribution to the reionisation budget.

The paper is structured as follows. In Section \ref{sec:obs_analysis} we detail our data reduction and Section \ref{sec:selection} describes our multi-step blind search for [\ion{O}{iii}] and the characterisation of its completeness and purity. The full catalogue and median stacked rest-frame optical spectrum of [\ion{O}{iii}] emitters is presented in Sections \ref{sec:complete_cat_O3} and \ref{sec:restframeopticalstack}, respectively. Section \ref{app:quality-checks} contrasts the spectroscopic redshifts to the \texttt{EAZY} photometric redshifts. In Section \ref{sec:results} we discuss the statistical properties of [\ion{O}{iii}] emitters including the [\ion{O}{iii}] luminosity function, the spectroscopic UVLF, the [\ion{O}{iii}]/UV luminosity ratio and the [\ion{O}{iii}] equivalent width distribution at $z\simeq 7,8$. We discuss the evolution of the  [\ion{O}{iii}] luminosity at $z>6$ and the contribution of [\ion{O}{iii}] emitters to reionisation in Section \ref{sec:discussion}. The full catalogue of emitters is made public in a machine-readable format at \url{https://github.com/rameyer/fresco/}. Throughout this paper, magnitudes are given in the AB system \citep[][]{Oke1974}{}{}, and we assume a concordance cosmology with $H_0 = 70\ \kms \rm{Mpc}^{-1},\ \Omega_m = 0.3, \Omega_\Lambda=0.7$.

\section{Observations}
\label{sec:obs_analysis}
\subsection{Imaging data and catalogues}
We make use of all FRESCO imaging data as well other public JWST/NIRCam imaging and HST ACS and WFC3/IR imaging in the two GOODS fields. This includes the original GOODS HST data \citep{Giavalisco2004} as well as the Cosmic Assembly Near-infrared Deep Extragalactic Legacy Survey data \citep[CANDELS;][]{Grogin2011, Koekemoer2011}, in addition to the deep data over the HUDF \citep{Illingworth2013,Ellis2013,Koekemoer2013}.  With \textit{JWST}, parts of these fields have been imaged very deeply by the JADES GTO team through different programs  \citep[][]{Rieke2023_JADESrelease,Eisenstein2023a_JADES,Eisenstein2023b_JADES} and we also make use of the public JEMS medium band data \citep{Williams2023}, a few pointings of the PANORAMIC survey (Williams et al. in prep) and pre-imaging from program 2198 \citep{Barrufet2024}. We use \texttt{SourceExtractor} \citep{Bertin1996} to derive a segmentation map as a basis for the grism extractions, as well as to obtain photometric fluxes in all the available HST and \textit{JWST} bands. To construct a catalog with homogeneous depth across the FRESCO footprint, we create custom reductions of the imaging in F210M and F444W obtained through FRESCO only, i.e., excluding any additional data that may be available in these filters from other surveys. Using an inverse-variance weighted stack of the F210M and F444W images as the detection image, we run \texttt{SourceExtractor} in dual image mode, measuring fluxes in circular apertures with a radius of $0.16$\arcsec\ in all the full stack of images in all available bands, PSF-matched to F444W. The aperture fluxes are first scaled to the fluxes measured in Kron apertures on a PSF-matched version of the detection image, and then to total fluxes by dividing by the encircled energy of the Kron aperture on the F444W PSF.

To estimate the detection completeness of our catalog, we use the GaLAxy survey Completeness AlgoRithm 2 (\texttt{GLACiAR2}) software \citep{Leethochawalit2022}. On a $1.5\arcmin\times1.5\arcmin$ cutout of the detection image, we inject galaxies in 34 magnitude bins ranging from 22.5 to 30.5 AB, assuming a Gaussian distribution in the logarithm of their sizes centered at R$_{\rm eff}=0.8\,{\rm kpc}$, a flat shape of their SED (i.e., the same AB-magnitude in both F210M and F444W). Performing 10 iterations with 500 galaxies per iteration, we thus measure the fraction of recovered sources as a function of the F444W magnitude. The photometric completeness is roughly constant at $90\%$ down to a magnitude of $\simeq 27$, and then declines to $0\%$ over the magnitude range $27-30.5$ (see further Appendix \ref{app:completeness} for details). With $\sim 50\%$ of [\ion{O}{iii}] emitters below a detection image magnitude of $27$ (e.g. where the completeness start declining), we need to take into account the photometric completeness in this work. The assumed SNR-cut in this work of SNR(F444W)$\geq3$ corresponds to an AB-magnitude of F444W$\simeq28.9$. The reason for applying a cut in SNR(F444W) rather than SNR(F210M+F444W) is that it directly corresponds to an emission line SNR$\gtrsim4$ cut for source without detected continuum due to the relative imaging and grism depths of FRESCO \citep[][]{Oesch2023}.
For more details on the photometric catalog production and the completeness simulation, see \citet[][]{Weibel2024}{}{}. We use the internal FRESCO data release v7.3. 

\subsection{NIRCam WFSS data}

 We reduce the FRESCO F444W WFSS data \citep[see][]{Oesch2023}{}{} using \texttt{GRIZLI} version $1.9.13$, and \texttt{JWST} pipeline version 1.12.0 and pmap 1123. Briefly, the \texttt{GRIZLI} pipeline starts from the MAST-downloaded rate files, applies step 1 of the \textit{JWST} reduction pipeline with custom snowball masking. \texttt{GRIZLI} then reduces the direct and grism-dispersed images in its own framework \citep[see further][]{Brammer2019, Brammer2021}. We use the standard CRDS grism dispersion files that were made available in September 2023 with pmap 1123 to predict the 2D trace and updated sensitivity functions from \texttt{GRIZLI}. Following \citet{Kashino2023}, we apply a median filter with a size of $71$ and central gap of $10$ pixels to the grism images to remove the continuum from all sources. The median filtering is run in two steps with a first pass necessary to identify positive features (defined as pixels with SNR$>3$), and then runs a new median filter subtraction with these pixels flagged, improving the continuum subtraction around lines. We also produce a version of the reduced grism data without median filtering to capture broad emission line features after the initial catalogue of [\ion{O}{iii}] emitters has been constructed.  For each source, we use the F444W+F210M segmentation map to extract a 1D spectrum using optimal extraction \citep[][]{Horne1986}{}{}. 

\section{Selection of high-redshift [\ion{O}{iii }] 4960,5008 \AA\ emitters}
\label{sec:selection}
\subsection{Optical non-detection selection}
\label{sec:dropout}
The high-redshift line emitters are selected in a three-step process. First, we discard objects that have detections in bands bluewards of Lyman-$\alpha$. As the Lyman-$\alpha$ line falls at the edge of the HST F814W band at $z=6.8$, our lowest redshift searched, and to remove objects with contamination in the F775W or F606W bands, we select objects using the following non-detection and colour cuts:
\begin{align}
\label{eq:lbg_cut}
    &F606W < 2\sigma, \ F775W < 2\sigma,\ F814W < 2\sigma \\
    \rm{OR} \nonumber \\
    &F606W< 3\sigma, F775W < 3\sigma ,  \ F814W-F160W > 1.75\ \rm{mag}
\end{align}
where the $1.75\ \rm{mag}$ color corresponds to that typically used in dropout searches for $z\sim7-8$ galaxies \citep[e.g.][]{Bouwens2015,Roberts-Borsani2016,Harikane2018}. We emphasize that the goal of the above cuts is not to isolate a good sample of high-redshift galaxies, rather it aims to remove objects that are unambiguously low-redshift contaminants. Our selection process is designed to avoid biases stemming from colour-cuts or photometric redshift estimates.

We also apply a SNR$\geq 3$ detection cut in F444W (FRESCO only) as the relative depth of the FRESCO imaging and grism data is chosen to ensure any significant emission line in the F444W WFSS data would correspond to a detection in the F444W image \citep[][]{Oesch2023}{}{}. We then extract all the 1D spectra for the sources selected as described above and run Gaussian-matched filters on the 1D spectra with $FWHM=50,100,200\ \kms$ to select emission lines from normal star-forming galaxies. Candidates are retained if the following conditions are met:
\begin{itemize}
    \item Two lines match the separation of  [\ion{O}{iii}]$\lambda\lambda 5008, 4960$ \AA\ at $6.8<z<9.0$, with a tolerance $\Delta v < 100\ \kms$ on the doublet separation 
    \item The strongest line of the doublet, i.e. [\ion{O}{iii}] 5008, is detected at SNR$>4$
    \item The observed ratio between the two lines of doublet candidates is $1<  [\ion{O}{iii}] 5008 /   [\ion{O}{iii}] 4960 < 10$ 
\end{itemize}
Whilst the two first conditions strictly enforce the line separation and significance of the doublet, the third one allows for a departure from the expected intrinsic ratio of $2.98$ \citep[][]{Storey2000}{}{}. This is because the \textit{observed} ratio is often found to be different, especially after median-filtering \footnote{The main issue affecting the observed line ratio is the median filter which uses running windows with fixed sizes in pixels to compute and subtract the median continuum. These are not adapted to each source and can thus over-subtract one or both lines, especially depending on the strength of H$\beta$ (we later improve the continuum subtraction of the final sample when stacking the spectra).}. In the final sample, we find that such cases are indeed driven by the low-SNR of the [\ion{O}{iii}] $4960$ \AA\ line. At SNR$([\ion{O}{iii}]  4960\ \text{\AA}) \gtrsim 3-4$, however, our sources have ratios consistent with the expected $2.98$ value.

The remaining sample is then visually inspected to remove contaminants and low-SNR objects. The contaminants can be primarily divided in three categories: a) residuals from the median filtering b) contamination by one or multiple adjacent sources c) lower-redshift single line emitters (most likely H$\alpha$ emitters at $z>5$ given the dropout cuts applied earlier). The latter, especially when they are SNR$\lesssim 5$, can be confused with low-SNR  [\ion{O}{iii}] doublet for which the second line is undetected due to the ratio of 2.98 between the lines. 

\subsection{Visual inspection}

The visual inspection was carried out by multiple team members using the custom developed tool \texttt{specvizitor}\footnote{\url{https://github.com/ivkram/specvizitor}}. To that end, the full sample was divided in overlapping halves such that each object was inspected by $4$ independent team members, whilst no one inspected exactly the same subset. Three team members also inspected the entire sample. Each inspector was assigned the same number of GN and GS objects to avoid biases due to the different quality of the data. Importantly, all team members also simultaneously inspected real candidates and mock emitters in order to reconstruct the completeness function of the visual inspection (see Section \ref{sec:completeness}and Appendix \ref{app:fake_emitters} for details on the simulated mock emitters). This was done in a blind trial fashion as inspectors did  not know which source was real and which source was fake, and both sets were mixed and presented in the same way (see Appendix \ref{app:fake_emitters} for an example of mock emitters and a comparison to real objects). 

\begin{figure*}
    \centering
   \includegraphics[width=0.85\textwidth,trim = 0 14.8cm 0cm 0, clip]{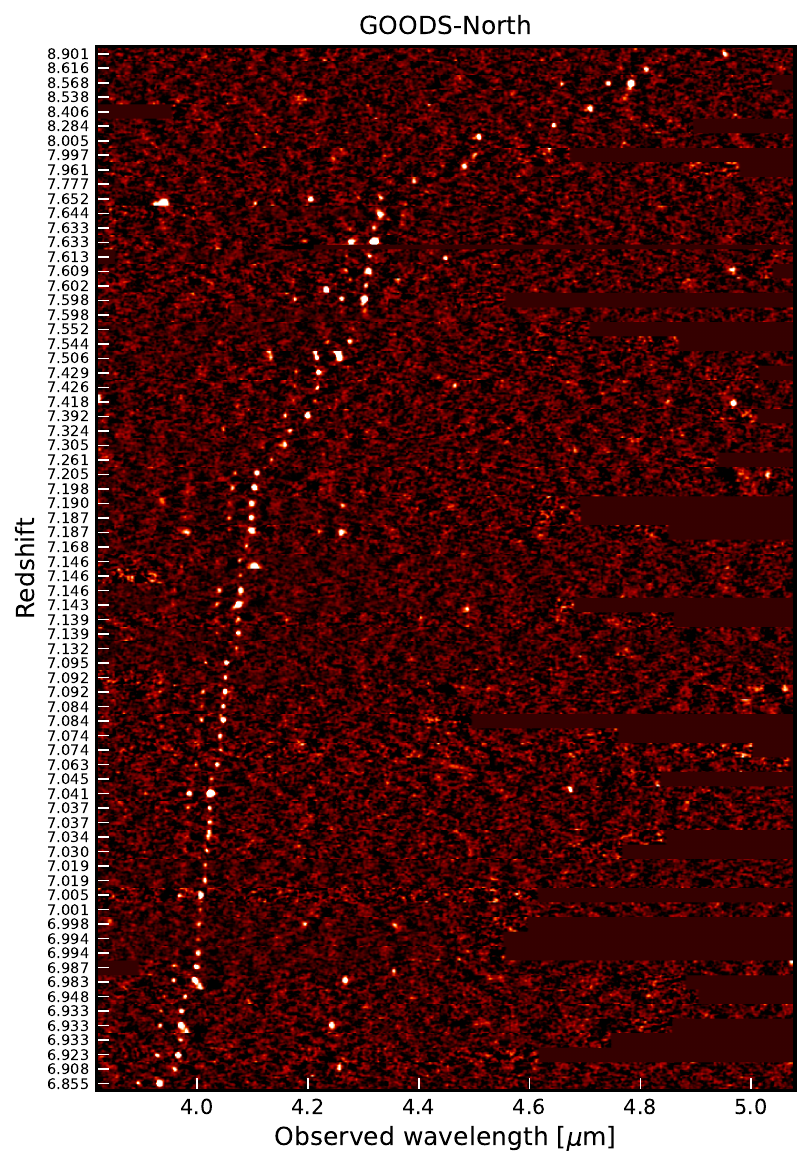}
    \caption{2D spectra of all the confirmed [\ion{O}{iii}] emitters in FRESCO in GOODS-North, ordered by redshift. A strong overdensity of sources is present at $6.9<z<7.2$. CO band heads contamination was masked in some extractions.}
    \label{fig:spectra_2D_GN}
\end{figure*}

\begin{figure*}
    \centering
   \includegraphics[width=0.85\textwidth,trim = 0 0cm 0cm 19.5cm, clip]{plots/FRESCO_O3emitters_2dspec_2column.pdf}
    \caption{2D spectra of all the confirmed [\ion{O}{iii}] emitters in FRESCO in GOODS-South, ordered by redshift. A small overdensity of sources is present at $z\sim7.25$. CO band heads contamination was masked in some extractions.}
    \label{fig:spectra_2D_GS}
\end{figure*}

\begin{table}
    \centering
    \begin{tabular}{l|r|r}\hline
    \hline
        & \multicolumn{2}{c}{Remaining sources} \\
       Selection & GOODS-N & GOODS-S  \\ \hline
       Full detection catalogue (F444W+F210M)  & 33539 & 31744 \\
       HST non-detection selection (Eq. 1 \& 2) & 2697 & 3915 \\ 
       Gaussian-matched filtering + Contaminants  & 309 & 345  \\ 
       Visual inspection, quality $q \geq 1.5$ & 78(6) & 59(7)  \\ \hline
       Visual inspection, quality $q \geq2.5$ & 38(4) & 14(1) \\ 
       Visual inspection, quality $2.0\leq q<2.5$ & 25(2) & 25(4)\\
       Visual inspection, quality $1.5\leq q<2.0$  & 15 & 20(2) \\ 
    \end{tabular}
    \caption{Summary of our search for high-redshift  [\ion{O}{iii}] emitters systems and the number of sources kept at each stage (see further Section \ref{sec:selection}). Numbers in parenthesis indicate emitters identified within $<1"$ of another source and merged into individual systems. }
    \label{table:search_steps}
\end{table}

During the visual inspection, team members assigned quality flags to the sources. Sources with a clear  [\ion{O}{iii}] doublet and a matching morphology between the direct image and 2D spectrum were given Quality $q=3$ ("definitely an [\ion{O}{iii}] emitter"). Sources with a lower SNR for the  [\ion{O}{iii}] 4960 but a good line ratio and a clear morphology match excluding the possibility of contamination were assigned Quality $q=2$ ("likely an [\ion{O}{iii}] emitter"). Unclear sources were given Quality 1 ("potentially an [\ion{O}{iii}] emitter"), objects without a line given Quality 0 ("no emission line"), and cases of contamination by neighbouring objects or continuum residuals were flagged as Quality $-$1. The scores were then averaged and rounded to the nearest half-integer. The catalogue was then cross-matched with the visually-inspected H$\alpha$ catalogue (Covelo-Paz et al., in prep.) to assign duplicates (9) which were then attributed to the most likely redshift, taking into account the location of the observed Lyman-$\alpha$ break in the ancillary photometry and the different quality grades in the H$\alpha$ and [\ion{O}{iii}] catalogues. The final catalogue comprises objects with quality $\geq 1.5$, i.e. objects for which a majority of the team believed them to be likely real. Whilst this approach is tailored to the statistical analysis performed in this paper, we however caution that for some uses such as single-object follow-up, a cut at $q=2$ (or even $q=2.5$ if high purity is needed) might be more appropriate. 

\subsection{Complete catalogue of [\ion{O}{iii}] emitters galaxies}
\label{sec:complete_cat_O3}
\begin{figure*}
    \centering
    \includegraphics[width=\textwidth]{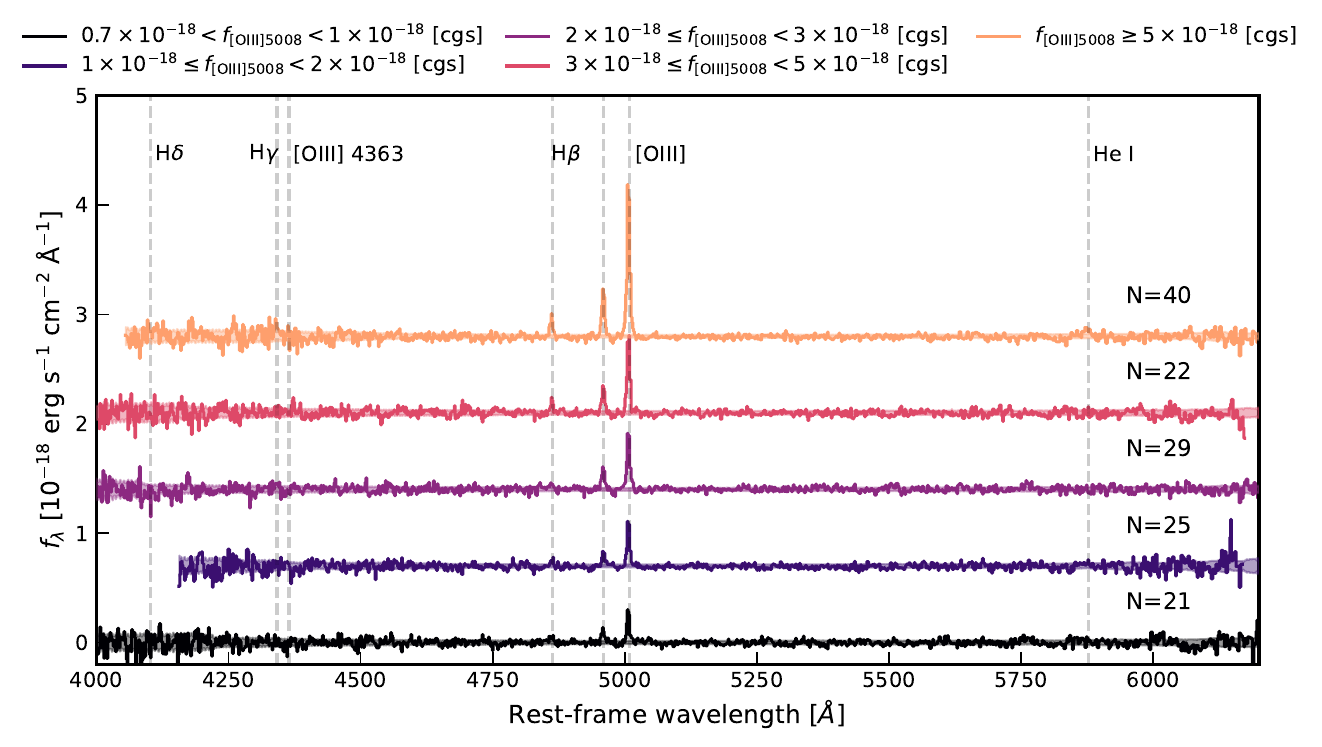}
    \caption{Median stacked rest-frame spectra of the [\ion{O}{iii}] emitters presented in this work, binned by [\ion{O}{iii}] flux. Each bin is shown in a distinct colour and is offset for clarity. The 1$\sigma$ error on the stacked spectra is shown with shaded areas. We do not detect rest-frame optical lines in the stacked spectra besides [\ion{O}{iii}] and H$\beta$, except a tentative \ion{He}{i} $5876$ \AA\ detection in the most luminous [\ion{O}{iii}] bin.}
    \label{fig:spectra_stacked}
\end{figure*}
\begin{figure*}
    \centering
    \includegraphics[width=\textwidth]{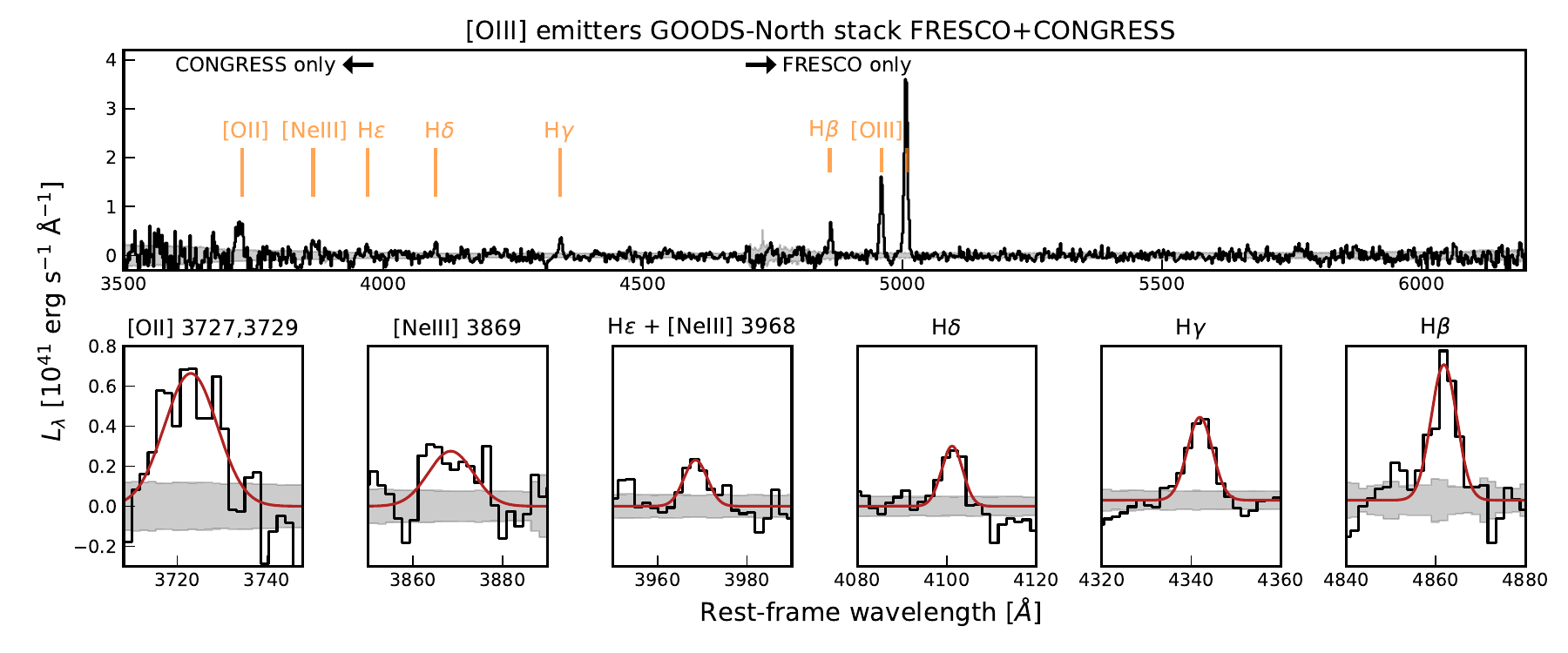}
    \caption{\textbf{Upper panel: }Rest-frame optical stacked spectrum of [\ion{O}{iii}] emitters in GOODS-North, using FRESCO (F444W) and CONGRESS (F356W) WFSS data. We indicate with black arrows which parts of the rest-frame stacked spectrum are \textit{only} by FRESCO or CONGRESS. Note that most coverage of the Balmer lines in the intermediate wavelength range comes from CONGRESS as there are more emitters at lower redshift. We detect Balmer lines down to H$\epsilon$, as well as [\ion{O}{ii}]$\lambda\lambda 3727,3729$ and [\ion{Ne}{iii}]$\lambda 3868$. \textbf{Lower panels: } Zoom-in on the [\ion{O}{ii}], [\ion{Ne}{iii}], and Balmer lines and their best-fit Gaussian (except for [\ion{O}{ii}] where we use a double Gaussian).}
    \label{fig:stack_GN_congress_fresco}
\end{figure*}

We summarize in Table \ref{table:search_steps} the steps of the search and the number of sources retained at each stage. The search results in 137 emitters with $q\geq1.5$ at $6.8<z<9.0$. We further merge the 137 emitters into 124 systems by grouping objects within $1"$ and at similar redshifts. Such systems are considered as one galaxy whose [\ion{O}{iii}] flux is the sum of its component galaxies for the purposes of computing luminosity functions and number densities \footnote{Such dual/triple systems are composed of galaxies at the same redshift, i.e., we checked that all $r<1"$ groupings do not merge serendipitously aligned galaxies at different redshifts}. We show the extracted 2D spectra of the selected [\ion{O}{iii}] emitters systems, sorted by redshift, in Figures \ref{fig:spectra_2D_GN} and \ref{fig:spectra_2D_GS}. Besides the [\ion{O}{iii}] doublet, only the H$\beta$ line is detected in a substantial fraction (38/137 at $>3\sigma$) of the sources. 

We show the median stacked 1D spectra at different [\ion{O}{iii}] line fluxes in Figure \ref{fig:spectra_stacked}. H$\beta$ is detected even in the faintest stack, and the [\ion{O}{iii}] 4960/5008 ratio is consistent with the expected value of $2.98$ \citep[][]{Storey2000}{}{}. This indicates that the emitters are real even at low-SNR. We do not detect any additional line beyond H$\beta$ and [\ion{O}{iii}] in the stacks, except perhaps a hint of \ion{He}{i} $5877 $\AA\ in the high luminosity stack (Fig. \ref{fig:spectra_stacked}). 
The full catalogue, the measured lines fluxes and the individual 1D and 2D spectra of selected objects are presented in Appendix \ref{app:full_cat_and_plots}. 

\subsection{The rest-frame optical spectrum of $6.8<z<9.0$ [\ion{O}{iii}] emitters}
\label{sec:restframeopticalstack}
It is worth noting that FRESCO covers the H$\gamma$ line only at $z\gtrsim 7.9$, whereas a majority of our [\ion{O}{iii}] emitters are at $z<7.9$. Therefore, the lack of emission lines bluewards of H$\beta$ and [\ion{O}{iii}] in the stacks presented in Figure \ref{fig:spectra_stacked} is mostly due to limited wavelength coverage for our sources. As a demonstration of the power of multi-filters NIRCam WFSS spectroscopy, we use the CONGRESS data (programme ID \#3577, PIs: E. Egami, F. Sun) to produce the median stacked rest-frame optical spectrum of the GOODS-North FRESCO [\ion{O}{iii}] emitters down to $\lambda_{\rm{rest}} \sim 3500$ \AA. CONGRESS targets the same area as FRESCO in GOODS-North using F356W grism spectroscopy at similar depths than FRESCO. The CONGRESS data was reduced following the same procedure as our FRESCO data (see Section \ref{sec:obs_analysis}). The median stacked spectrum is produced using the continuum-filtered extractions, with each galaxy weighted according to its luminosity distance (in order to cancel the (1+z) term when redshifting and stacking the spectral flux densities in the rest-frame). We checked that the wavelengths and fluxes are consistent for objects with [\ion{O}{iii}] covered both by FRESCO (F444W) and CONGRESS (F356W). However, in order to keep the H$\beta$ and [\ion{O}{iii}] properties consistent with that of a FRESCO-only stack, we only use the CONGRESS data up to a rest-frame wavelength $\lambda=4700$ \AA.

The median stacked spectrum is presented in Figure \ref{fig:stack_GN_congress_fresco} and different measurements from the stack are listed in Table \ref{tab:properties_of_stack}. In addition to the already detected H$\beta$ and [\ion{O}{iii}] lines, we detect H$\delta$, H$\gamma$ at SNR$>5$, H$\epsilon$+[\ion{Ne}{III}] 3986 and [\ion{Ne}{III}] 3869 at SNR$>3$, and [\ion{O}{ii}] at SNR$>10$ \footnote{We note that the wavelength of [\ion{O}{ii}] is slightly lower than expected, which we hypothesize could be an issue with the wavelength calibration. This does not affect the measured flux or any of the conclusion in this work.}. The detection of the multiple lines in the stack further evidences the high purity of our sample. We note that a significant fraction of [\ion{O}{iii}] / H$\alpha$ confusions in the visual inspection would have produced a spurious feature at rest-frame wavelength $\sim 3820\ $\AA, corresponding to the [\ion{O}{iii}] $5008$ line of H$\alpha$ emitters wrongly assigned to be  [\ion{O}{iii}] emitters at higher redshift. The numerous lines detected also enable us to globally characterise the [\ion{O}{iii}] emitters in this work. The various Balmer lines ratios are slightly higher ($1-1.5\sigma$)\footnote{We note that these slight discrepancies occur both when comparing H$\beta$ detected primarily with FRESCO to other Balmer lines detected by CONGRESS (H$\gamma$/H$\beta$, H$\delta$/H$\beta$, H$\epsilon$/H$\beta$) and when comparing CONGRESS-detected lines only  H$\gamma$/H$\delta$, H$\epsilon$/H$\gamma$).}, but consistent within $1-1.5 \sigma$ errors with that expected for Case B recombination with $n_e=100\ \rm{cm}^{-3}$ and $T_e=10^4\ \rm{K}$. Under the assumption that Case B recombination is valid \citep[but see ][]{Scarlata2024}{}{} and those higher ratios are due to noise and/or potential contamination of the lines, we conclude that dust attenuation is negligible in our [\ion{O}{iii}] emitters. The R23, O32 and Ne3O2 ratios point to relatively low metallicities $12+\log(O/H)= 7.2-7.7$ \citep[][]{Bian2018, Nakajima2022} and high-ionisation parameter $\log_{10}U \simeq -2.5$ \citep{Witstok2021}. Future work will examine in details the physical properties of the [\ion{O}{iii}] emitters discovered with FRESCO, including their emission line properties.

\begin{table}
    \centering
    \setlength{\tabcolsep}{6pt} 
\renewcommand{\arraystretch}{1.2}
    \begin{tabular}{l|r}
 Property & Observed Value \\  \hline
$M_{\rm{UV}}$ & $-19.65^{\tabularrel+0.59}_{\tabularrel-1.05}$ \\ 
$L_{\rm{[OII] 3727,3729}} [10^{42}\ \rm{erg\ s}^{-1}]$ & $0.99\pm0.08$ \\ 
$L_{\rm{[NeIII] 3869}} [10^{42}\ \rm{erg\  s}^{-1}]$ & $0.35\pm0.08$ \\ 
$L_{\rm{H\epsilon + [NeIII] 3969}} [10^{42}\ \rm{erg\  s}^{-1}]$ & $0.15\pm0.04$ \\ 
$L_{\rm{H\delta}} [10^{42}\ \rm{erg\  s}^{-1}]$ & $0.18\pm0.04$ \\ 
$L_{\rm{H\gamma}} [10^{42}\ \rm{erg\  s}^{-1}]$ & $0.27\pm0.03$ \\ 
$L_{\rm{H\beta}} [10^{42}\ \rm{erg\  s}^{-1}]$ & $0.44\pm0.06$ \\ 
$L_{\rm{[OIII] 4960,5008}} [10^{42} \rm{erg\  s}^{-1}]$ & $3.95\pm0.07$ \\ 
 \hline
H$\gamma$ / H$\beta$  & $0.60\pm0.11$ \\
H$\delta$ / H$\beta$  & $0.40 \pm 0.10$ \\
(H$\epsilon$+[NeIII] 3969)/H$\beta$ & $0.34 \pm 0.11$ \\
H$\delta$ / H$\gamma$  & $0.67\pm0.16$ \\
(H$\epsilon$+[NeIII] 3969)//H$\gamma$ & $0.57\pm0.18$ \\ \hline
O32$=$[\ion{O}{iii}]/[\ion{O}{ii}] & $4.0\pm0.3$  \\
R23$=$([\ion{O}{iii}]+[\ion{O}{ii}])/\rm{H}$\beta$ &  $11.14\pm1.53$ \\
R3$=$([\ion{O}{iii}] $5008$)/\rm{H}$\beta$ &   $6.38\pm0.85$ \\
Ne3O2=[\ion{Ne}{III}\ 3869]/[\ion{O}{ii}] & $0.34\pm0.08$ \\ 
$12+\log_{10}(\rm{O/H})(R3) ^{a}$ & $7.21^{\tabularrel+0.10}_{\tabularrel-0.10}$ \\
$12+\log_{10}(\rm{O/H})(R3) ^{b}$ & $7.68^{\tabularrel+0.11}_{\tabularrel-0.09}$ \\
$\log_{10}U(\rm{Ne3O2}) ^{c}$ & $-2.48\pm0.1$ \\
$\log_{10}U(\rm{O32}) ^{c}$ & $-2.54\pm0.01$ \\
\\
    \end{tabular}
    \caption{Rest-frame optical properties of the median stack of 78 [OIII] emitters in GOODS-North (FRESCO+CONGRESS, see Fig. \ref{fig:stack_GN_congress_fresco}). Line ratio diagnostics: $^{a}$ \citet[][]{Bian2018}{}{} $^{b}$ \citet[][]{Nakajima2022}{}{} $^{c}$ \citet[][]{Witstok2021}{}{}}
    \label{tab:properties_of_stack}
\end{table}

\subsection{Completeness of the line emitter search}
\label{sec:completeness}

Understanding the purity and completeness of our [\ion{O}{iii}] search is crucial to accurately constrain number densities, luminosity functions and scaling relations for our whole sample (or subsets thereof). In order to assess the completeness of our search, we use the mock emitters inspected by the inspectors. The mock emitters are created using a grid of [\ion{O}{iii}] fluxes at various redshifts and then inserted at the location of sources which passed our photometric cuts but did not present any emission line. Specifically, we created $168$ mock emitters with redshifts spanning $6.8<z<9.0$ in steps of $\Delta z = 0.2$ (12 bins), [\ion{O}{iii}] fluxes $-18.5 < \log f_{\rm{ [\ion{O}{iii}] 5008}}/ [\rm{erg\ s}^{-1}\rm{\ cm}^{-2}] <-16.5$ (in steps of $0.14$ dex - 14 bins). For each of the resulting $168$ mock emitters, we assigned H$\beta$ fluxes by randomly sampling the range $-1 < \log \rm{ [\ion{O}{iii}]/H\beta} < 0.5$ ratios (in steps of $0.3$). The mock emitters have an [\ion{O}{iii}] FWHM of $100\ \rm{km\ s}^{-1}$. The Gaussian-matched filtering performance was found to be constant when choosing fiducial FWHMs for the kernels between $50-200\ \rm{km\ s}^{-1}$. Half of the emitters were inserted in the GOODS-South grism mosaic and the other half in the GOODS-North mosaic. From the selection function of these emitters by the Gaussian-matched filter and the visual inspection, we can compute the completeness function of our [\ion{O}{iii}] emitter search for all potential redshifts and line ratios.  

\begin{figure}
    \centering
    \includegraphics[width=0.46\textwidth]{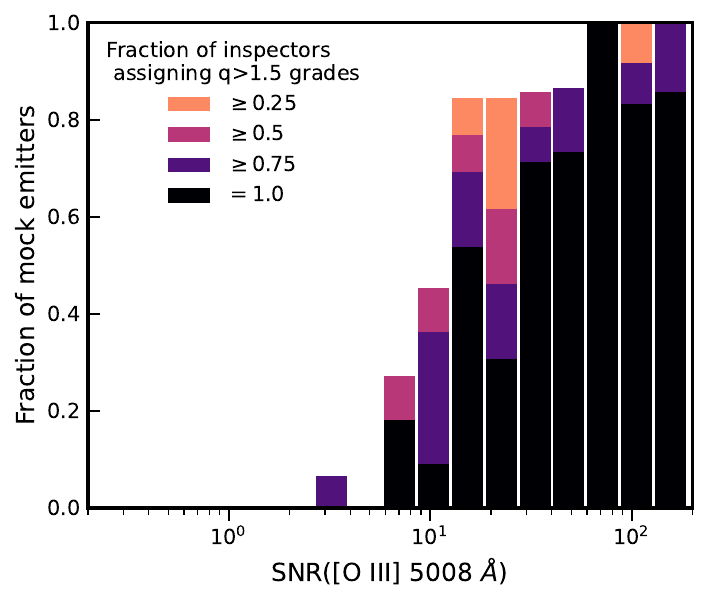}
    \caption{Fraction of mock emitters for which a weak or strong consensus is reached during visual inspection by the team as a function of [\ion{O}{iii}] $5008$ \AA \ SNR. Whilst a majority ($>75\%$) of objects above a SNR$\gtrsim10$ are identified as [\ion{O}{iii}] emitters by a majority  of the team, unanimity on all objects is only reached at SNR$\gtrsim 50$. This is indicative of a conservative visual inspection approach where alignment, morphology and contamination play an important role besides the emission line.}
    \label{fig:agreement-checkers}
\end{figure}
\begin{figure}
    \centering
    \includegraphics[width=0.46\textwidth]{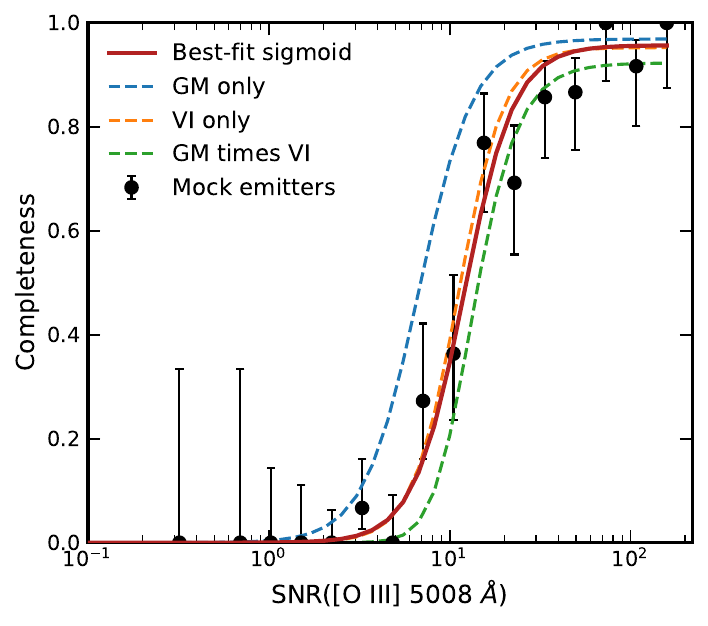}
    \caption{Combined completeness of the Gaussian-matched filtering and the visual inspection. The best-fit completeness (with a fixed upper end point $C=1$) is shown in dark red. We also show the completeness for either step, as well as the multiplication of both completeness functions, which is significantly lower than the combined one.}
    \label{fig:completeness_vis_and_GM}
\end{figure}

We first quantify the level of agreement between different team members in Fig. \ref{fig:agreement-checkers}. The majority ($>80\%$) of objects with a SNR([\ion{O}{iii}] $5008$)$>10$ are identified as real by a sizeable fraction of the team ($\geq 50\%$). This high SNR threshold can be understood as the identification of the doublet relying mostly on the SNR of the [\ion{O}{iii}] line $4960$, which is 2.98 times fainter than the $5008$ \AA \ line. Majority consensus on the nature of an object is thus reached at SNR([\ion{O}{iii}] $4960$)$\gtrsim 3.4$.

We compute the end-to-end completeness function by combining the mock emitters recovered by the Gaussian-matched (GM) filtering algorithm and the visual inspection (VI). We show the combined completeness on Fig. \ref{fig:completeness_vis_and_GM}. The separate completeness functions for each step are detailed in Appendix \ref{app:completeness}. We find that the completeness of the [\ion{O}{iii}] emitters is not simply the multiplication of the two completeness functions for each step, which indicates significant correlation between the set of emitters selected by the algorithm and the visual inspection. The completeness used throughout this work is the best-fit function 
\begin{equation}
\label{eq:completeness}
    C_{\rm{VI}\odot\rm{GM}}(x) =\frac{ 0.96\pm 0.03}{1+\exp{-(7.20\pm1.3)[x-(1.08\pm0.04)]}} \rm{\ \ \ \ ,}
\end{equation}
where $x=\log_{10}$SNR$([\ion{O}{iii}]\ 5008)$. The total completeness of the line selection is thus $21\%,48\%,68\%$ at SNR$([\ion{O}{iii}]\ 5008)=8,12,16$. The completeness in Equation \ref{eq:completeness} has been calculated as a function of the [\ion{O}{iii}] $5008$ SNR. We are then able to determine the completeness of any location and wavelength in the FRESCO data for a given line flux by combining the empirical 3D noise cube (see Appendix \ref{app:rms}) and the completeness function (which is a function of SNR).

\subsection{Purity of the line emitter search}
\label{sec:purity}

Having characterised the completeness of our line emitter search in detail in the previous section, we now turn to the purity of the line emitter catalogue. Given that the visual inspection removes obvious contaminants such as median-filtering residuals and contamination by neighbouring sources, the remaining contaminants are mostly dominated by noise fluctuations in the 1D extractions (including correlated noise due to the median filtering).  We therefore run the Gaussian-matched filtering process on the ´inverted' or ´negative' spectra (i.e. $f' = -f_\lambda$) for sources where no lines have been detected (the same as those used to create mock emitters), which by construction only contain noise. We use 6451 spectra extracted and inverted for sources passing the HST non-detection selection (see Section \ref{sec:dropout}) but without any significant emission features, and find no emitters satisfying the SNR cuts after the Gaussian-matched filtering. We thus conclude that our initial LBG+doublet selection is pure.

The remaining type of contaminant is the misidentification of a faint [\ion{O}{iii}] doublet with a single line (real, but at a different redshift than expected) and a noisy feature mimicking the fainter $4960$ \AA\ line. In this scenario, we estimate the purity by simply computing the probability of the [\ion{O}{iii}] 4960 being due to noise using the measured SNR of the line (e.g. $2.27\% (0.13\%)$ for a SNR$=2(3)$ line, and so forth). As demonstrated above however (see Fig. \ref{fig:agreement-checkers}), most objects have $SNR(4960)>2$, and the contamination rate in our whole sample can be estimated to $<3.1$ objects. We note that, due to the HST non-detection performed on the imaging catalogue, most of these contaminants (noise+single line) should be H$\alpha$ emitters at a slightly lower redshift. It is also expected that H$\alpha$ emitters will suffer from the same problem: low-SNR H$\alpha$ lines could be mimicked by a higher redshift [\ion{O}{iii}] 5008 line, with the 4960 feature too faint to be detected or suppressed by noise.

\subsection{Comparison with photometric redshift estimates}
\label{app:quality-checks}
\begin{figure}
    \centering
    \includegraphics[width=0.48\textwidth]{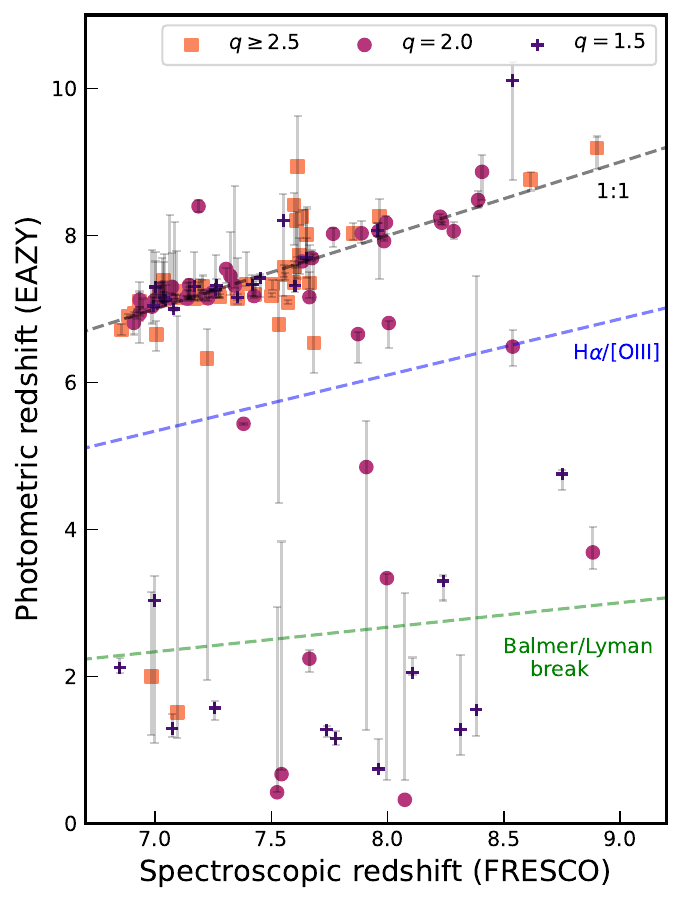}
    \caption{Spectroscopic redshift (FRESCO) against the maximum likelihood photometric redshift (\texttt{EAZY}) for the sources in the final catalogue, indicated with symbols depending on the quality of the source. We also plot as an errorbar the 16-84th percentiles of the \texttt{EAZY} redshift posterior. Coloured lines highlight from top to bottom; the 1:1 line, the H$\alpha$/[\ion{O}{iii}] ambiguity line, the Lyman/Balmer break confusion. }
    \label{fig:specz_photz}
\end{figure}

\begin{figure}
    \centering
    \includegraphics[width=0.45\textwidth]{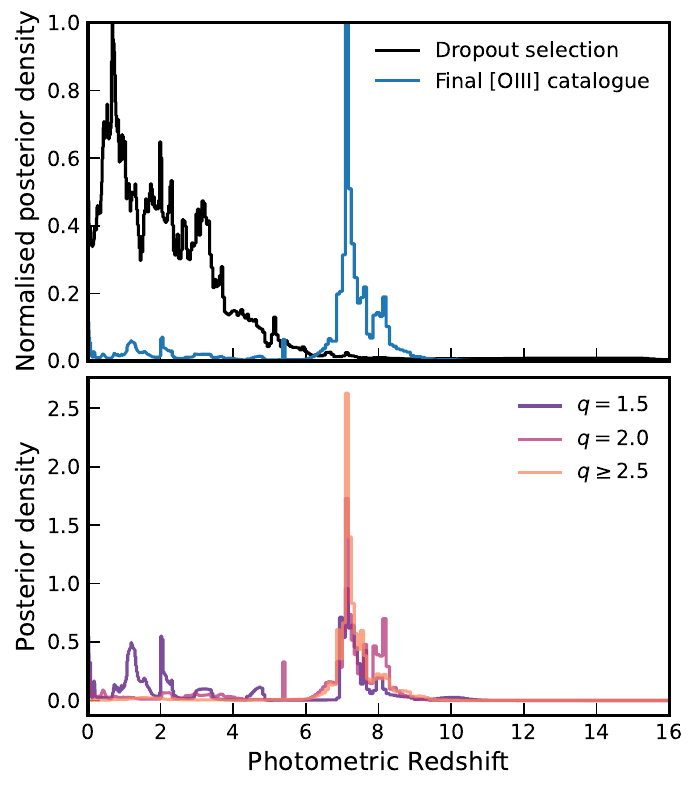}
    \caption{EAZY posterior photometric redshift distributions for different subsamples at various stages of the selection process (see further Table \ref{table:search_steps} and text). \textit{Top panel:} Sample photometric redshift distribution for the Lyman-break selected sources (black) and the final FRESCO [\ion{O}{iii}] sample (blue). The HST-dropout sample is dominated by faint objects undetected in blue HST bands and detected in F444W.  \textit{Lower panel:} Sample photometric redshift distribution for subsamples of the full catalogue divided by quality (see further text).  }
    \label{fig:photz_distribution}
\end{figure}

This work leverages JWST grism observations to obtain an unbiased census of high-redshift galaxies in the GOODS, in contrast with spectroscopic follow-up campaigns of objects selected based on their photometry. We now compare the photometric redshift and spectroscopic redshifts of our sources in order to evaluate the accuracy of photometric redshifts and to determine whether our method can recover $z>6.8$ sources classified as intermediate redshift based on photometry only. The photometric redshifts were obtained using \texttt{EAZY} with the photometric catalogues described in Section \ref{sec:obs_analysis}, which we stress were not used in the selection of our final sample. To compute the \texttt{EAZY} redshift, we used the \emph{agn\_blue\_sfhz} templates which include an AGN template as well as bluer templates tailored to young star-forming high redshift galaxies and run \texttt{EAZY} using standard parameters.

We first show the spectroscopic redshift against the photometric redshift estimates for all individual sources in the catalogue in Figure \ref{fig:specz_photz}. Overall, we find good agreement between the \texttt{EAZY} redshift and the spectroscopic redshifts, especially for high-confidence objects ($q\geq2.5$). Unsurprisingly, we find that objects with large spectroscopic versus photometric redshift differences are preferentially those with the lowest F182M or F444W fluxes in the sample. This indicates either fainter objects whose photometric redshifts are more uncertain, and/or objects with fainter emission lines for which the visual inspection quality is unclear (as seen in Fig. \ref{fig:specz_photz}). Outliers can be broadly separated into two categories: ambiguity between H$\alpha$ and [\ion{O}{iii}] as discussed previously, and objects for which \texttt{EAZY} likely assigned the Lyman-Break to be a Balmer break instead. The latter objects often have large uncertainties associated to their photometric redshifts. Additionally, the number of filters and depth of \textit{HST}/\textit{JWST} imaging data, and therefore the photometric redshift uncertainties, vary significantly across the FRESCO footprint. The recovery of objects with such large photometric redshift uncertainties is one of the goals of our selection. 

We find that that the overall fraction of galaxies with spectroscopic redshift within the 16-84 percentile range of the \texttt{EAZY} photometric redshift is $0.29_{\tabularrel-0.04}^{\tabularrel+0.04}$, e.g. well below the $0.68$ fraction expected. This ratio varies only modestly with the visual inspection quality flag (and thus luminosity). We find a fraction of $0.36_{\tabularrel-0.06}^{\tabularrel+0.07}$ at $q\geq2.5$, $0.26_{\tabularrel-0.06}^{\tabularrel+0.07}$ at $q=2$ and $0.21_{\tabularrel-0.07}^{\tabularrel+0.08}$ at $q=1.5$. This simple result, combined with the relatively low fraction of catastrophic outliers in Fig. \ref{fig:specz_photz}, suggests that the posterior photometric redshift distributions are likely too narrow for high-redshift galaxies in our sample.

We further show the stacked \texttt{EAZY} photometric redshift posterior of our sample at various stages of the selection in Figure \ref{fig:photz_distribution}. In the top panel, we clearly see that the optical non-detection selection on the \textit{HST}/\textit{JWST} photometry mostly selects low-redshift objects. The final sample however has a strongly peaked distribution around the spectroscopic redshift range probed by FRESCO (lower panel). [\ion{O}{iii}] emitters with the lowest quality assigned at the visual inspection stage have higher likelihood to be at lower redshift. This is expected because such objects are often fainter, and thus the photometric redshift posterior is often broader and less narrowly peaked. 

Overall this analysis demonstrates the ability of our method to recover high-redshift galaxy samples in good agreement with photo-z-selected samples on the population level, but without the inherent and intractable biases associated with the latter method.

\subsection{Comparison with other published catalogues}
\label{ref:comparison_litterature}

The public JADES data releases provide MSA PRISM and grating spectroscopy for a large number of sources in GOODS-South and GOODS-North  \citep[][]{Bunker2023_JADESrelease, DEugenio2024_JADESrelease3}{}{}. We find a total of $11(12)$ matches with our catalogue in GS(GN) using a maximum separation of $1"$, and find that 21 sources have the same redshifts within $\Delta z \lesssim 0.005$. Two additional sources have no redshifts in the JADES release, but we identify them with a quality flag $q=2.5$ (JADES NIRSpec IDs 4613,41905, respectively FRESCO GS-12363 and GS-4357). This \textit{a posteriori} comparison with the much higher sensitivity NIRSpec spectroscopy underscores the high purity and quality of our [\ion{O}{iii}] search. 

\citet[][]{Helton2023}{}{} use FRESCO and JADES to study an overdensity at $z=5.4$ in GOODS-South.
We find only one match, GS-23077. For this source \citet[][ID 70]{Helton2023}{}{} report a spectroscopic redshift of $z=5.392$, whereas we find instead $z=7.38$ with $q=2$. The strongest line identified is the same, but we interpret it as [\ion{O}{iii}] 5008 rather than H$\alpha$ at lower redshift. We note that our \texttt{EAZY} redshift for this source is $z=5.438^{\tabularrel+0.010}_{\tabularrel-0.010}$.

\section{Redshift distribution, overdensities and luminosity functions of the [\ion{O}{iii}] emitters}
\label{sec:results}
\begin{figure}
    \centering
    \includegraphics[width=0.5\textwidth,trim = 0 0 0cm 0, clip]{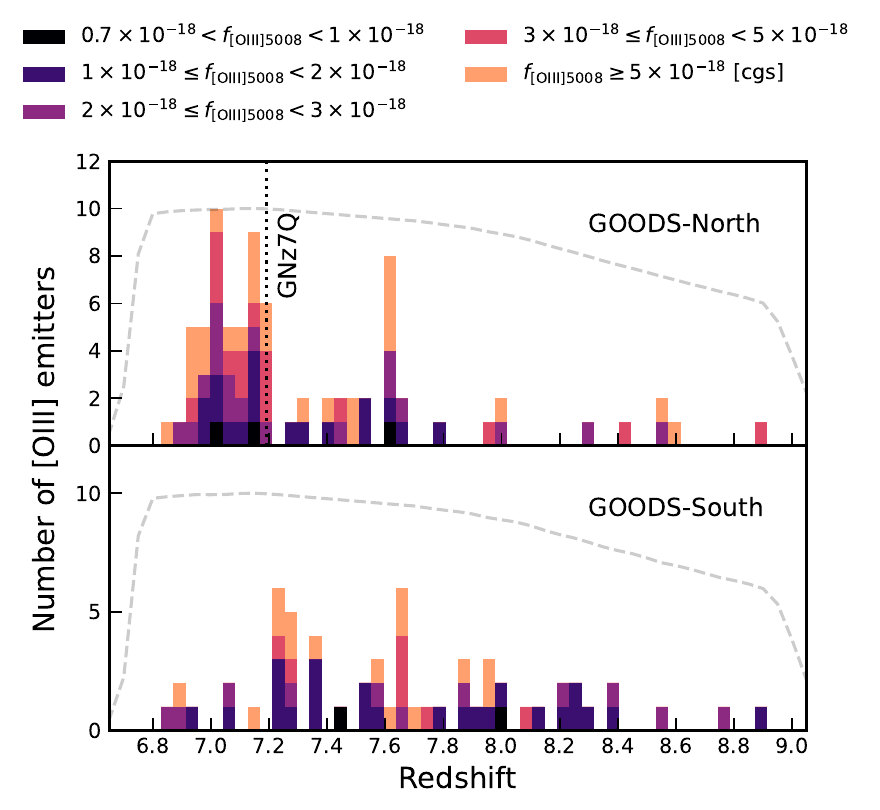}
    \caption{Redshift distribution of the FRESCO [\ion{O}{iii}] line emitting systems in the GOODS-North and South fields. The emitters are further subdivided by the strength of the [\ion{O}{iii}] 5008 line (coloured shades). The (relative) effective volume surveyed as a function of redshift is indicated with a gray dotted line. The presence of overdensities in both fields is immediately clear. An excess of galaxies is seen in GOODS-N at $z\sim7-7.2$, which might be correlated with the presence of a faint quasar at $z=7.19$ \citep[GNz7Q, ][]{Fujimoto2022}{}{}. }
    \label{fig:redshift_distribution}
\end{figure}

\begin{figure*}
    \centering
    \includegraphics[width=0.96\textwidth]{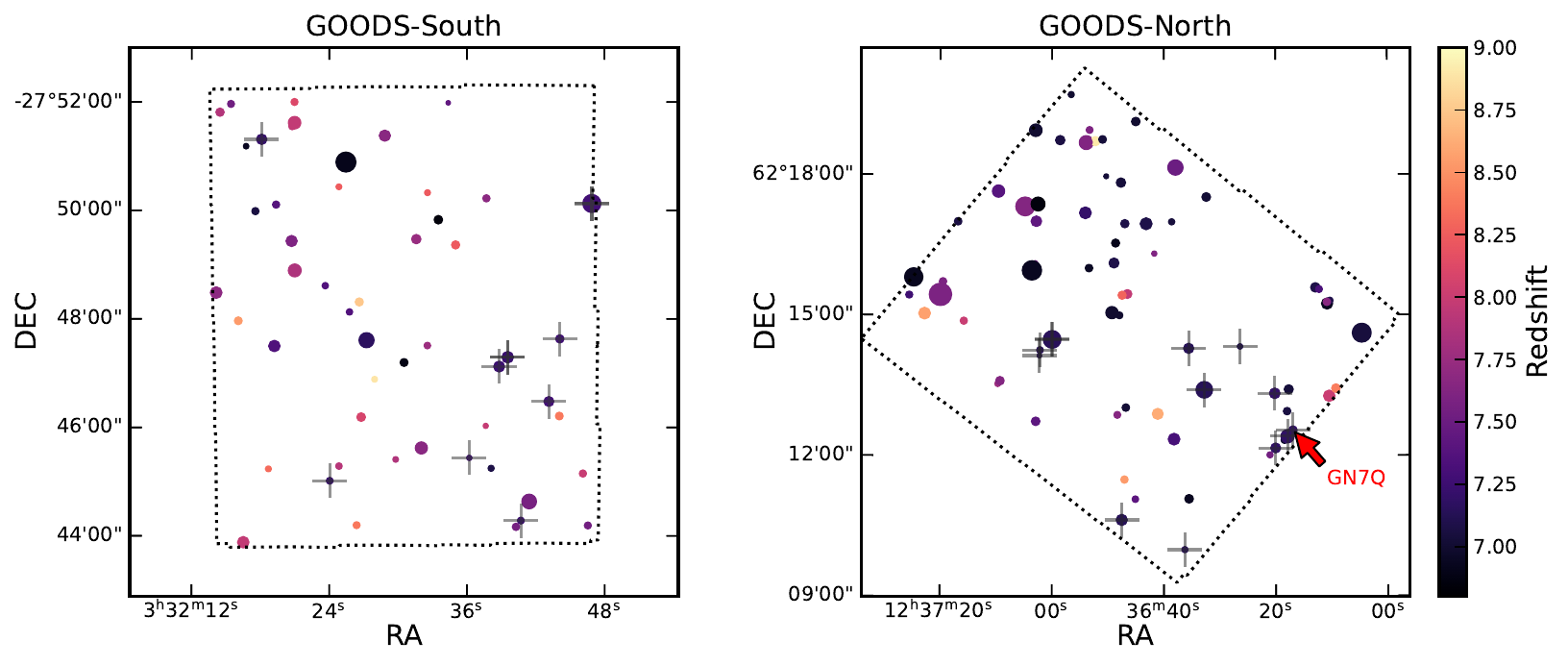}
    \caption{Distribution of [\ion{O}{iii}] emitters in the GOODS-North and GOODS-South fields. The colour of the points indicates the redshift of the galaxy, whereas the size of the circles increases with the measured [\ion{O}{iii}] $5008$ flux. In GOODS-North, we have highlighted objects at  $7.1<z<7.2$ with gray crosses. Similarly, in GOODS-South we highlight the members of the overdensity at $7.2<z<7.3$. Clearly, both overdensities are not only delimited in redshift space but also spatially. The FRESCO footprint is indicated in each field in dotted lines.}
    \label{fig:sky_distribution}
\end{figure*}
\subsection{Redshift distribution and overdensities}

We show the redshift distribution of the emitters in Fig. \ref{fig:redshift_distribution} as well as their distribution on the sky in Fig. \ref{fig:sky_distribution}. Interestingly, the redshift distributions between GOODS-North and GOODS-South are strikingly different. In GOODS-North we find a large number of sources at $6.9\lesssim z\lesssim 7.2$, potentially clustered with the $z=7.19$ faint quasar discovered by \citet[][]{Fujimoto2022}{}{}. Note that this object is detected with a quality of 2 and an [\ion{O}{iii}] redshift of $z=7.1868$  (GN-27541 in our catalogue). We find another over-density at $z\simeq7.6$ in GOODS-North. In GOODS-South we only find two smaller overdensities at $z=7.23-7.26$ and $z= 7.65$. We show the distribution of the FRESCO [\ion{O}{iii}] emitters on the sky in Figure \ref{fig:sky_distribution}, highlighting the GN $z=7.1-7.2$ and GS $z=7.2-7.3$ overdensities. Clearly, the overdensities are not simply located in redshift-space but also in the sky-projected plane as emitters are clustered in one corner of (and probably extend beyond) the FRESCO footprint. This suggests that we are only seeing a fraction of more extended structures which must be investigated in 3D space and using larger NIRCam Grism mosaics.

The GN $z\simeq 7.0-7.2$ and GS $z=7.23-7.26$ overdensities were already presented in great detail in \citet[][]{Helton2023}{}{}. Interestingly, they do not report the GN $z\simeq 7.6$ and GS $z\simeq 7.65$ overdensities. The latter case can be explained as it is composed of $6$ galaxies grouped in $3$ systems, and thus does not qualify as an overdensity if the merging systems are counted as one object as in \citet[][]{Helton2023}{}{}. Similarly, the GN $z\simeq 7.6$ overdensity might not be clustered tightly enough to be detected by the Friends-of-Friends approach of \citet[][]{Helton2023}{}{}.

In total, the number of sources in GOODS-North is $\sim 1.3\times$ higher than in GOODS-South in the redshift range considered and we find that this discrepancy is solely explained by the GNz7Q-associated overdensity,
as we will discuss in the following section.

\subsection{The [\ion{O}{iii}] luminosity function at $6.8<z<9.0$}
\label{sec:O3_lf}
We now compute the [\ion{O}{iii}] luminosity function using the full catalogue of emitters and the completeness function (Eq. \label{eq:completeness}) presented in the previous sections. For each logarithmic bin in \rm{[\ion{O}{iii}]} $5008$ luminosity, we compute the number density of emitters using the \citet[][]{Schmidt1968} V$_\mathrm{max}$ estimator
\begin{equation}
    \Phi(L) \text{d}\log L = \Sigma_i \frac{1}{ V^{}_{\rm{Total}} C_{\rm{VI}+\rm{GM}} (f^i) C_{\rm{det}} (m_{\rm{det}}^i)}
\end{equation}
where the summation index $i$ runs over all sources in a given luminosity bin. The total completeness function $C_{\rm{VI}+\rm{GM}}$ is a function of SNR of the [\ion{O}{iii}] line (see Section \ref{sec:completeness}, Eq. \label{eq:completeness}), which is itself dependent on the detected flux ($f^i$) and the RMS at the position of the emitter and wavelength of the line (determined from our master RMS cube). We also include a detection completeness correction $C_{\rm{det}}$ determined using source injection in the F210M+F444W detection image, as detailed in Section \ref{sec:obs_analysis}. The best-fit completeness function (Eq. \label{eq:completeness}) parameters can be found in Appendix \ref{app:completeness}. The uncertainty budget is composed of three terms: the error on the image detection obtained by sampling the covariance matrix of the best-fit completeness function (see Appendix \ref{app:completeness}), the visual inspection completeness error obtained by resampling the line flux using the measured line flux error, and a Poisson noise term following the number of sources in each luminosity bin. We do not add a cosmic variance uncertainty to our error budget (see further below for a discussion of cosmic variance in the FRESCO fields).
 
\begin{figure*}
    \centering
    \includegraphics[width=0.8\textwidth]{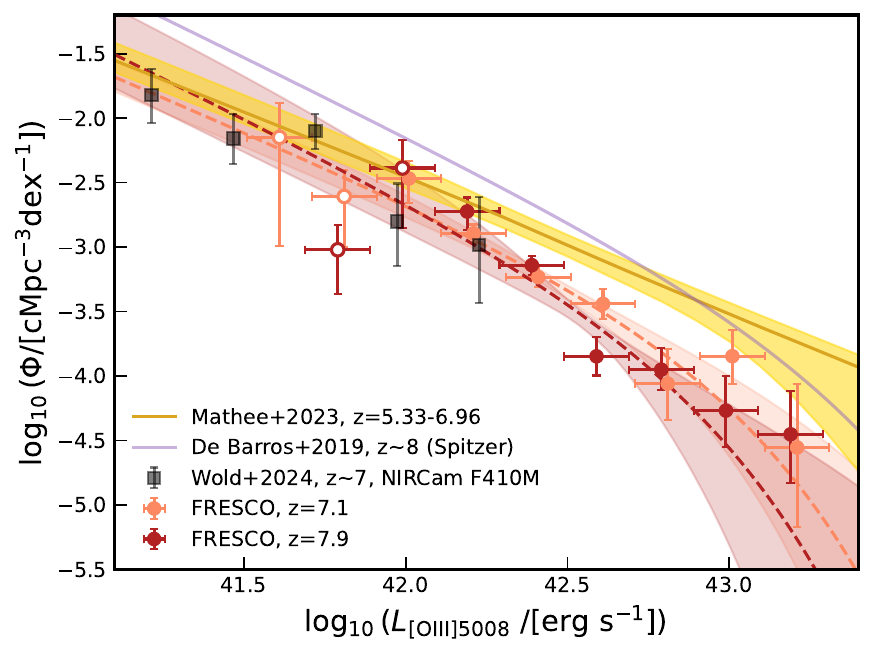}
    \caption{The evolution of the observed high-redshift [\ion{O}{iii}] luminosity function. The measurements at $z=7.1$ and $z=7.9$ from FRESCO are indicated in orange, dark red points alongside with their best-fit Schechter functions. The uncertainties include a contribution from the image detection and visual inspection completeness as well as a Poisson term. Luminosity bins with completeness $<25\%$ are shown with empty symbols, which are not used in the Schechter fits.  We also plot the $z\sim 8$ [\ion{O}{iii}] LF derived from Spiter/IRAC photometry \citep[][]{DeBarros2019}{}{} in purple, and the more recent $z\sim 7$ results using medium band NIRCam photometry from \citet[][]{Wold2024} in black squares. The \textit{JWST}/NIRCam WFSS constraints \citet[][]{Matthee2023_EIGER}{}{} at z=5.33-6.96 are shown with a yellow line and shaded area. Relative to the latter, our observations show a decline of the overall [\ion{O}{iii}] (luminosity) density at $z\gtrsim 6$.}
    \label{fig:O3_LF}
\end{figure*}

\begin{figure*}
    \centering
    \includegraphics[width=\textwidth]{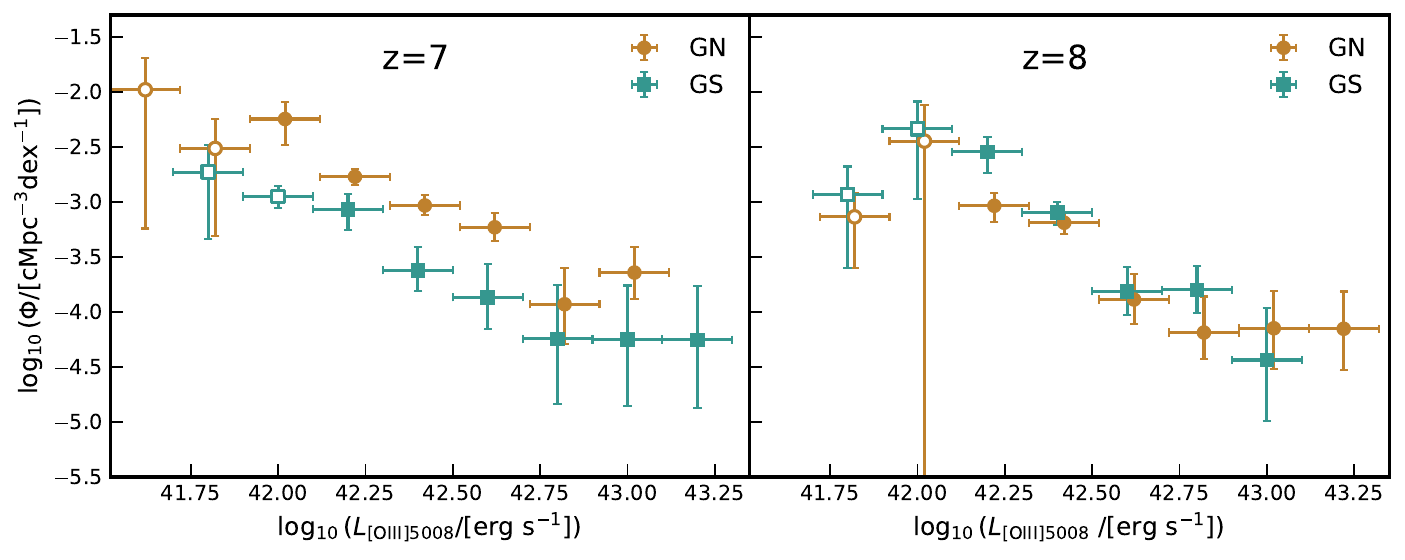}
    \caption{The [\ion{O}{iii}] luminosity function at $z\sim7$ and $z=8$ in GOODS-North and GOODS-South. Luminosity bins with completeness $<25\%$ are shown with empty symbols. The amplitude difference at $z=7$ can be explained by the large overdensity in GN at $z=7.0-7.2$. Nonetheless, the shape of the luminosity function is consistent between the two fields in the two redshift bins.}
    \label{fig:o3lf_differentfields}
\end{figure*}

For the purpose of computing the [\ion{O}{iii}] luminosity function (LF), we split our [\ion{O}{iii}] emitter catalogue in two redshifts bins: $6.75<z<7.5$ and $7.5<z<9.0$. The corresponding median redshift of the subsamples is $\Bar{z}=7.1$ and $\Bar{z}=7.9$. We fit the observed [\ion{O}{iii}] LF using a Schechter function with a slope fixed at the value measured for the UV LF at the same redshift $\alpha=-2.06(-2.23)$ at $z\simeq7(8)$ \citep[][]{Bouwens2021}. Datapoints with completeness $<25\%$ are not used in the fit. The posterior 50th percentile value and errors (16-84 percentiles) of the Schechter function parameters can be found in Table \ref{tab:o3_lf_parameters}, and the tabulated measurements and corrections of the FRESCO [\ion{O}{iii}] LF can be found in Table \ref{tab:o3_lf_values}.

We show the FRESCO [\ion{O}{iii}] LFs alongside other measurements from the literature in Fig. \ref{fig:O3_LF}. We find a decrease in the number density of [\ion{O}{iii}] emitters at $z>6$ when comparing our measurements to that of EIGER \citep[][]{Matthee2023_EIGER} at  $5.33<z<6.96$. The FRESCO and EIGER measurements reveal little evolution of the shape of the [\ion{O}{iii}] LF between $z=5$ and $z=8$. At face-value the [\ion{O}{iii}] LF is steeper at $z\geq7$ than $z=5-6$, but this might partially driven by the fixed LF slopes ($\alpha=-2.0$ at z=$5-6$ and $\alpha=-2.06(-2.23)$ at $z\simeq7(8)$ and/or an excess of bright source in the field studied by \citet[][]{Matthee2023_EIGER}{}{}. At the fainter end we find a good agreement with the recent results of \citet[][]{Wold2024} using NIRCam medium band photometry to constrain the [\ion{O}{iii}] LF at $z\sim 7$. Improved constraints at the bright- and faint-end of the [\ion{O}{iii}] LF ought to constrain a potential evolution of the LF slope in the future. We find that the [\ion{O}{iii}] LF from \citet[][]{DeBarros2019}{}{}, based on \textit{Spitzer}/IRAC photometry of high-redshift candidates overestimated the numbers of  [\ion{O}{iii}] emitters. Given that the \citet[][]{DeBarros2019}{}{} [\ion{O}{iii}] LF was inferred using the observed UV LF and the observed $L_{[\ion{O}{iii}]} - L_{\rm{UV}}$ relation, we investigate the $L_{[\ion{O}{iii}]} - L_{\rm{UV}}$ relation in Section \ref{sec:O3_UV}. 

Finally, we show the [\ion{O}{iii}] LFs in the GOODS-N and GOODS-S fields separately in Figure \ref{fig:o3lf_differentfields}. The [\ion{O}{iii}] LF shapes are consistent across the two fields and two redshift bins, and the number density is similar at $z=7.9$. At $z=7.1$ we find an offset in the number density between GOODS-N and GOODS-S, which we attribute to the overdensity at $7.0<z<7.2$ discussed previously. This overdensity introduces a constant offset of $\sim 0.45\ \rm{dex}$ (e.g. a factor $2.8$) between the two fields over the redshift range $6.8<z<7.5$. This value can be considered as a lower limit on cosmic variance or field-to-field variance of the [\ion{O}{iii}] luminosity function for similar cosmological volumes at high-redshift. The amplitude of the field-to-field variation is roughly consistent with the results of \citet[][]{Bouwens2015}{}{} who finds the normalisation of the UVLF in GN to be $\sim 2\times$ that of GS at the redshift of interest $z\simeq 7-8$ using \textit{HST} imaging. More generally, the JWST CANUCS survey also find differences of $\sim 2\times$ in the UV number densities between several of their fields \citep[][]{Willott2023}{}{}. We note however that as cosmic variance scales with area and volume, we cannot directly compare with these previous studies. Instead, we use the estimator of \citet[][]{Ucci2021} to calculate the cosmic variance of the UVLF over the redshift range $6.8<z<7.5$ in a $64$ arcmin$^{2}$ field (corresponding to one of the FRESCO GOODS mosaics) using their `photoionisation' model. The predicted cosmic variance increases from $25\%$ to $78\%$ for galaxies with $M_{\rm{UV}}=-17$ to $-21$. Assuming that the UVLF and [\ion{O}{iii}] LF cosmic variance are similar, and assuming the $z=7.1$ GOODS-South [OIII] LF is close to the cosmic median, the number of [\ion{O}{iii}] emitters in GOODS-N at $6.8<z<7.5$ would be a $\approx 4-6\sigma$ outlier \citep[see also][]{Helton2023}. This can be reduced if the median number density lies in between that observed by FRESCO in GOODS-North and South. Additionally, the assumption of similar cosmic variance for UV- and [\ion{O}{iii}]-selected galaxies might be incorrect given the large scatter in the [\ion{O}{iii}]/UV ratio (see Section \ref{sec:O3_UV}). Given the absence of any significant cosmic variance between the two fields at $z=8$ (Fig. \ref{fig:o3lf_differentfields}, we cannot immediately determine whether estimates for the UVLF are applicable to the [\ion{O}{iii}] LF and do not include these in our error budget for this study. Future studies using a large number of fields observed with JWST slitless spectroscopy and theoretical estimates are necessary to conclude on this matter.

\begin{table}
    \centering
    \setlength{\tabcolsep}{6pt} 
\renewcommand{\arraystretch}{1.3} 
    \begin{tabular}{c|c|c|c|c}
         Emitter sample & $\bar{z}$ & $N_{\rm{gal}}$ & $\log \phi^{*}$ & $\log L_{\rm{[\ion{O}{iii}] 5008}}^{*}$  \\ \hline
        $6.75<z<7.50$ & 7.1 & 72 & $-3.93_{\tabularrel-0.59 }^{\tabularrel+0.40}$ & $42.88^{\tabularrel+0.46}_{\tabularrel-0.27}$\\
        $7.50<z<9.0$ & 7.9 & 52 & $-3.92_{\tabularrel-1.36}^{\tabularrel+0.86}$ & $42.78^{\tabularrel+0.88}_{\tabularrel-0.41}$   \\
    \end{tabular}
    \caption{Parameter posterior results for the FRESCO $z=7.1$ and $z=7.9$ [\ion{O}{iii}] LF. We give the median value and the $16-84$ percentiles as errors. The Schechter slope $\alpha$ is fixed to the UVLF values $\alpha=-2.06(-2.23)$ at $z\simeq7(8)$ \citep[][]{Bouwens2021} as the data do not constrain the knee of the luminosity function.}
    \label{tab:o3_lf_parameters}
\end{table}

\begin{table*}
\centering
\caption{Number of [\ion{O}{iii}] emitters and number densities at $z=7.1$ (first part) and $z=7.9$ (second part) computed for GOODS-North, GOODS-South and the combined GOODS fields. We also indicate the average completeness (Eq. \label{eq:completeness}) for the [\ion{O}{iii}] emitters in each luminosity bin, which combines the detection completeness and that of the [\ion{O}{iii}] doublet search (see further Appendix \ref{app:completeness}). As these vary across the field and the [\ion{O}{iii}]/UV ratio scatter is large, the average completeness of a few sources can vary non-monotonically across contiguous luminosity bins.}
\setlength{\tabcolsep}{6pt} 
\renewcommand{\arraystretch}{1.3} 
\begin{tabular}{lcccccccccc}
& & \multicolumn{3}{c}{GOODS-North} & \multicolumn{3}{c}{GOODS-South}  & \multicolumn{3}{c}{Combined} \\
$z=7.1$ &$\log L_{\rm{[\ion{O}{iii}]}5008}$ & N & $<c>$ & $\log\Phi_{corr}$  & N & $<c>$ & $\log\Phi_{corr}$ & N & $<c>$ & $\log\Phi_{corr}$ \\  \hline 
& 41.6 & 2 & 0.01 & $ -1.98^{\tabularrel+0.29}_{\tabularrel-1.26} $ & 1 & 0.01 & $ -2.43^{\tabularrel+0.46}_{\tabularrel-10} $ & 3 & 0.01 & $ -2.15^{\tabularrel+0.27}_{\tabularrel-0.84} $ \\ 
& 41.8 & 1 & 0.02 & $ -2.51^{\tabularrel+0.26}_{\tabularrel-0.80} $ & 2 & 0.08 & $ -2.73^{\tabularrel+0.24}_{\tabularrel-0.61} $ & 3 & 0.06 & $ -2.61^{\tabularrel+0.20}_{\tabularrel-0.39} $ \\ 
& 42.0 & 10 & 0.26 & $ -2.25^{\tabularrel+0.15}_{\tabularrel-0.24} $ & 6 & 0.43 & $ -2.95^{\tabularrel+0.09}_{\tabularrel-0.10} $ & 16 & 0.32 & $ -2.47^{\tabularrel+0.13}_{\tabularrel-0.19} $ \\ 
& 42.2 & 11 & 0.44 & $ -2.77^{\tabularrel+0.07}_{\tabularrel-0.08} $ & 4 & 0.41 & $ -3.07^{\tabularrel+0.14}_{\tabularrel-0.18} $ & 15 & 0.43 & $ -2.89^{\tabularrel+0.07}_{\tabularrel-0.07} $ \\ 
& 42.4 & 11 & 0.64 & $ -3.03^{\tabularrel+0.10}_{\tabularrel-0.09} $ & 3 & 0.68 & $ -3.62^{\tabularrel+0.21}_{\tabularrel-0.19} $ & 14 & 0.65 & $ -3.23^{\tabularrel+0.08}_{\tabularrel-0.08} $ \\ 
& 42.6 & 9 & 0.78 & $ -3.23^{\tabularrel+0.13}_{\tabularrel-0.13} $ & 2 & 0.75 & $ -3.87^{\tabularrel+0.30}_{\tabularrel-0.29} $ & 11 & 0.77 & $ -3.44^{\tabularrel+0.12}_{\tabularrel-0.11} $ \\ 
& 42.8 & 2 & 0.86 & $ -3.93^{\tabularrel+0.33}_{\tabularrel-0.36} $ & 1 & 0.89 & $ -4.24^{\tabularrel+0.49}_{\tabularrel-0.59} $ & 3 & 0.87 & $ -4.06^{\tabularrel+0.27}_{\tabularrel-0.28} $ \\ 
& 43.0 & 4 & 0.88 & $ -3.64^{\tabularrel+0.23}_{\tabularrel-0.24} $ & 1 & 0.90 & $ -4.25^{\tabularrel+0.49}_{\tabularrel-0.61} $ & 5 & 0.89 & $ -3.85^{\tabularrel+0.21}_{\tabularrel-0.21} $ \\ 
& 43.2 & 0 & -- & -- & 1 & 0.91 & $ -4.25^{\tabularrel+0.49}_{\tabularrel-0.62} $ & 1 & 0.91 & $ -4.55^{\tabularrel+0.49}_{\tabularrel-0.62} $ \\ 


\hline
$z=7.9$ &$\log L_{\rm{[\ion{O}{iii}]}5008}$ & N & $<c>$ & $\log\Phi_{corr}$  & N & $<c>$ & $\log\Phi_{corr}$ & N & $<c>$ & $\log\Phi_{corr}$ \\  \hline 
& 41.8 & 1 & 0.04 & $ -3.13^{\tabularrel+0.22}_{\tabularrel-0.46} $ & 1 & 0.03 & $ -2.93^{\tabularrel+0.25}_{\tabularrel-0.67} $ & 2 & 0.03 & $ -3.02^{\tabularrel+0.19}_{\tabularrel-0.34} $ \\ 
& 42.0 & 4 & 0.23 & $ -2.45^{\tabularrel+0.33}_{\tabularrel-10.0} $ & 7 & 0.21 & $ -2.33^{\tabularrel+0.25}_{\tabularrel-0.64} $ & 11 & 0.22 & $ -2.39^{\tabularrel+0.22}_{\tabularrel-0.47} $ \\ 
& 42.2 & 5 & 0.34 & $ -3.03^{\tabularrel+0.11}_{\tabularrel-0.15} $ & 7 & 0.38 & $ -2.54^{\tabularrel+0.13}_{\tabularrel-0.20} $ & 12 & 0.36 & $ -2.72^{\tabularrel+0.11}_{\tabularrel-0.14} $ \\ 
& 42.4 & 4 & 0.23 & $ -3.19^{\tabularrel+0.10}_{\tabularrel-0.11} $ & 7 & 0.49 & $ -3.10^{\tabularrel+0.10}_{\tabularrel-0.11} $ & 11 & 0.40 & $ -3.14^{\tabularrel+0.07}_{\tabularrel-0.08} $ \\ 
& 42.6 & 3 & 0.75 & $ -3.89^{\tabularrel+0.23}_{\tabularrel-0.22} $ & 4 & 0.82 & $ -3.81^{\tabularrel+0.22}_{\tabularrel-0.22} $ & 7 & 0.79 & $ -3.85^{\tabularrel+0.15}_{\tabularrel-0.15} $ \\ 
& 42.8 & 1 & 0.48 & $ -4.19^{\tabularrel+0.33}_{\tabularrel-0.24} $ & 4 & 0.79 & $ -3.80^{\tabularrel+0.21}_{\tabularrel-0.21} $ & 5 & 0.73 & $ -3.95^{\tabularrel+0.17}_{\tabularrel-0.16} $ \\ 
& 43.0 & 2 & 0.88 & $ -4.15^{\tabularrel+0.34}_{\tabularrel-0.37} $ & 1 & 0.86 & $ -4.44^{\tabularrel+0.48}_{\tabularrel-0.55} $ & 3 & 0.87 & $ -4.27^{\tabularrel+0.27}_{\tabularrel-0.28} $ \\ 
& 43.2 & 2 & 0.89 & $ -4.15^{\tabularrel+0.34}_{\tabularrel-0.38} $ & 0 & -- & --  & 2 & 0.89 & $ -4.45^{\tabularrel+0.34}_{\tabularrel-0.38} $ \\ 

\end{tabular} \label{tab:o3_lf_values}
\end{table*}

\subsection{The UV luminosity function of spectroscopically-confirmed [\ion{O}{iii}] emitters}
\label{sec:O3_UV}

\begin{figure}
    \centering
    \includegraphics[width=0.49\textwidth]{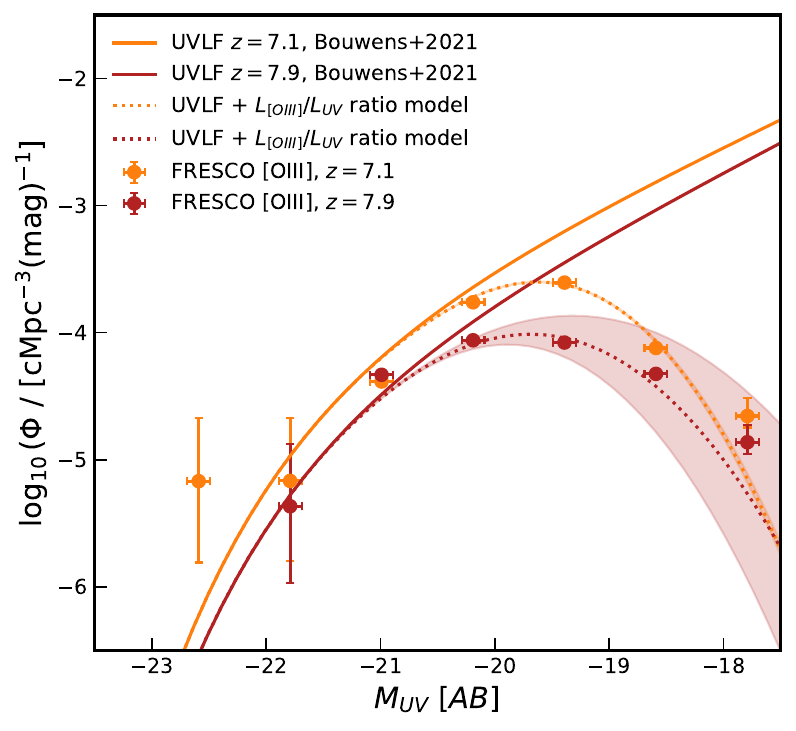}
    \caption{The spectroscopically confirmed UV luminosity function of FRESCO [\ion{O}{iii}] emitters. We show the measurements at $z=7.1$ and $z=7.9$ (circles and errorbars), as well as the UVLF interpolated to $z=7.1$ and $z=7.9$ \citep[full lines][]{Bouwens2021}{}{}. The bright end of the UV LF, where all galaxies are detected by FRESCO in [\ion{O}{iii}], is consistent with the photometric UVLF. The decline at $M_{\rm{UV}}$ is due to the sensitivity of FRESCO and the scatter in the  $L_{[\ion{O}{iii}]}/L_{\rm{UV}}$ ratio, which we can fit to infer the $L_{[\ion{O}{iii}]}/L_{\rm{UV}}$ ratio distribution (dotted lines and shaded areas). }
    \label{fig:O3_UVLF}
\end{figure}

Our unbiased, line-flux-limited sample of $z>6.8$ galaxies enables us to revisit the UV luminosity function of high-redshift galaxies, which is typically computed using photometrically-selected objects \citep[e.g.][]{Ellis2013,Oesch2014, Bouwens2015,Livermore2017,Atek2018, Bouwens2021}. For simplicity, we compare our results only to \citet[][]{Bouwens2021}{}{} as they use the largest sample of $z=6-8$ galaxies selected from \textit{HST} imaging, and provide fits to the UVLF parameters which we can interpolate at the median redshift of our [\ion{O}{iii}] sample. To compute the UV LF of the [\ion{O}{iii}] emitters we only consider the detection completeness function and do not correct for the [\ion{O}{iii}] line detection completeness as the $L_{[\ion{O}{iii}]}/L_{\rm{UV}}$ ratio distribution is uncertain. We show the spectroscopically-confirmed UV LF against the HST photometric-only UV LF in Figure \ref{fig:O3_UVLF} and give the datapoints for reference in Table \ref{tab:UVLF_O3}. We find excellent agreement at the bright end ($M_{\rm{UV}}\lesssim-20.5$). This is an important validation of previous decades of work using \textit{HST} to measure the number densities of bright $z>6$ galaxies. 

\begin{table}
\centering
\caption{Number of [\ion{O}{iii}] emitters and number densities at $z=7.1$ (first part) and $z=7.9$ (second part) as a function of rest-frame UV magnitude for the combined GOODS fields, as shown in Fig. \ref{fig:O3_UVLF}. Note that the number densities are only corrected for the imaging detection completeness as we aim to model the [\ion{O}{iii}]-UV ratio scatter (see further text). Cosmic variance is not added in the error budget.}
\setlength{\tabcolsep}{6pt} 
\renewcommand{\arraystretch}{1.3} 
\begin{tabular}{llrr}
$z=7.1$ & $M_{\rm{UV}}$ & $N$ & $log\Phi_{corr}$ \\ \hline 
& -17.8  & 2 & $ -4.656^{\tabularrel+0.139}_{\tabularrel-0.090} $ \\ 
& -18.6 & 9 & $ -4.121^{\tabularrel+0.017}_{\tabularrel-0.014} $ \\ 
& -19.4 & 33 & $ -3.606^{\tabularrel+0.003}_{\tabularrel-0.003} $ \\ 
& -20.2 & 25 & $ -3.761^{\tabularrel+0.005}_{\tabularrel-0.004} $ \\ 
& -21.0 & 6 & $ -4.385^{\tabularrel+0.038}_{\tabularrel-0.028} $ \\ 
& -21.8 & 1 & $ -5.167^{\tabularrel+0.496}_{\tabularrel-0.634} $ \\ 
& -22.6 & 1 & $ -5.169^{\tabularrel+0.497}_{\tabularrel-0.640} $ \\ \hline
$z=7.9$  & $M_{\rm{UV}}$ & $N$  & $log\Phi_{corr}$ \\  \hline 
& -17.8 & 2 & $ -4.863^{\tabularrel+0.139}_{\tabularrel-0.090} $ \\ 
& -18.6 & 9 & $ -4.325^{\tabularrel+0.017}_{\tabularrel-0.013} $ \\ 
& -19.4 & 17 & $ -4.078^{\tabularrel+0.007}_{\tabularrel-0.006} $ \\ 
& -20.2 & 20 & $ -4.060^{\tabularrel+0.006}_{\tabularrel-0.005} $ \\ 
& -21.0 & 11 & $ -4.332^{\tabularrel+0.015}_{\tabularrel-0.012} $ \\ 
& -21.8 & 1 & $ -5.367^{\tabularrel+0.489}_{\tabularrel-0.601} $ \\
    \end{tabular}\label{tab:UVLF_O3} 
\end{table}

\begin{figure}
    \centering
    \includegraphics[width=1\linewidth]{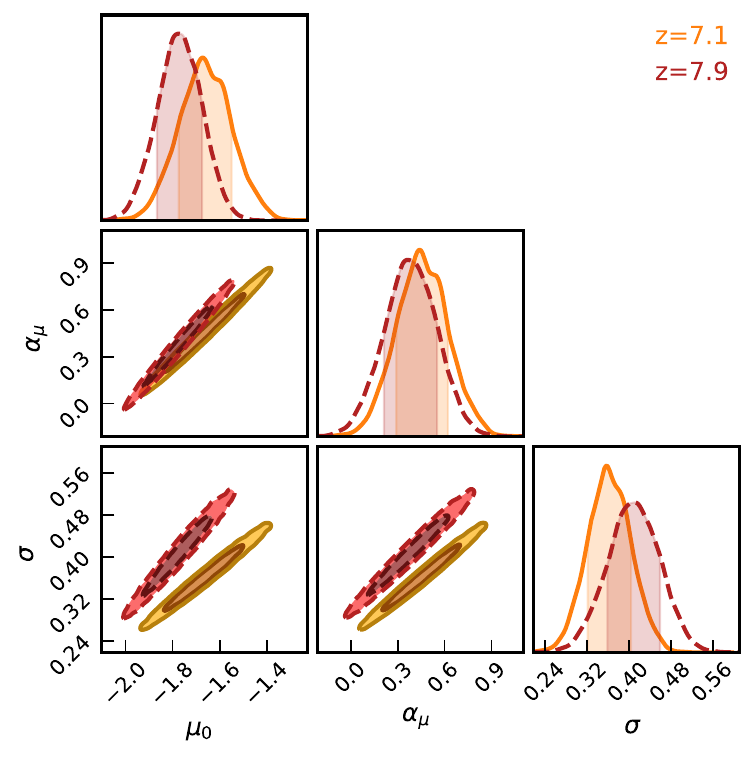}
    \caption{Posterior distribution of the three parameters for our [\ion{O}{III}]/UV ratio model at $z=7.1$ and $z=7.9$ (see further Section \ref{sec:O3_UV}). The 2D contours show the 1- and 2-$\sigma$ level. The three quantities are highly correlated as expected, but show clear evidence for a positive slope $\alpha_\mu$ (e.g. higher [\ion{O}{III}]/UV ratios at higher UV luminosities).  }
    \label{fig:posterior_oiii_uv_params}
\end{figure}

At the faint end however ($M_{\rm{UV}}\gtrsim-20.5$), the observed number density drops, which can easily be explained by the $L_{[\ion{O}{iii}]}/L_{\rm{UV}}$ scatter and the sensitivity limit of FRESCO. Using a constant limiting line sensitivity of $1\times10^{-18}\ \rm{erg\ s}^{-1}\ \rm{cm}^{-2}$, and assuming the \citet[][]{Bouwens2021}{}{} luminosity function is correct, we can simply fit for the $L_{[\ion{O}{iii}]}/L_{\rm{UV}}$ distribution necessary to reproduce the observed decline of the spec-z UVLF at $M_{\rm{UV}}\gtrsim-20.5$. We assume a model\footnote{We also tested a simple model with unique Gaussian distribution for the $\log_{10}(L_{[\ion{O}{iii}]}/L_{\rm{UV}})$ ratio at all luminosities, and find best-fit parameters for the mean and standard deviations $(\mu=-1.97\pm0.01, \sigma=0.25\pm0.02)$ at $z=7.1$, and $(\mu=-1.98\pm0.03, \sigma=0.30\pm0.03)$ at $z=7.9$. The predicted UVLF from this simple model is in excellent agreement with the observed spec-z UVLF. However, the inferred $L\_{[\ion{O}{iii}]}/L\_{\rm{UV}}$ ratio is in strong tension with the observed distribution at the bright end, where the sample is complete, thus motivating the use of the more complex model discussed in the main text.} where the mean of the $L_{[\ion{O}{iii}]}/L_{\rm{UV}}$ ratio evolve linearly with UV luminosity, matching the slow evolution seen in simulations (see next section), and with a fixed scatter around the mean. Specifically, we fit for the observed UVLF of [\ion{O}{iii}] emitters assuming that the  $L_{[\ion{O}{iii}]}/L_{\rm{UV}}$ ratio follows a Gaussian distribution $\mathcal{N(\mu, \sigma)}$ with $\sigma$ constant and $\mu=\mu_0+a(\log10(L_{\rm{UV}} / [L_\odot]) - L^0_{\rm{UV}})$. At the bright-end where the sample is complete in UV luminosity ($\log10(L_{\rm{UV}} / [L_\odot])>10.7, L^0_{\rm{UV}}=10.84)$, we measure a median ratio $\mu=-1.61\pm0.113$ (with errors computed using bootstrap resampling). We then use a Gaussian prior for $\mu_0$ using these values. Fitting the linear model we find the best-fit relation at $z=7.1$
\begin{equation}
\log_{10}(L_{[\ion{O}{iii}]}/L_{\rm{UV}}) = -1.66 + 0.45(\log_{10}L_{\rm{UV}}- 10.84) \pm 0.36  \\
\end{equation}
and at $z=7.9$
\begin{equation}
\log_{10}(L_{[\ion{O}{iii}]}/L_{\rm{UV}}) = -1.77 + 0.38(\log_{10}L_{\rm{UV}}- 10.84) \pm 0.41
\end{equation}
where $L_{\rm{UV}}$ in units of solar luminosities. We show the posterior distributions of the parameters ($\mu_0,\mu,\sigma$) in Figure \ref{fig:posterior_oiii_uv_params}. The three parameters are highly correlated, but the posterior is well-defined and clearly different from the prior (see above). We find evidence at the $2\sigma$ level for a positive $\alpha_\mu$ slope (e.g. higher [\ion{O}{III}]/UV ratios at higher UV luminosities) regardless of the other parameters values. 
We note that in principle, the scatter $\sigma$ could also evolve with luminosity, but the current observations do not provide enough statistical evidence to constrain this evolution. We find the same trend of increasing \citep[][]{Endsley2023}{}{} $L_{[\ion{O}{iii}]}/L_{\rm{UV}}$ ratio with UV luminosity. Additionally, the inferred $L_{[\ion{O}{iii}]}/L_{\rm{UV}}$ ratios are significantly lower at low luminosities than that used by \citet[][]{DeBarros2019}{}{} to convert the UVLF to the [\ion{O}{iii}] LF at $z\sim8$, explaining why their [\ion{O}{iii}] luminosity function is higher than our measurement (Fig. \ref{fig:O3_LF}).  

\subsection{Comparison to simulations}

\begin{figure*}
    \centering
    \includegraphics[width=\textwidth]{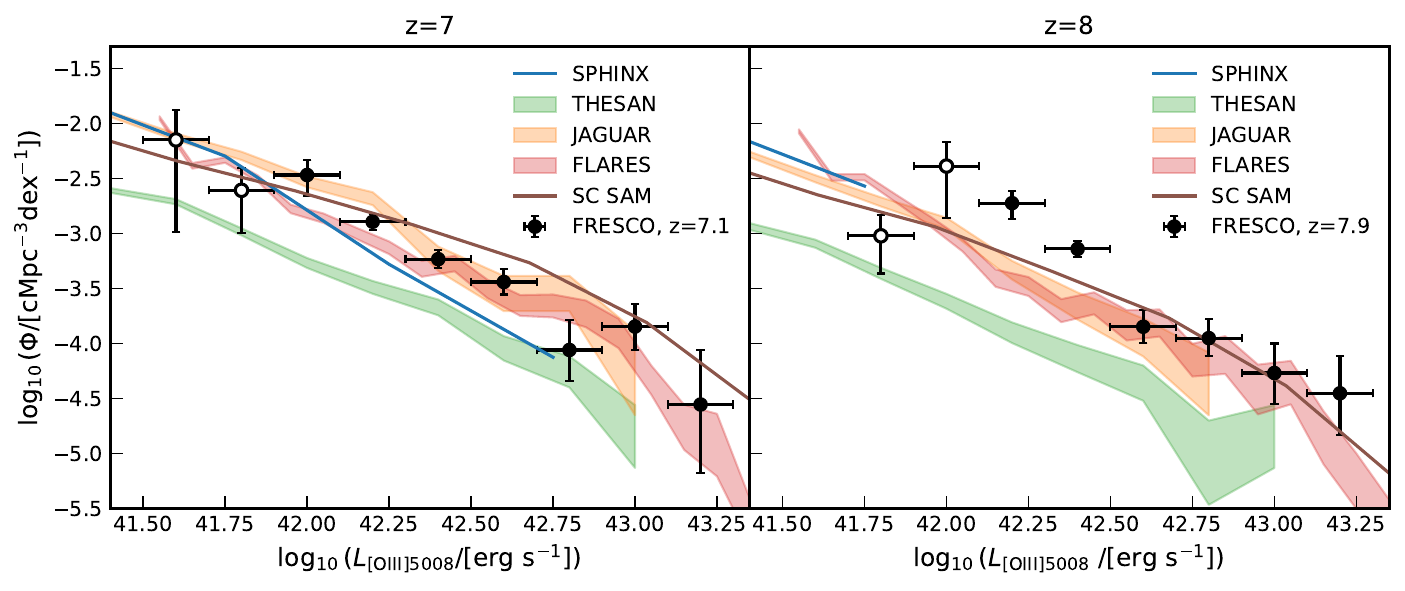}
    \caption{Comparison between the observed [\ion{O}{iii}] LF at $z\sim7$ and $z\sim8$ (left and right panels, respectively) in FRESCO and simulations. Luminosity bins with completeness $<25\%$ are shown with empty symbols. FLARES, JAGUAR and SC SAM match the [\ion{O}{iii}] LF, whereas the observations are in tension with SPHINX and THESAN predictions. The excess at the faint-end of the $z\sim 8$ [\ion{O}{iii}] LF is subject to high completeness corrections ($10\%-50\%$, see further Table \ref{tab:o3_lf_values}). }
    \label{fig:O3_LF_sims}
\end{figure*}

We now compare our results with predictions for the [\ion{O}{iii}] luminosity function from a variety of models and simulations in Fig. \ref{fig:O3_LF_sims}. In particular we use predictions from THESAN \citep[][]{Kannan2022a,Kannan2022}{}{}, SPHINX \citep[][]{Katz2023_sphinx}{}{}, JAGUAR \citep[][]{Williams2018}{}{}, FLARES \citep[][]{Lovell2021,Wilkins2023_o3}{}{} and for the SantaCruz Semi-Analytical Model galaxies \citep[SC SAM][]{Yung2019}{}{} for which the emission line modelling is done following \citet[][]{Hirschmann2017,Hirschmann2023}{}{}, and Yung (in prep.). On the one hand, we find that FLARES, JAGUAR and the SC SAM match well the [\ion{O}{iii}] LF at $z=7$ and at the bright-end at $z=8$. The discrepancy at $\log_{10}(L_{\rm{[\ion{O}{iii}]}} / \rm{erg\ s}^{-1}) < 42.5$ is harder to assess as the completeness declines below $<50\%$. On the other hand, SPHINX and THESAN underpredict the [\ion{O}{iii}] LF. The discrepancy is smaller for SPHINX but still significant ($\sim 0.3$ dex at $>3\sigma$). THESAN underpredicts the [\ion{O}{iii}] LF by about half a dex at all luminosities and redshifts. As all these models and simulations match the UVLF at the redshifts of interest, this suggests that their different modelling of the line emission has a strong impact on the inferred line luminosity function. Broadly, we can separate between FLARES, JAGUAR and SC SAM which use Cloudy \citep[][]{Ferland1998,Chatzikos2023}{}{} models to predict [\ion{O}{iii}] (as well as other line) fluxes from the properties of the stars (analytical, e.g. JAGUAR) or stellar particles (hydrodynamical simulation, e.g. FLARES). SPHINX and THESAN however model the high-redshift interstellar medium using radiative transfer and self-consistently determine the nebular emission. 

\begin{figure*}
    \centering
    \includegraphics[width=1\textwidth]{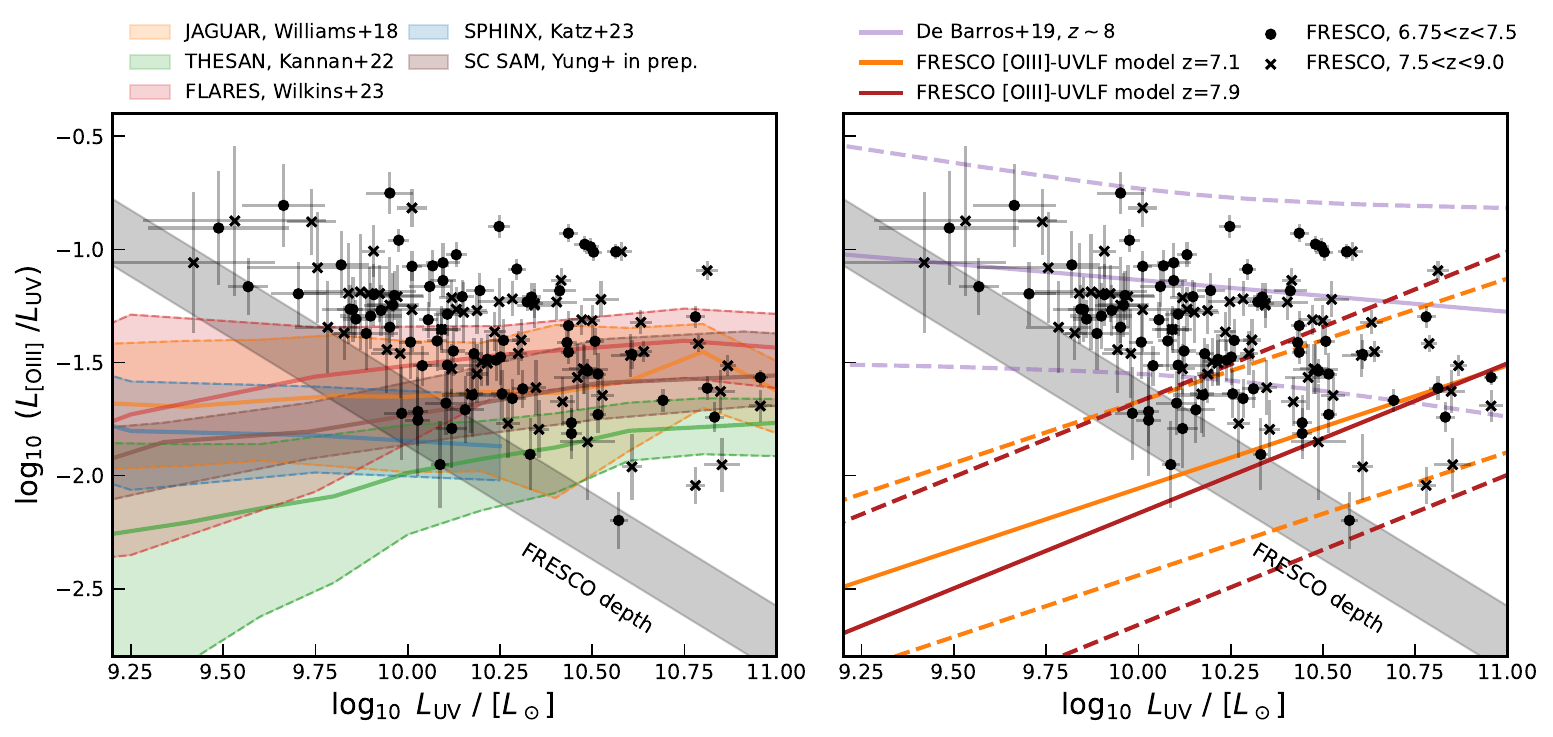}
    \caption{ [\ion{O}{iii}] / UV luminosity ratio as a function of UV luminosity. \textbf{Left panel:} The FRESCO measurements are shown in black circles ($6.75<z<7.5$) and crosses ($7.5<z<9$). Predictions from simulations and models are shown with full lines and shaded areas (16-84 percentiles). The diagonal grey band shows the FRESCO detection limit (set by the WFSS line sensitivity). \textbf{Right panel:} FRESCO measurements and [\ion{O}{iii}] / UV relation inferred from the UVLF of [OIII] emitters (full lines), with dotted lines showing the best-fit scatter (see further text and Fig. \ref{fig:O3_UVLF}). The \citet[][]{DeBarros2019}{}{} best-fit relation to \textit{HST} and \textit{Spitzer} observations is shown in purple. 
    }
    \label{fig:O3_UV_relation}
\end{figure*}

\begin{figure*}
    \centering
    \includegraphics[width=0.8\textwidth]{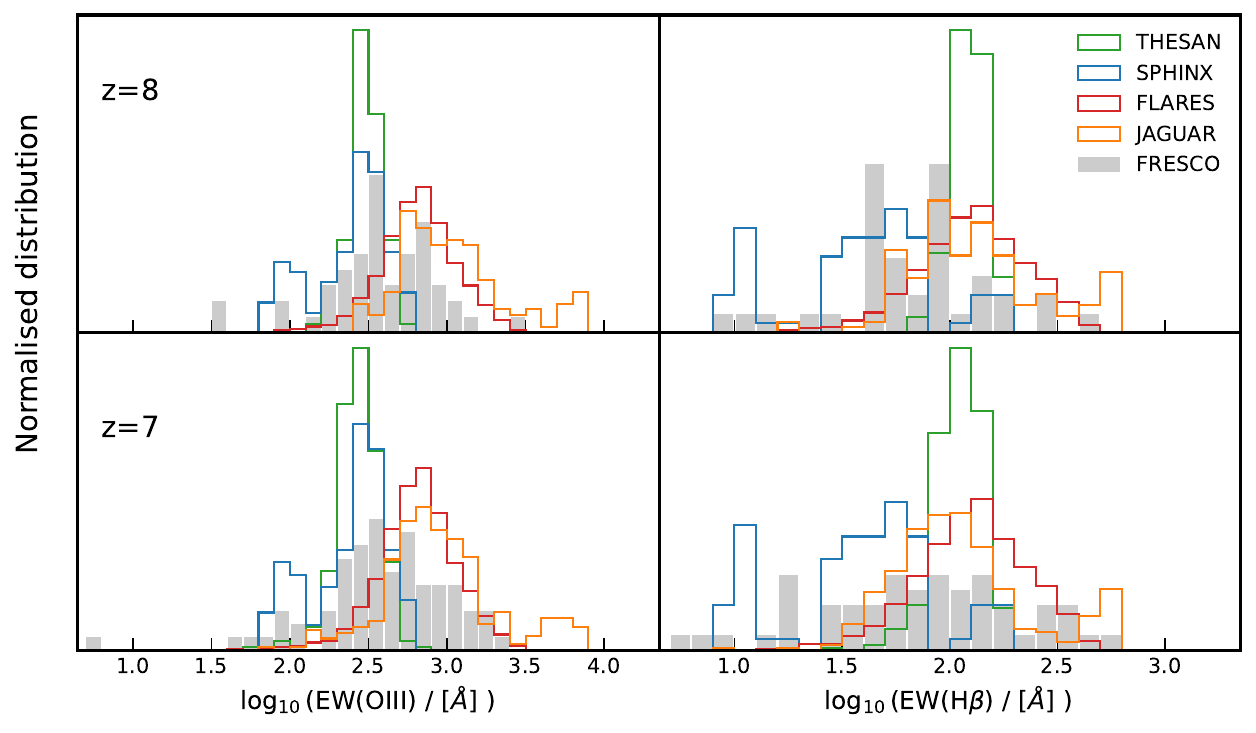}
    \caption{Observed (shaded grey) and predicted (coloured lines) EW distribution of [\ion{O}{iii}] $5008+4960\ \text{\AA}$ (left) and H$\beta$ (right). The histograms for the simulations only include objects that would be bright enough in [\ion{O}{iii}] to be detected with FRESCO. The bin size of all histograms is $0.1$ dex, and the histograms are normalised by the integrated numbers for each dataset and redshift. }
    \label{fig:EW_O3_Hb}
\end{figure*}

To investigate this further, we compare the [\ion{O}{iii}] / UV luminosity ratio as a function of UV luminosity for the models and observations in Figure \ref{fig:O3_UV_relation}. We first note that the \citet[][]{DeBarros2019}{}{} best-fit relation is indeed biased high compared to observed values in FRESCO. At the bright-end ($\log10(L_{\rm{UV}} / [L_\odot])\gtrsim10.6)$, where the sample is complete, JAGUAR and FLARES are in good agreement with the mean values in FRESCO, whereas THESAN, SPHINX and SC SAM  predict a lower mean [\ion{O}{iii}] / UV ratio than observed at the UV-bright end. None of the models seem to capture accurately the scatter in [\ion{O}{iii}] / UV  observed with \textit{JWST} at the bright-end. To quantify this further, we look at the fraction of $>1\sigma$ outliers for objects above the median prediction of the various models at luminosities $\log L_{\rm{UV}} / [L_\odot]>10.4$. If a model accurately models the scatter in the data above the median, the fraction of $1\sigma$ outliers should be $16/34=0.47$. However, we find much higher $1\sigma$ outlier fractions of $0.56^{\tabularrel+0.08}_{\tabularrel-0.08}$, $0.75^{\tabularrel+0.06}_{\tabularrel-0.07}$, $0.78^{\tabularrel+0.06}_{\tabularrel-0.07}$, $0.78^{\tabularrel+0.08}_{\tabularrel-0.11}$, $0.55^{\tabularrel+0.09}_{\tabularrel-0.09}$ for JAGUAR, THESAN, SPHINX, FLARES and SC-SAM, respectively. THESAN, SPHINX and FLARES are in clear tension with the observed scatter. JAGUAR and SC-SAM are in agreement within $1\sigma$, but this could change at fainter luminosities where the scatter seems higher (see Fig. \ref{fig:O3_UV_relation}, left).

Pushing this comparison below $\log10(L_{\rm{UV}} / [L_\odot])\lesssim 10.4)$ requires taking into account a key factor: the directly observed [\ion{O}{iii}] / UV ratios sample only the maximum of the intrinsic distribution, especially at the UV-faint end. Indeed, the sensitivity of our observations only enables us to observe the brightest [\ion{O}{iii}] emission lines at almost all UV luminosities (as shown by the grey shaded area in Fig \ref{fig:O3_UV_relation}). We therefore show the inferred distribution of the [\ion{O}{iii}] / UV determined from the UVLF (see previous section) in the right panel of Fig \ref{fig:O3_UV_relation}. The inferred distribution is in better agreement with the THESAN predictions at the faint end, despite the latter model underpredicting the [\ion{O}{iii}] luminosity function. Detailed comparisons between simulations as well as deeper observations with higher completeness below $\log10(L_{\rm{UV}} / [L_\odot])\lesssim 10.4)$ would be necessary to understand the source of these tensions betwen observations and simulations.

In summary, no model reproduces exactly both the mean value and scatter of the [\ion{O}{iii}]/UV ratio. The FLARES and JAGUAR prescription for the [\ion{O}{iii}]/UV ratio is biased high, probably enabling these two models to better reproduce the [\ion{O}{iii}] luminosity function as observations only sample from the extreme of the intrinsic distribution of luminosity ratios. THESAN and the SC SAM predict a steeper evolution of the [\ion{O}{iii}]/UV relation in good agreement with that inferred from observations, but seem to lack the higher scatter necessary to reproduce the observed number of strong [\ion{O}{iii}] emitters at high UV luminosities. The increased scatter could stem from more varied metallicities, ionising parameters, dust attenuation and/or star-formation histories in the observed Universe compared to existing simulations. Other studies have already pointed out the discrepancy between the simulated and directly measured metallicities \citep[e.g.][]{Nakajima2023}{}{} and burstier than expected star-formation histories \citep[][]{Endsley2023, Looser2023,Dressler2023,Cole2023}{}{}. This work opens a new statistical avenue to investigate the first galaxies using global statistics of line number densities and line ratios. Investigating the discrepancies between the observations and the different models is a promising way to put constraints on the formation and evolution of the first galaxies, but is outside of the scope of this work.

\subsection{Equivalent width distribution of [\ion{O}{iii}] and H$\beta$ at $6.8<z<9.0$}

We compute equivalent widths (EW) for the sample by inferring the continuum around [\ion{O}{iii}] with the \texttt{Prospector} inference framework (Naidu et al. in prep.). All available photometry and line fluxes are simultaneously fit to derive physical parameters using the fixed spectroscopic redshift. For $\approx60\%$ of the sample the continuum fits are constrained by multiple medium-bands at $4-5\mu$m, whereas for the rest they are determined by the combination of the line-fluxes and F444W photometry. We find a median [\ion{O}{iii}]5008 \AA\ EW of $380^{\tabularrel+47}_{\tabularrel-34}$ \AA \ (the errors are computed using bootstrap resampling). We verify that simply subtracting the line-fluxes from F444W (assuming ([\ion{O}{iii}]5008\AA+[\ion{O}{iii}]4959\AA)/H$\beta$ = 8.48 when H$\beta$ is not detected, cf. Section \ref{sec:discussion}) produces a consistent result of $407^{\tabularrel+51}_{\tabularrel-40}$ \AA. 

Our observed median H$\beta$+[OIII] equivalent width EW([\ion{O}{iii}]$5008+4960+H\beta) =  597^{\tabularrel+154}_{\tabularrel-65}\ \text{\AA}$ is in good agreement with photometric estimates such as that of \citet[][]{DeBarros2019}{}{} who find $EW([\ion{O}{iii}]5008+4960+H\beta) =  649^{\tabularrel+92}_{\tabularrel-49}\ \text{\AA}$ and \citet[][]{Endsley2023} who report EWs of $520-780\ \text{\AA}$ over the redshift ($z=7-9$) and magnitude range considered ($M_{\rm{UV}}\lesssim 18$). We do not find a trend of higher equivalent width of [\ion{O}{iii}]$5008+4960+H\beta$ with UV magnitude in our sample as \citet[][]{Endsley2023} do. However, our selection is biased towards high EWs in fainter objects in order for them to be spectroscopically confirmed and thus be part of our selection. Indeed, when considering the selection biases and comparing the [\ion{O}{iii}] and UV LFs, we find increased [\ion{O}{iii}]/UV ratio (and presumably EW([\ion{O}{iii}]) with UV luminosity (see Section \ref{sec:O3_UV}). Our median equivalent width distribution is closer to the the non-Lyman-$\alpha$ emitter (LAE) stack of \citet[][]{RobertsBorsani2024}{}{} ($734\pm 17\ \text{\AA}$) than to their LAE stack ($1573\pm51$ \AA), although $25$ objects in the sample have $EW([\ion{O}{iii}]5008+4960+H\beta)>1500 \ \text{\AA}$ and should be in principle more likely to show Lyman-$\alpha$. Deep spectroscopic follow-up of the Lyman-$\alpha$ line of the FRESCO sample would test whether observed Lyman-$\alpha$ in high-redshift galaxies is primarily  linked to the intrinsic properties of the galaxy (e.g. high EW([\ion{O}{iii}]+H$\beta$)) or the progress of reionisation in its vicinity.

We finally compare our observed [\ion{O}{iii}] and H$\beta$ equivalent width distribution with that of simulations in Figure \ref{fig:EW_O3_Hb}. To do so, we only include objects with [\ion{O}{iii}] fluxes above the approximate $1\times10^{-18}$ cgs sensitivity of FRESCO (corresponding to luminosities of $\sim 10^{41.5}, 10^{42}\ \rm{erg\ s}^{-1}$ at $z=7,8$). Overall, we find similar results as for the [\ion{O}{iii}]/$L_{\rm{UV}}$ ratio distribution: FLARES and JAGUAR are biased high compared to the observed EW distribution, whereas SPHINX and THESAN have better median values but a smaller dispersion than observed. The equivalent width distribution of H$\beta$ is mostly well reproduced for all simulations (within uncertainties) except for THESAN which predicts a much tighter distribution than observed and JAGUAR which predicts an exteded tail of EW([\ion{O}{iii}])>$3\times10^3\ $\AA\ which we do not observe in our sample. Again, we speculate that lower metallicities and burstier star-formation histories could help reduce the discrepancy, with specific corrections needed for each different model.

\section{Discussion}
\label{sec:discussion}
\subsection{The decline of the [\ion{O}{iii}] luminosity density at $z>6$: the role of SFR and metallicity}

\begin{figure}
    \centering
    \includegraphics[width=0.5\textwidth]{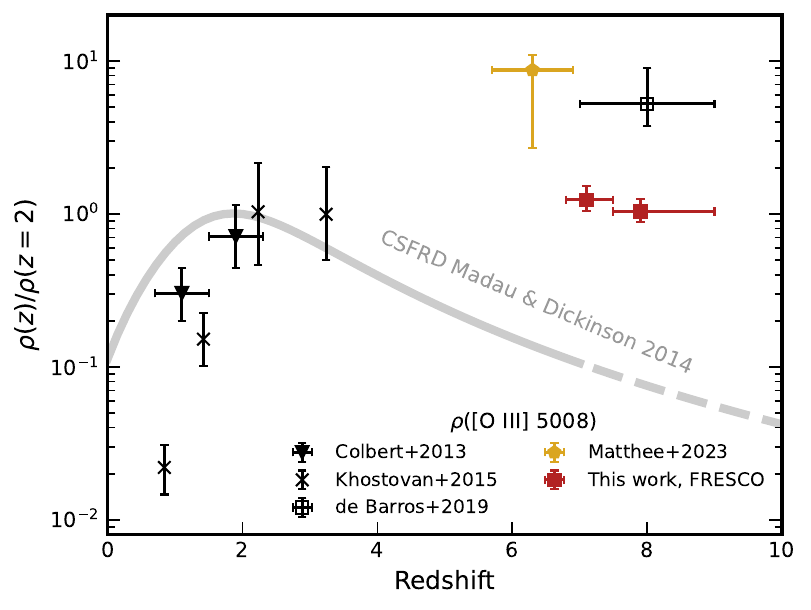}
    \caption{The evolution of the [\ion{O}{iii}] luminosity density (points, all integrated down to $L_{[\ion{O}{iii}]}=10^{42}\ \rm{erg\ s}^{-1}$) and the cosmic star-formation rate density \citep[grey line, CSRFD][]{Madau2014}{}{}. The [\ion{O}{iii}] luminosity density is normalised to the value at $z=2$, assuming $\rho(z=2)=3\times10^{39}\ \rm{erg s}^{-1}\rm{cMpc}^{-3}$ close the \citet[][]{Colbert2013}{}{} and \citet[][]{Khostovan2015}{}{} measurements. The CSFRD is also normalised at $z=2$. }
    \label{fig:rho_oiii_csfrd}
\end{figure}

In Section \ref{sec:O3_lf} we showed that the normalisation of the [\ion{O}{iii}] luminosity function significantly declines over the redshift range $6.25<z<7.9$ covered by this work and \citet[][]{Matthee2023_EIGER}{}{}. Prior to this work, literature studies \citep[][]{Colbert2013,Pirzkal2013,Khostovan2015, DeBarros2019} showed no decline of the [\ion{O}{iii}] luminosity density $\rho_{[\ion{O}{iii}]}$ all the way to $z=8$. Our revised measurement at $z=8$ (see Section \ref{sec:O3_lf}), and the new \textit{JWST} constraints at $z=5-7$ indicate that $\rho_{[\ion{O}{iii}]}$ reaches a maximum between $z\sim 3$ and $z\sim 6$ (see Fig. \ref{fig:rho_oiii_csfrd}). We note that the \citet[][]{Matthee2023_EIGER}{}{} measurement relies on a single field. Future measurements less impacted by cosmic variance could therefore show the [\ion{O}{iii}] luminosity density plateau between $z=3$ and $z=6$. Ongoing wide and deep NIRCam/WFSS surveys will determine the position and amplitude of the peak. Nonetheless, the constraints from FRESCO at $z>7$ demonstrate a decline in the [\ion{O}{iii}] luminosity density at high-redshift.

At $z=6.3$, the best-fit [\ion{O}{iii}] luminosity from \citet[][]{Matthee2023_EIGER}{}{} results in a luminosity density of $\rho_{[\ion{O}{iii}]}=1.1\times10^{40}\ \rm{erg\ s}^{-1}\rm{cMpc}^{-3}$ when integrated down to $L_{[\ion{O}{iii}]}=10^{42}\ \rm{erg\ s}^{-1}$, whereas we find $\rho_{[\rm{\ion{O}{iii}}]} = 1.55^{\tabularrel+0.36}_{\tabularrel-0.24} \times 10^{39}, 1.30^{\tabularrel+0.26}_{\tabularrel-0.19} \times 10^{39}\rm{erg\ s}^{-1}\rm{cMpc}^{-3}$ at $z=7.1, z=7.9$, respectively. The decline between $z=7.1$ and $z=7.9$ (factor $1.19\pm0.09$) is in perfect agreement with the decline in CSFRD ($1.3$). However, the significant drop between the $z=6.25$ and $z=7.1$ (factor $\sim 7$, consistent with the $\sim 1$ dex drop at the bright-end of the [\ion{O}{iii}] LF, see Fig. \ref{fig:O3_LF}) does not follow the mild decline of the SFRD ($\times 1.6$). Notwithstanding the caveats posed by the different selection in \citet[][]{Matthee2023_EIGER}{}{} and this study, and the small number of fields studied, this could hint at additional evolutionary effects such as a rapid decrease in metallicity and thus the line strength.

\begin{figure*}
    \centering
    \includegraphics[width=\textwidth]{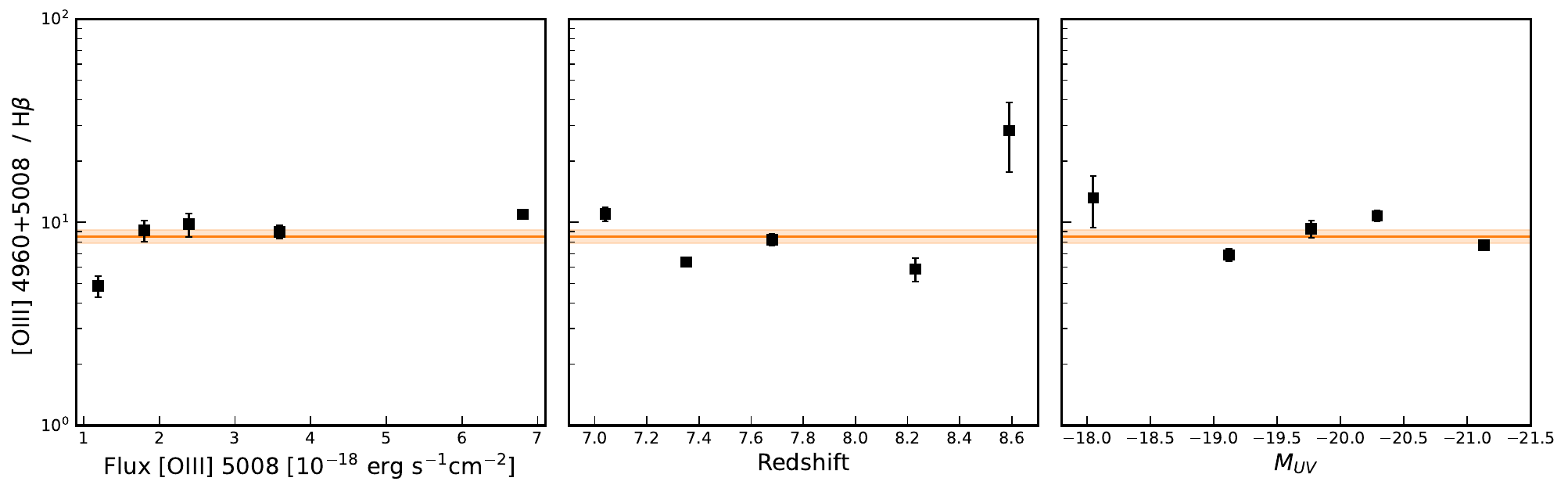}
    \caption{The [\ion{O}{iii}]/H$\beta$ flux ratio in median-stacked spectrum across [\ion{O}{iii}] flux, redshift and rest-frame UV magnitude. We find no evidence for an evolution with any of these parameters. We show the measured ratio for the full sample stack in orange. }
    \label{fig:hb_o3_ratio}
\end{figure*}

As discussed in \ref{sec:restframeopticalstack}, the R3 = [\ion{O}{iii}] 5008 /  H$\beta$ ratio in the GN median stack is $6.38\pm0.85$. We find a similar value when stacking all the FRESCO data in the GN and GS fields. We also split and stack our sample not only with respect to the line flux, but with respect to $M_{\rm{UV}}$ or redshift. We show the [\ion{O}{iii}] / H$\beta$ ratio dependence on these various parameters in Fig. \ref{fig:hb_o3_ratio}. The R3 value is close to the extremum found in low-z SDSS analogues of high-z emitters \citep[][]{Bian2018, Nakajima2022}{}{}. Assuming we are on the low-metallicity branch of the \citet[][]{Bian2018}{}{} strong line metallicity calibrations, and the `all' model of \citep{Nakajima2022},  we infer a metallicities $12+\log(\rm{O/H})_{\rm{O3H}\beta} = 7.2-7.7$ (not taking into account calibration uncertainties). The higher value of [\ion{O}{iii}]5008/H$\beta$ at $z>7,8$ compared to  $5.33<z<6.9$ \citep[][]{Matthee2023_EIGER}{}{} is consistent with a decline in metallicity at $z>6$. The measurements are however still subject to calibration uncertainties, and deeper spectroscopic observations (e.g. NIRSpec spectroscopy) are necessary to conclude on this topic.

One caveat to the above interpretation of the [\ion{O}{iii}] luminosity density evolution is the unknown contamination of AGN. Our observations and that of \citep[][]{Matthee2023_EIGER}{}{} cannot unambiguously separate AGN from starburst activity in individual objects due to the lack of wavelength coverage and sensitivity. However, we note that we do not find evidence for a broad ($>2000-3000\ \rm{km\ s}^{-1}$) H$\beta$ component and do not detect [\ion{O}{iii}]$ 4363$\AA \ \citep[see e.g.][]{Mazzolari2024}{}{} in the stacked spectrum (Section \ref{sec:restframeopticalstack}), indicating that such AGN contamination is at worst marginal.

\subsection{The contribution of [\ion{O}{iii}] emitters to reionisation}
\begin{figure*}
    \centering
    \includegraphics[width=\textwidth]{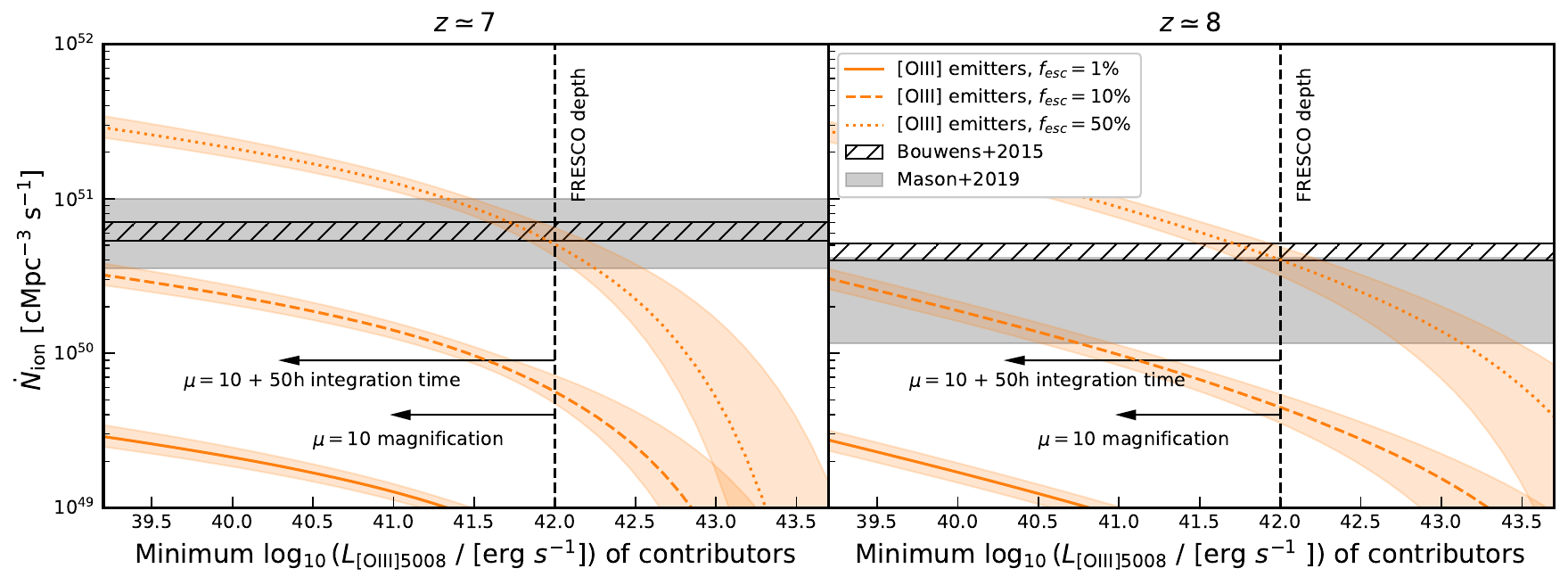}
    \caption{Cumulative ionising photon emissivity per unit volume as a function of the minimum [\ion{O}{iii}] luminosity. We show three models with escape fraction $f_{\rm{esc}}=1\%,10\%,50\%$ in orange lines and dashed envelopes derived from our [\ion{O}{iii}] luminosity functions at $z=7.1$ and $z=7.9$. The emissivity constraints derived from the IGM opacity measurements and the UV LF are shown in (hatched) grey  \citep[][]{Bouwens2015a,Mason2019a}{}{}. Assuming $f_{\rm{esc}}=10\%$, FRESCO detects the galaxies contributing to $\sim 10\%$ of the ionising budget in the GOODS fields. For reference, we indicate the depth reached by a survey similar to FRESCO ($\sim 2$h integration time with NIRCam WFSS F444W) with the assistance of gravitational lensing with magnifications $\mu=10$ as well as deeper integration time.}
    \label{fig:reionisation_budget}
\end{figure*}
We now focus on the [\ion{O}{iii}] / H$\beta$ ratio, with the aim of deriving global estimates of SFR and ionising output from the [\ion{O}{iii}] luminosity function. 
In the previous section, we have established that the [\ion{O}{iii}]/ H$\beta$ ratio is constant as a function of redshift and [\ion{O}{iii}] luminosity in our sample. This result opens an opportunity to derive the contribution of [\ion{O}{iii}] emitters to the cosmic SFRD, as well as the ionising budget with minimal assumptions. 

We have shown in Section \ref{sec:restframeopticalstack} that the Balmer line ratios are consistent with negligible dust attenuation. We can now simply determine the ionising budget from the [\ion{O}{iii}] LF as 
\begin{align}
    \dot N_{ion} &= \int_{L_{\rm{min}}}^{\infty} Q_{\rm{ion}}(L_{\rm{[\ion{O}{iii}]}}) \phi(L_{\rm{[\ion{O}{iii}]}}) \rm{d}L_{\rm{[\ion{O}{iii}]}}  \\
    &= \int_{L_{\rm{min}}}^{\infty} f_{\rm{esc}} \frac{2.79}{c_\alpha (1-f_{\rm{esc})}} \Big\langle \frac{L_{\rm{[\ion{O}{iii}]}}}{L_{\rm{H}\beta}}\Big\rangle ^{-1}  \phi(L_{\rm{[\ion{O}{iii}]}}) \rm{d} L_{\rm{[\ion{O}{iii}]}}
\end{align}
where $L=L_{\rm{[\ion{O}{iii}]} 5008}$, $f_{\rm{esc}}$ is the escape fraction of ionising photons, and we have used $L_{H\alpha} = Q_{ion} c_\alpha(1-f_{\rm{esc}})$, the unattenuated Balmer decrement for case B recombination assuming $T_e=10^4\ \rm{K}$ and $n_e=100\ \rm{cm}^{-3}$ $L_{\rm{H}\alpha}/ L_{\rm{H}\beta}=2.79$ \citep[e.g.][]{Ferland2017,Reddy2023}{}{}, the average [\ion{O}{iii}] 5008/H$\beta$ ratio observed in our sample $\Big\langle \frac{L_{\rm{[\ion{O}{iii}]\ 5008}}}{L_{\rm{H}\beta}} \Big\rangle  = 6.35$, and a recombination coefficient $c_\alpha=1.37 \times 10^{-12}\ \rm{erg}$ \citep[e.g.][ assuming $T_e=10^4\ \rm{K}$ and $n_e=100\ \rm{cm}^{-3}$]{Kennicutt1998,Schaerer2003}{}{}. Notably, this approach does not use the photometrically-determined UV luminosity function nor the $\xi_{ion}$ parameter and its associated uncertainties.

We show the contribution of [\ion{O}{iii}] emitters to the ionising budget as a function of the minimum [\ion{O}{iii}] luminosity on Figure \ref{fig:reionisation_budget} for different assumed escape fractions. We compare our inferred ionising output to the constraints derived by \citet[][]{Bouwens2015a,Mason2019a}{}{} from the UV LF, Planck Thomson optical depth and Lyman-$\alpha$ fraction measurements. We find that if the escape fraction of [\ion{O}{III}] galaxies detected by FRESCO were close to $50\%$, then their ionising output is enough to reionise the Universe at $z=7-8$. An assumed average escape fraction of $50\%$ is in tension with indirect measurements at high redshift \citep[e.g.][]{Meyer2020,Mascia2023}{}{} and values generally derived from analytical models matching the ionising budget \citep[e.g.][]{Robertson2013,Robertson2015, Finkelstein2019,Dayal2020, Naidu2020}{}{}. Furthermore, the O32 ratio measured in the median stack (see Section \ref{sec:restframeopticalstack}) is correlated with a lower escape fraction of $f_{\rm{esc}}\sim 5-10\%$ in low-redshift Lyman Continuum leakers \citep[][]{Faisst2016, Izotov2018a,Flury2022b}{}{}. Nonetheless, this result showcases the leap in sensitivity \textit{JWST} provides, bringing closer the possibility of establishing a complete spectroscopic census of the sources of reionisation.

At a nominal $f_{\rm{esc}}=10\%$, we find that the [\ion{O}{iii}] emitters detected with FRESCO account for $\sim 10\%$ of the ionising budget at $z=7$ and $z=8$ assuming the \citet[][]{Bouwens2015a}{}{} ionising emissivity constraints. The updated \citet[][]{Mason2019a}{}{} constraints yields the same result at $z=7$, but suggest that FRESCO is capturing a higher fraction of the ionising emissivity at z=8 ($18_{\tabularrel-8}^{\tabularrel+37}\ \%$).

This already sizeable fraction of photons accounted for could grow when adding fainter sources detected with NIRSpec/MSA in the GOODS fields \citep[e.g. JADES, ][]{Bunker2023_JADESrelease, Rieke2023_JADESrelease, DEugenio2024_JADESrelease3}{}{}. This result raises the imminent prospect of detecting a majority of the sources of reionisation in a given volume with \textit{JWST}. Detecting all the sources of reionisation would require luminosity limit of $L_{\rm{[\ion{O}{iii}] 5008}}\sim 10^{39.5} \ \rm{erg\ s}^{-1}$ is necessary,  i.e. $300\times$ fainter than achieved with FRESCO in 7034s integrations. However, at a luminosity limit of $L_{\rm{[\ion{O}{iii}] 5008}}\sim 10^{41} \ \rm{erg\ s}^{-1}$ readily achieved with the aid of gravitational lensing ($\mu\gtrsim10$) by foreground clusters, spectroscopically-confirmed [OIII] emitters would account for $20\% (20-40)\%$ of the ionising budget at $z=7(8)$. Ongoing Cycle 2 programmes \# 2883, 3516 and 3538 should therefore detect a sizeable fraction of the sources of reionisation, albeit in small volumes subject to larger variance due gravitational lensing magnification. Further hypothetical large surveys pushing to $50\rm{h}$ integration time could in principle detect the sources responsible for the majority ($>50\%$) of the ionising budget at $z\gtrsim 7$. The remaining question will be whether their ionising output and escape fraction will indeed match the values assumed here and elsewhere in the literature.

\section{Summary and outlook}
We have presented the results of a search for H$\beta$ and [\ion{O}{iii}] emitters at $6.8<z<9.0$ in the GOODS-South and GOODS-North fields using the \textit{JWST} FRESCO programme \#1895. The galaxies were selected independently from their photometric redshift estimates using NIRCam/WFSS spectroscopy. We have carefully characterised the completeness of our search (Eq. \label{eq:completeness}), from the imaging detection, the gaussian-matched filtering spectroscopic search to the final visual inspection selection. We report the discovery of $137$ individual sources at $6.8<z<9.0$, which constitutes the largest unbiased sample of line-emitting galaxies at these redshifts to date. We report the following findings:
\begin{itemize}
    \item The median stacked rest-frame optical spectrum of $6.8<z<9.0$ [\ion{O}{iii}] emitters at a median UV magnitude $M_{\rm{UV}}=-19.65^{\tabularrel+0.59}_{\tabularrel-1.05}$ indicate negligible dust attenuation, low metallicity ($12+\log(\rm{O/H})= 7.2-7.7$) and a high ionisation parameter $\log_{10}U \simeq -2.5$.
    \item We find a discrepancy in the number of [\ion{O}{iii}] emitters in the GOODS-South and GOODS-North fields.  The shape of the [\ion{O}{iii}] LF is consistent across fields in different redshift intervals, with only a change in amplitude at $z=7.1$. The $\sim 30\%$ excess of [\ion{O}{iii}]-emitting galaxies in GOODS-N is solely explained by the presence of a strong overdensity at $7.0\lesssim z\lesssim 7.2$, potentially associated with a faint, red quasar \citep[GNz7Q][]{Fujimoto2022}{}{}.
    \item Our spectroscopic selection of [\ion{O}{iii}] emitters confirms the accuracy of photometric redshift at the statistical level. We find a good agreement with the predicted \texttt{EAZY} photometric redshifts, although we find that the photometric redshift uncertainties are underestimated when comparing to the final spectroscopic vs. photometric redshift differences. Additionally, the spectroscopically-confirmed UV LF is consistent (at the bright end where all sources are detected) with that established from dropout-selected samples prior to \textit{JWST}.
    \item We compute for the first time the [\ion{O}{iii}] 5008 luminosity function at $z=7.1$ and $z=7.9$ using a spectroscopically-confirmed sample, revising previous measurements with \textit{HST}/ \textit{Spitzer} \citep[][]{DeBarros2019}{}{}. We find no significant evolution in the shape of the [\ion{O}{iii}] LF, and the decrease in luminosity density between $z=7.1$ and $z=7.9$ is in perfect agreement with the evolution of the cosmic star-formation rate density.
    \item Simulations and models of reionisation-era galaxies can only reproduce the [\ion{O}{iii}] LF or the [\ion{O}{iii}]/UV luminosity ratio, but not both. We conclude that a larger scatter in the [\ion{O}{iii}]/UV luminosity ratio is missing in current models, which might be driven (in part) by variations in the metallicity or bursty star-formation histories at high redshift.
    \item By comparing our results to the [\ion{O}{iii}] luminosity density at $z=6.25$ from the first EIGER survey results \citep[][]{Matthee2023_EIGER}{}{}, we find a strong drop in amplitude which could signal by a change in the properties of [\ion{O}{iii}] emitters at $z>6$. Taken at face value, this sudden drop between $z=6.25$ and $z=7.1$ does not follow the cosmic SFR density evolution, and could be explained by a decline in the metallicity of high-redshift [\ion{O}{iii}] emitters. However, we note that the \citet[][]{Matthee2023_EIGER}{}{} measurement is based on a single field, and ongoing NIRCam/WFSS surveys might provide updated constraints at $z=5-6$ in the near future.
    \item Under the assumption of negligible dust attenuation and a constant escape fraction of $10\%$, we show that FRESCO (or any NIRCam WFSS observations with $\sim 2$h integration time) detects star-forming galaxies accounting for $10\%$ of the ionising budget at $z=7.1$ and $10-20\%$ at $z=7.9$, showcasing the potential of dedicated \textit{JWST} observations to chart the distribution of a sizeable fraction of the sources of reionisation in small volumes lensed by foreground clusters.
\end{itemize}

The release of the FRESCO [\ion{O}{iii}] emitter sample will likely foster new science results and follow-up observations. Overall, this work demonstrates the efficiency of NIRCam/WFSS in providing statistical samples at high-redshift in order to push the frontier of our understanding of early galaxy evolution.

\label{sec:conclusion}
\section*{Acknowledgements}
The authors thank the anonymous referee for comments and suggestions which improved this paper. 

RAM thanks R. Kannan for sharing emission line luminosities from THESAN and H. Katz for similar data from an early version of the SPHINX20 data release (we use the final data release in this paper). The authors thank the CONGRESS team for proposing and designing their program with a zero exclusive access period.

RAM, PA, ACP, AW acknowledge support from the Swiss National Science Foundation (SNSF) through project grant 200020\_207349. PA, AW, EG, MX acknowledges support from  the Swiss State Secretariat for Education, Research and Innovation (SERI) under contract number MB22.00072. YF acknowledges support by JSPS KAKENHI Grant Number JP22K21349 and JP23K13149. RPN acknowledges support for this work provided by NASA through the NASA Hubble Fellowship grant HST-HF2-51515.001-A awarded by the Space Telescope Science Institute, which is operated by the Association of Universities for Research in Astronomy, Incorporated, under NASA contract NAS5-26555. MS acknowledges support from the European Research Commission Consolidator Grant 101088789 (SFEER), from the CIDEGENT/2021/059 grant by Generalitat Valenciana, and from project PID2019-109592GB-I00/AEI/10.13039/501100011033 by the Spanish Ministerio de Ciencia e Innovaci\'on - Agencia Estatal de Investigaci\'on. The Cosmic Dawn Center (DAWN) is funded by the Danish National Research Foundation under grant No.\ 140. Cloud-based data processing and file storage for this work is provided by the AWS Cloud Credits for Research program. RJB and MS acknowledges support from NWO grant TOP1.16.057.

This work is based on observations made with the NASA/ESA/CSA James Webb Space Telescope. The raw data were obtained from the Mikulski Archive for Space Telescopes at the Space Telescope Science Institute, which is operated by the Association of Universities for Research in Astronomy, Inc., under NASA contract NAS 5-03127 for \textit{JWST}. These observations are associated with \textit{JWST} Cycle 1 GO program \#1895. Support for program JWST-GO-1895 was provided by NASA through a grant from the Space Telescope Science Institute, which is operated by the Associations of Universities for Research in Astronomy, Incorporated, under NASA contract NAS5-26555. 

\section*{Data Availability}

The results presented in this work are mostly based on public \textit{JWST} GO1 data (Programme \#1895), reduced with the publicly-available code \texttt{GRIZLI} (\url{grizli.readthedocs.io}). The reduced imaging data is publicly available at \url{https://s3.amazonaws.com/grizli-v2/JwstMosaics/v7/index.html} or through MAST: \url{https://archive.stsci.edu/hlsp/fresco} (DOI:10.17909/gdyc-7g80). The  [\ion{O}{iii}] emitters catalogues presented in this work, as well as subsequent updates, is publicly-available at the following link \url{https://github.com/rameyer/fresco/}. 

\section*{Affiliations}
\noindent
{\it 
$^{1}$Department of Astronomy, University of Geneva, Chemin Pegasi 51, 1290 Versoix, Switzerland\\
$^{2}$Cosmic Dawn Center (DAWN), Niels Bohr Institute, University of Copenhagen, Jagtvej 128, K\o benhavn N, DK-2200, Denmark\\
$^{3}$Institute of Science and Technology Austria (ISTA), Am Campus 1, 3400 Klosterneuburg, Austria \\
$^{4}$MIT Kavli Institute for Astrophysics and Space Research, 77 Massachusetts Ave., Cambridge, MA 02139, USA\\
$^{5}$Leiden Observatory, Leiden University, NL-2300 RA Leiden, Netherlands\\
$^{6}$Department of Astronomy, The University of Texas at Austin, 2515 Speedway, Stop C1400, Austin, TX 78712-1205, USA\\
$^{7}$Center for Frontier Science, Chiba University, 1-33 Yayoi-cho, Inage-ku, Chiba 263-8522, Japan\\
$^{8}$Department of Astronomy, University of Wisconsin-Madison, 475 N. Charter St., Madison, WI 53706 USA\\
$^{9}$Department for Astrophysical and Planetary Science, University of Colorado, Boulder, CO 80309, USA\\
$^{10}$ Centro de Astrobiolog\'{i}a (CAB), CSIC-INTA, Carretera de Ajalvir km 4, Torrej\'{o}n de Ardoz, 28850, Madrid, Spain\\
$^{11}$Department of Astronomy and Astrophysics, University of California, Santa Cruz, CA 95064, USA\\
$^{12}$Kapteyn Astronomical Institute, University of Groningen, P.O. Box 800, 9700 AV Groningen, The Netherlands\\
$^{13}$Centre for Astrophysics and Supercomputing, Swinburne University of Technology, Melbourne, VIC 3122, Australia\\
$^{14}$GRAPPA, Anton Pannekoek Institute for Astronomy and Institute of High-Energy Physics\\
$^{15}$University of Amsterdam, Science Park 904, NL-1098 XH Amsterdam, the Netherlands\\
$^{16}$Department of Physics and Astronomy, University of California, Riverside, 900 University Avenue, Riverside, CA 92521, USA\\
$^{17}$Department of Physics \& Astronomy, University of California, Los Angeles, 430 Portola Plaza, Los Angeles, CA 90095, USA\\
$^{17}$Departament d'Astronomia i Astrof\`isica, Universitat de Val\`encia, C. Dr. Moliner 50, E-46100 Burjassot, Val\`encia,  Spain\\
$^{18}$Unidad Asociada CSIC "Grupo de Astrof\'isica Extragal\'actica y Cosmolog\'ia" (Instituto de F\'isica de Cantabria - Universitat de Val\`encia)\\
$^{20}$Department of Physics, University of Bath, Claverton Down, Bath, BA2 7AY, UK\\ 
}



\bibliographystyle{mnras}
\bibliography{library} 



\appendix

\section{Mock and True emitters comparison}
\label{app:fake_emitters}

\begin{figure*}
    \centering
    \includegraphics[width=0.49\textwidth]{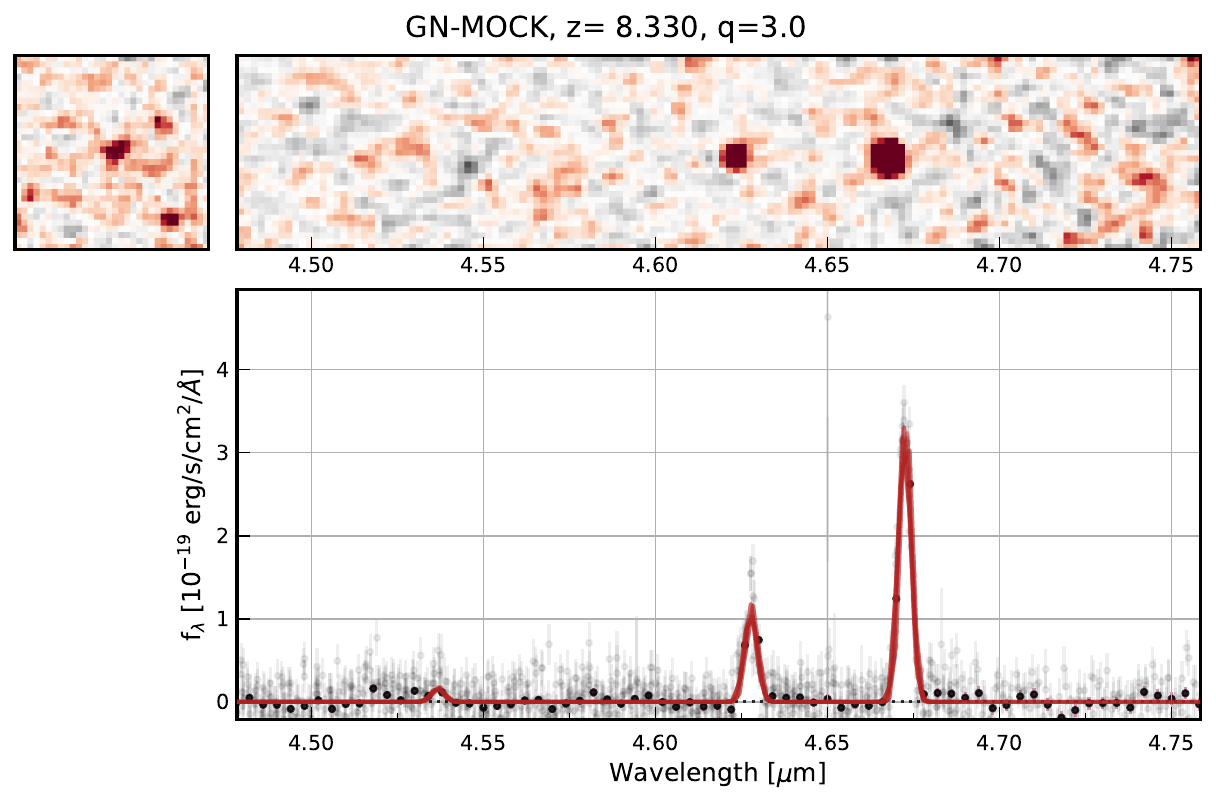}
    \includegraphics[width=0.49\textwidth]{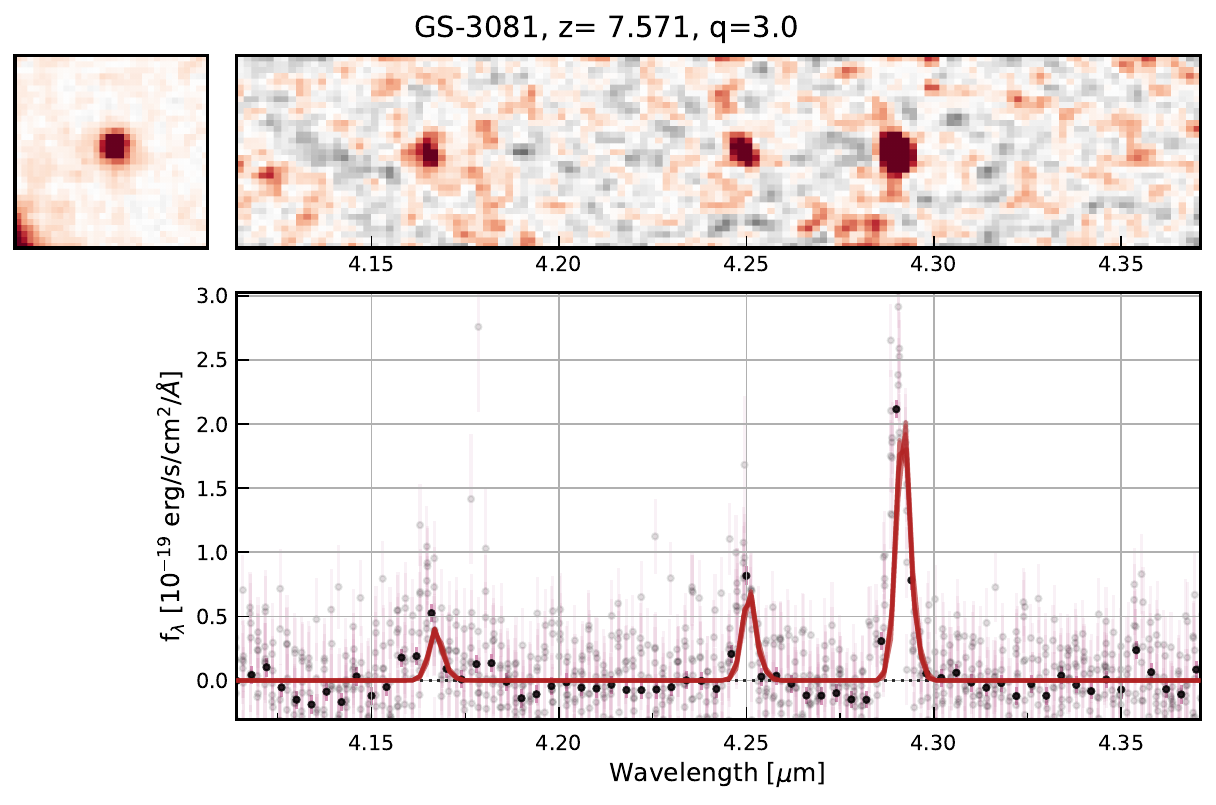}
    \includegraphics[width=0.49\textwidth]{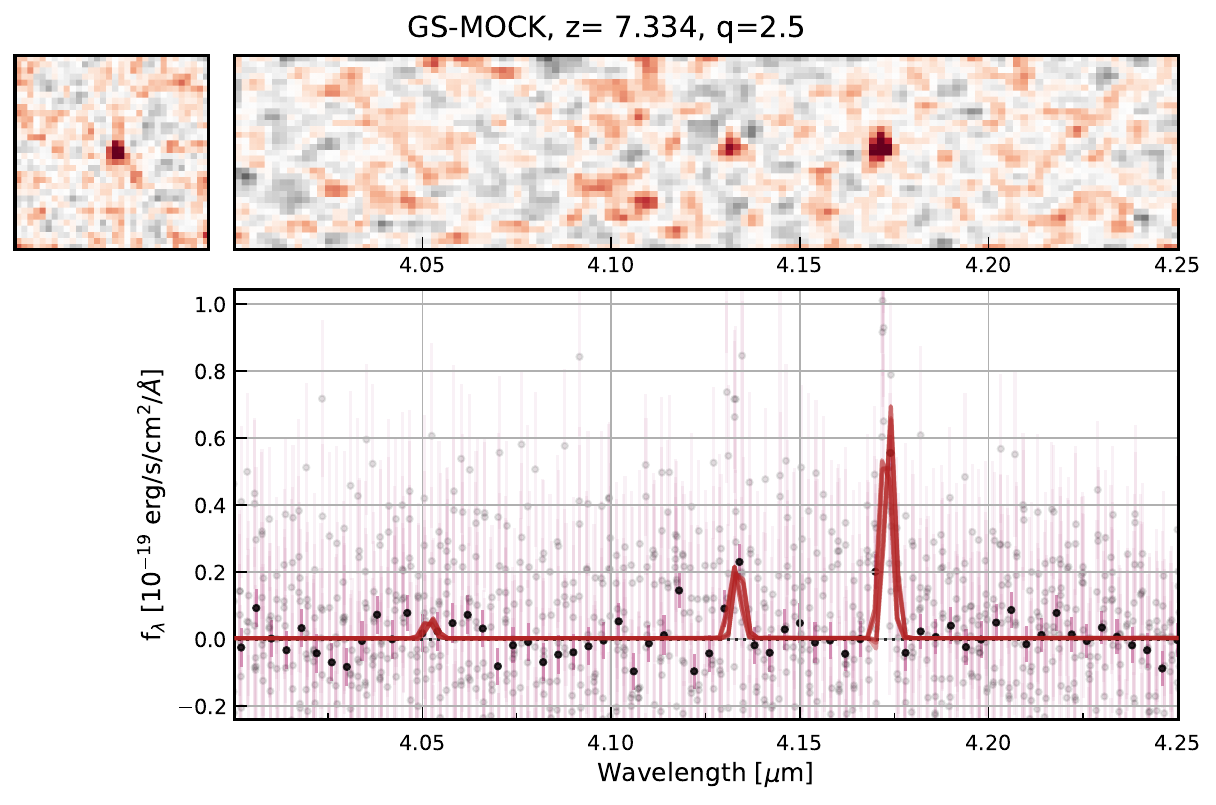}
    \includegraphics[width=0.49\textwidth]{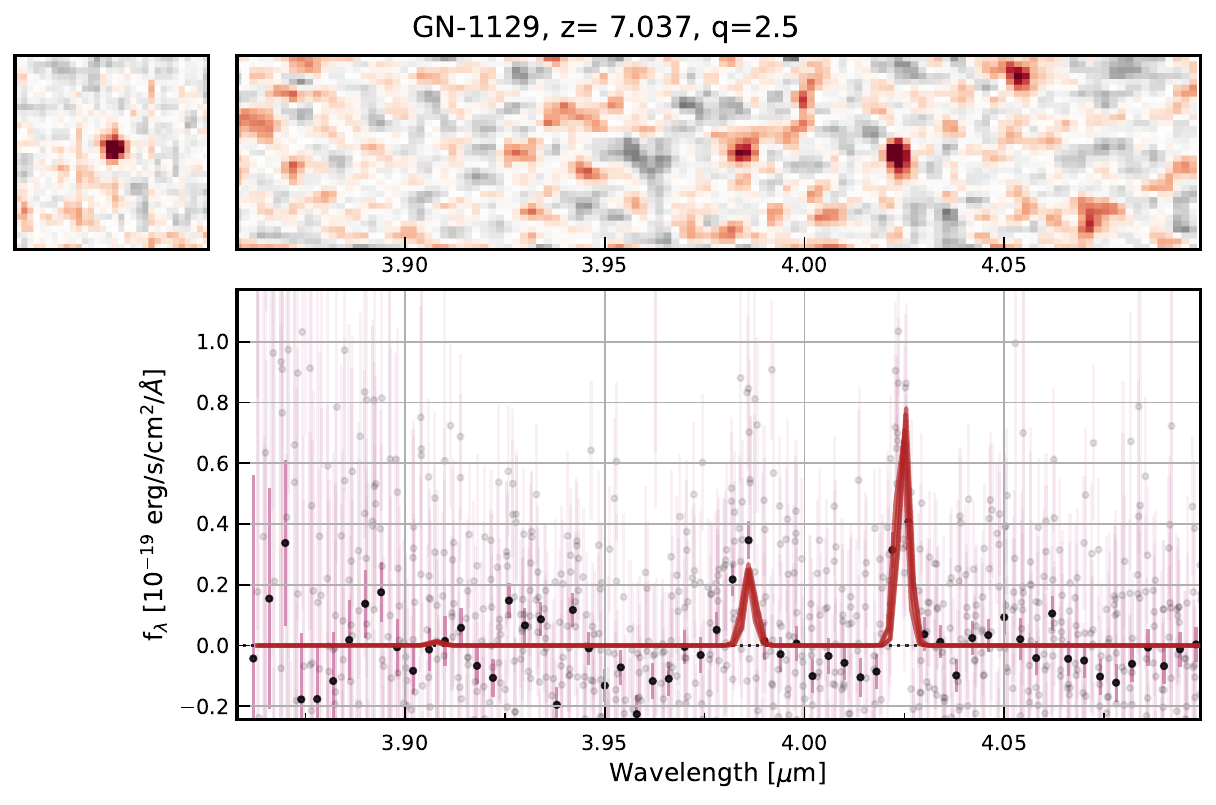}
    \includegraphics[width=0.49\textwidth]{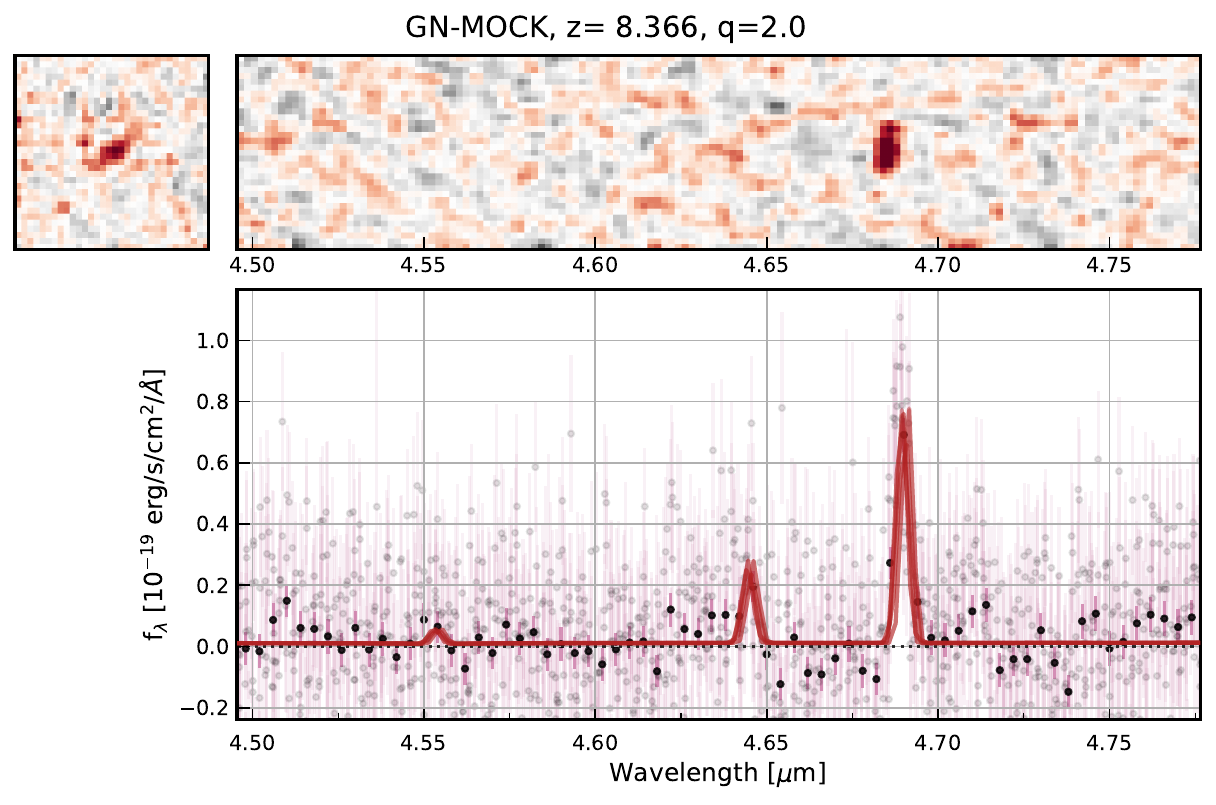}        \includegraphics[width=0.49\textwidth]{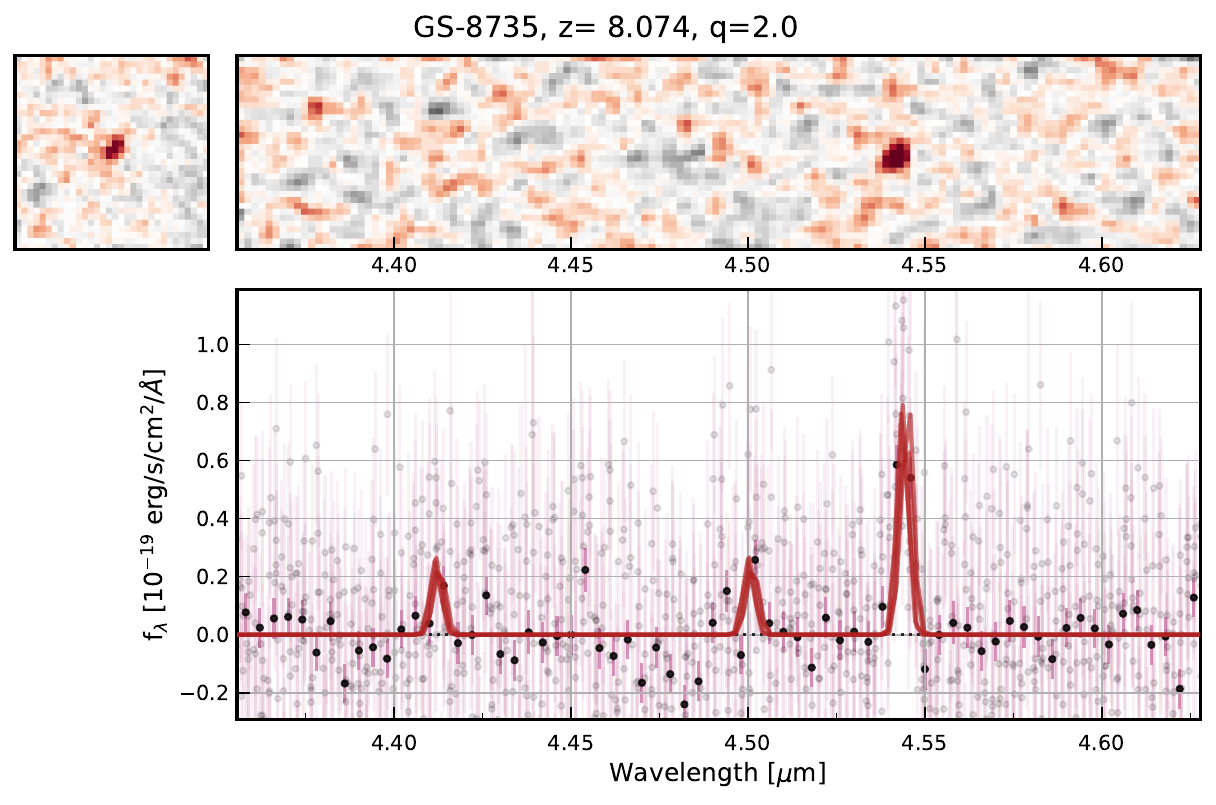}
    \includegraphics[width=0.49\textwidth]{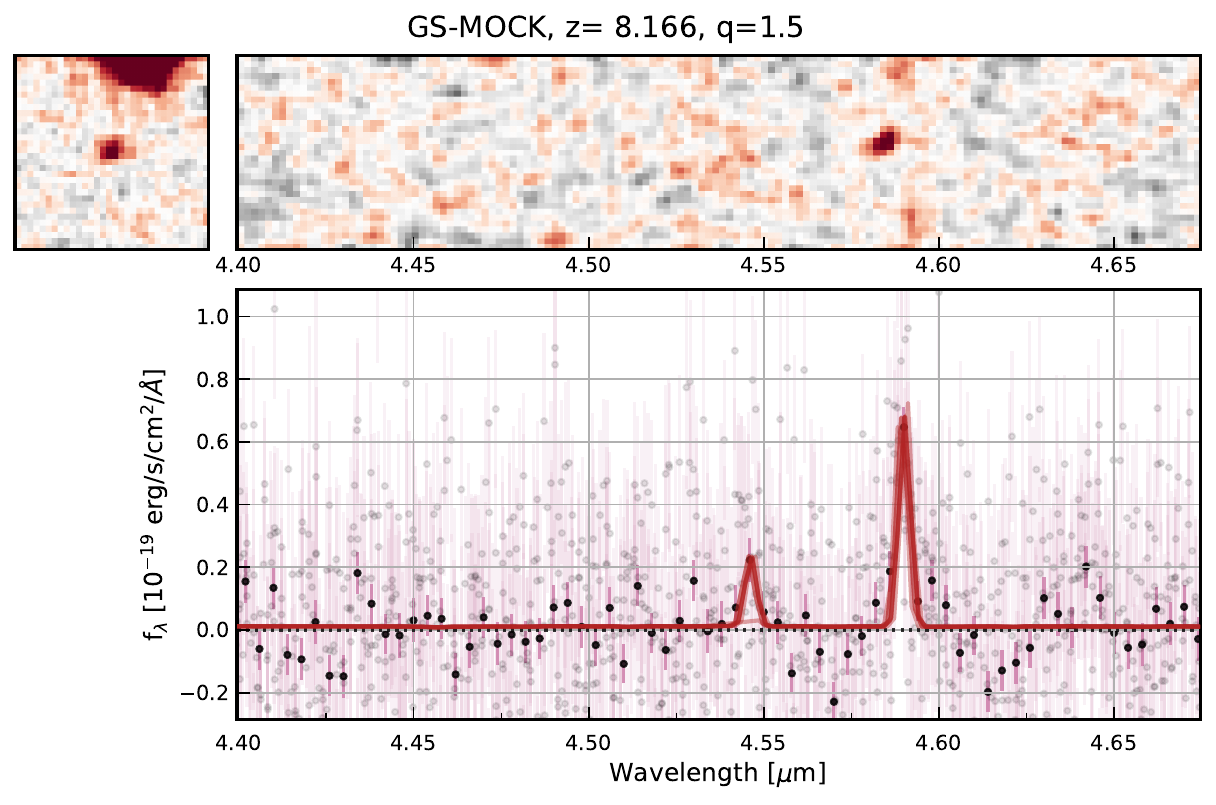}
    \includegraphics[width=0.49\textwidth]{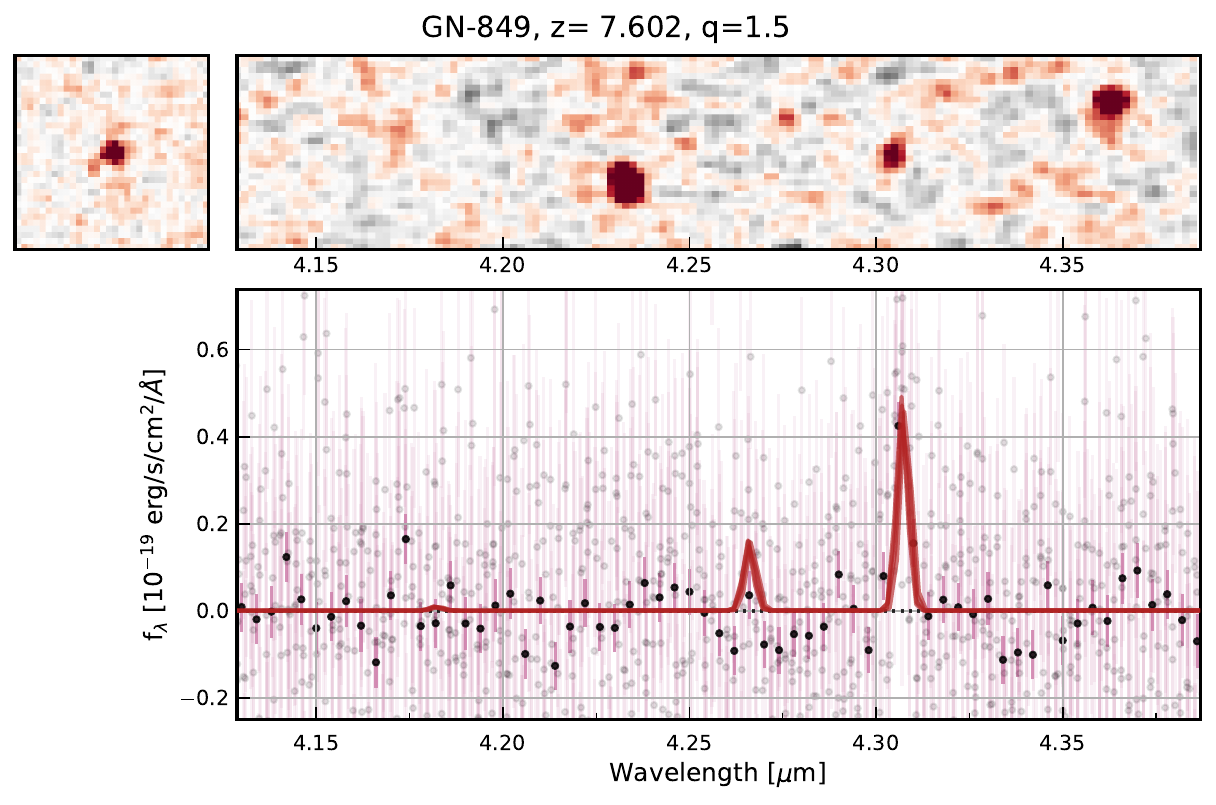}
    \caption{Comparison of mock (left column) and real (right) [\ion{O}{iii}] emitters with similar line fluxes and visual inspection grades that were graded simultaneously and blindly by the visual inspection team.}
    \label{fig:fake_emitters}
\end{figure*}

We present in Figure \ref{fig:fake_emitters} mock and real line emitters graded by the visual inspection team to illustrate their high similarity.

\section{Full catalogue of [\ion{O}{iii}] emitters and individual spectra}
\label{app:full_cat_and_plots}
\begin{figure*}
\includegraphics[width=0.49\textwidth]{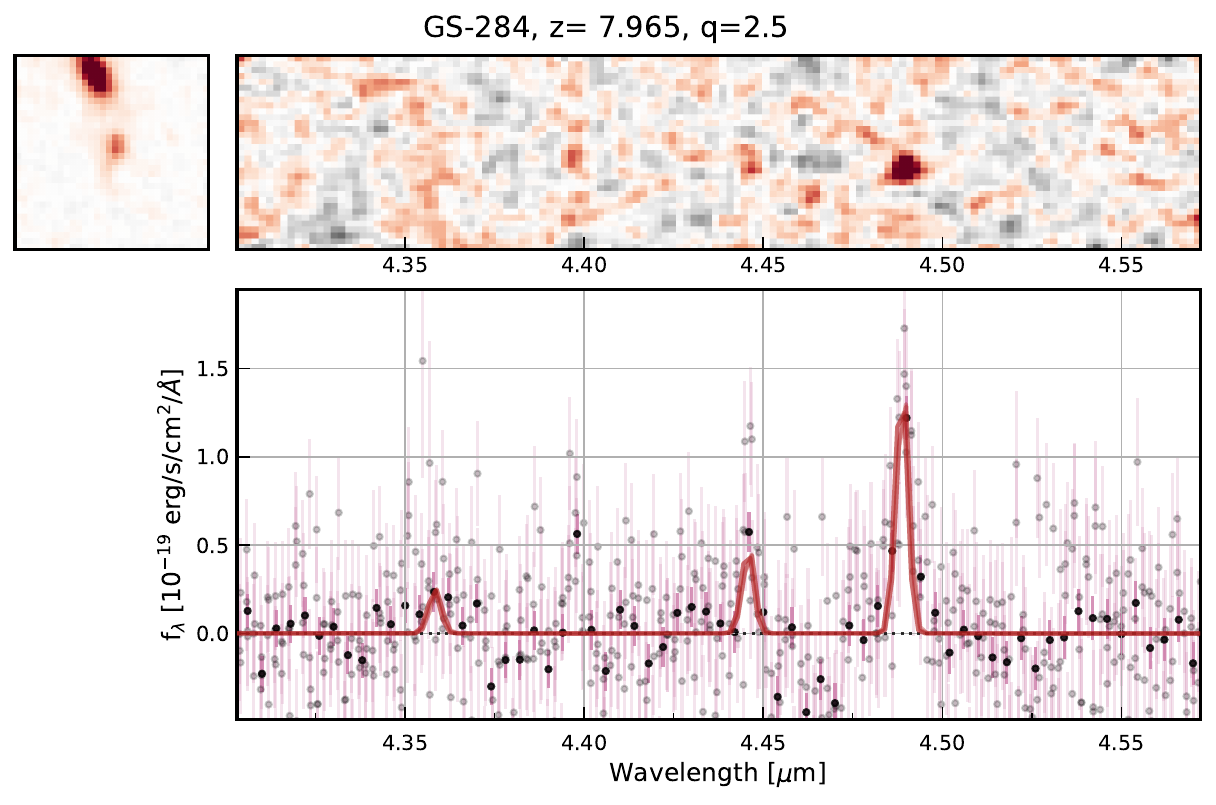}
\includegraphics[width=0.49\textwidth]{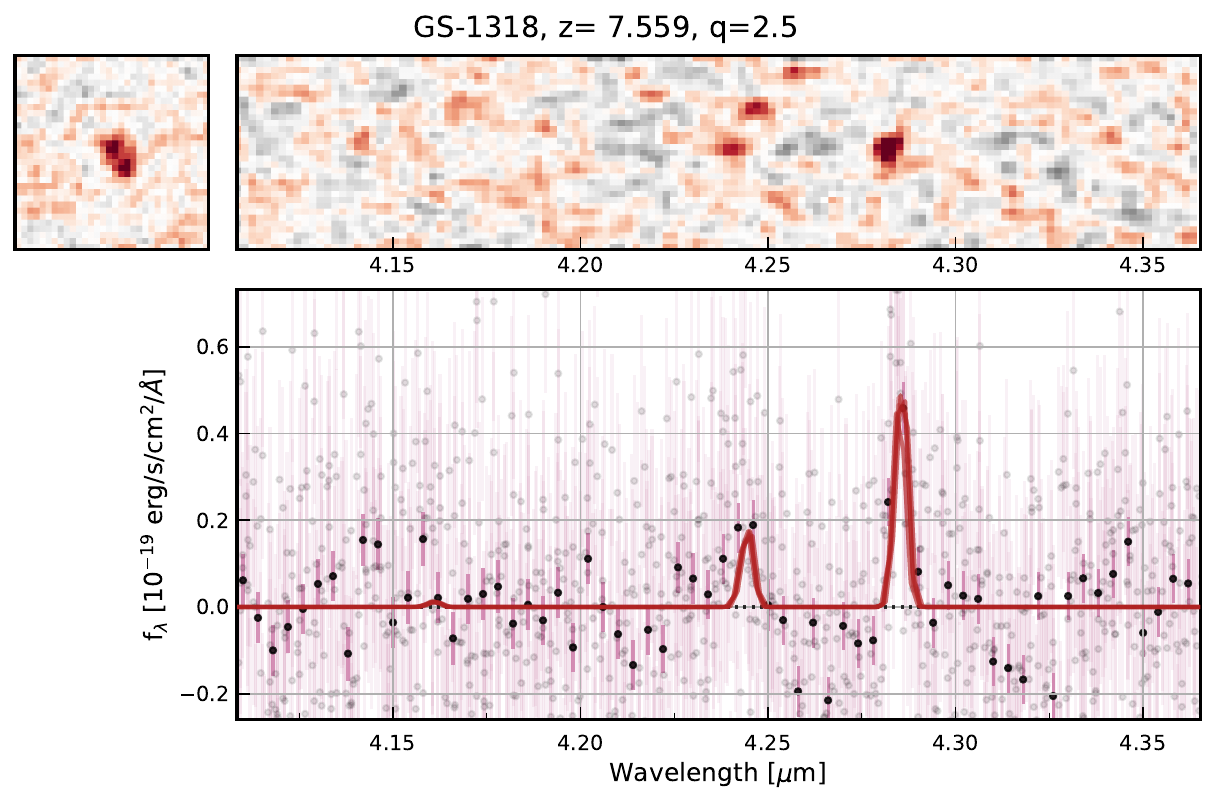} \\
\includegraphics[width=0.49\textwidth]{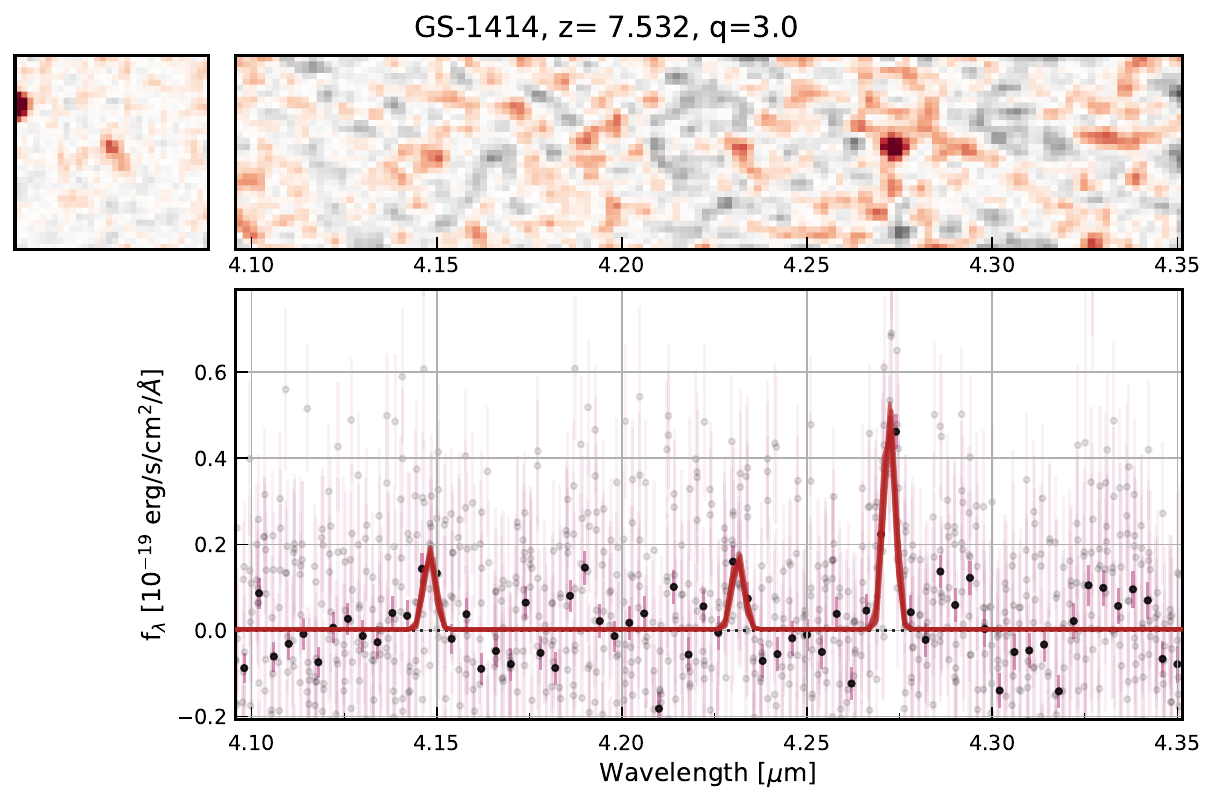} 
\includegraphics[width=0.49\textwidth]{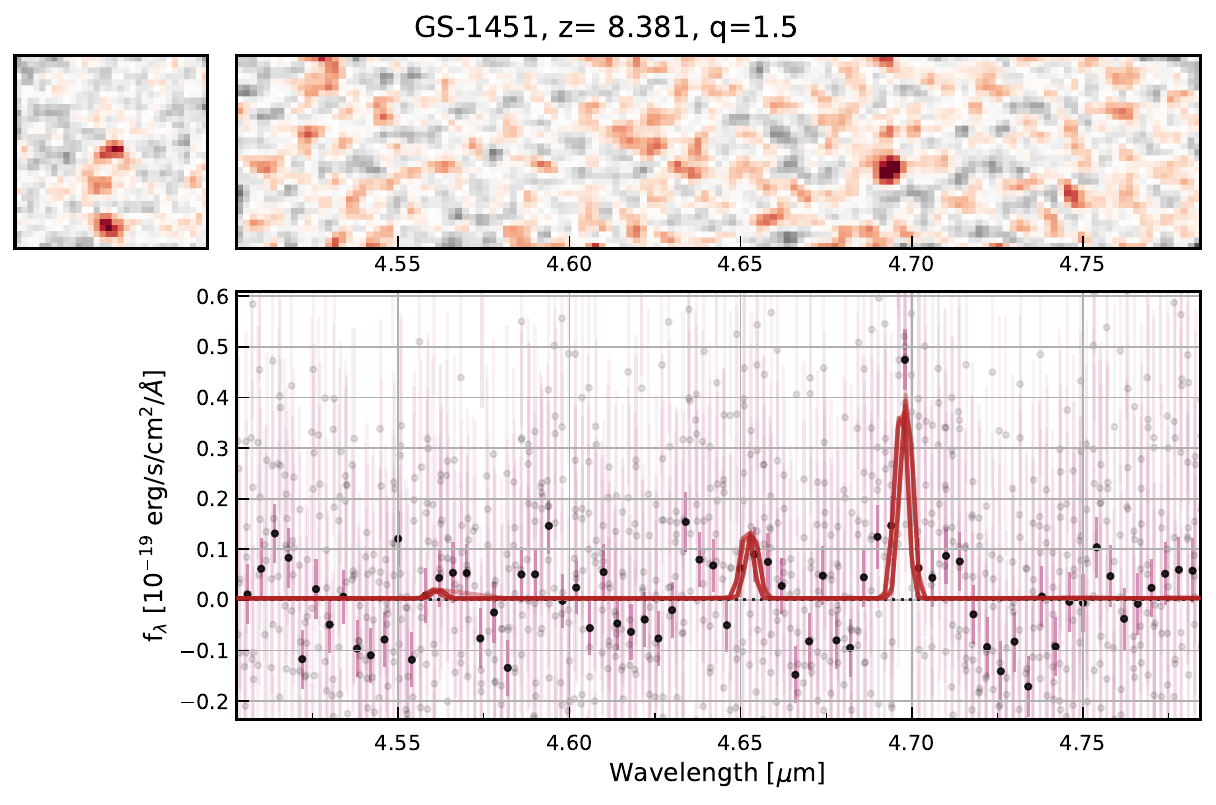}
\includegraphics[width=0.49\textwidth]{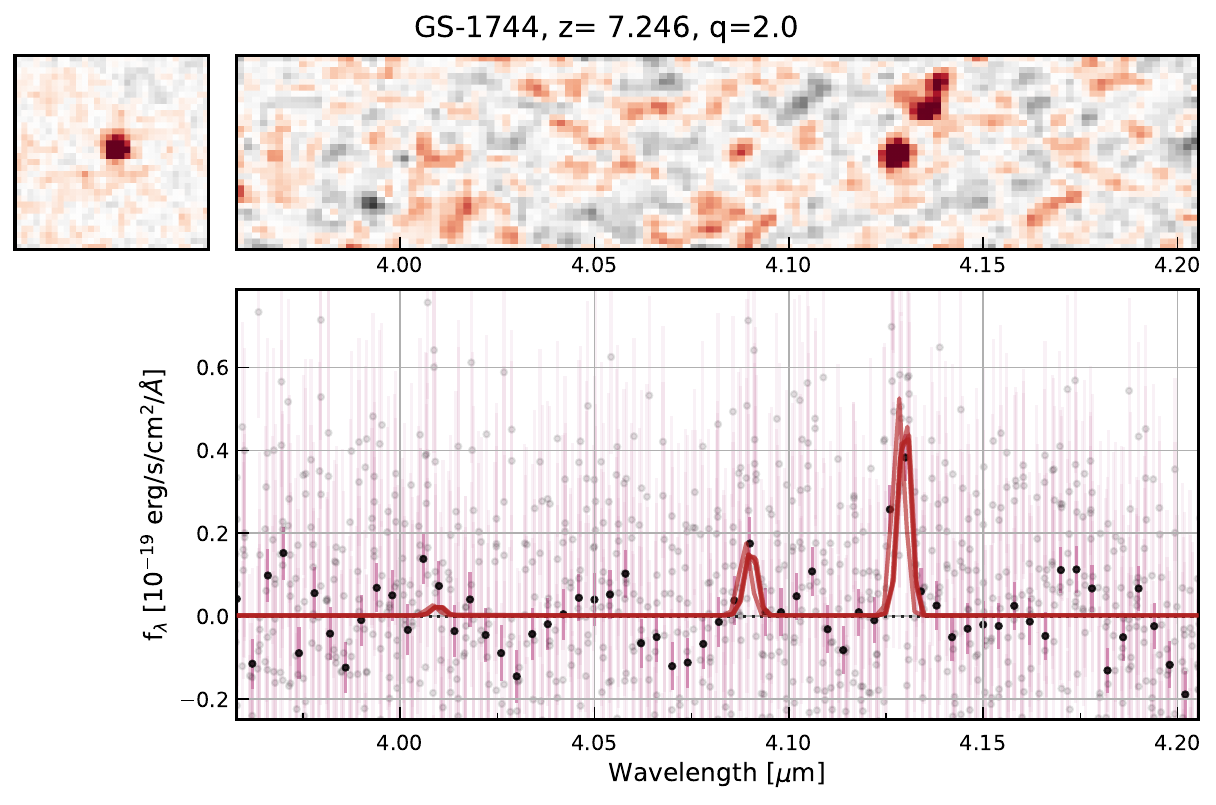} 
\includegraphics[width=0.49\textwidth]{plots_emitters/plot_direct1d2d_GS3081.pdf}
\includegraphics[width=0.49\textwidth]{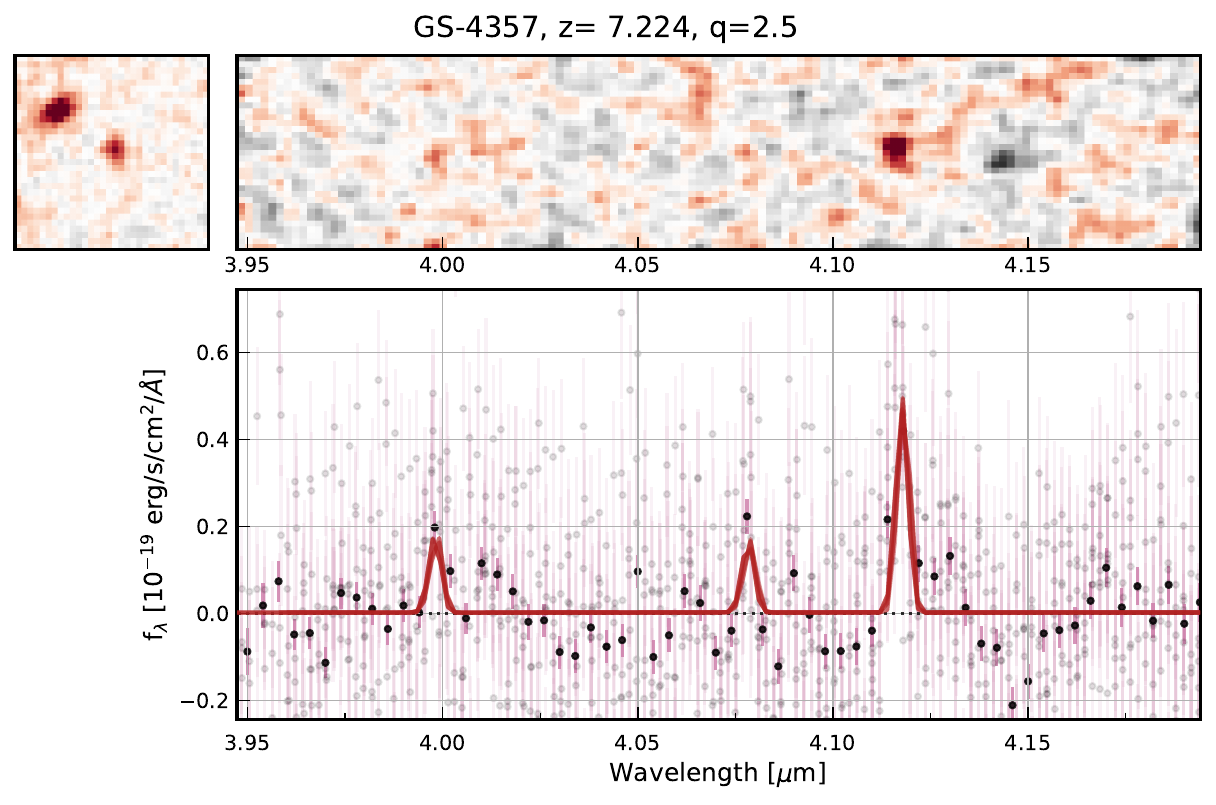}
\includegraphics[width=0.49\textwidth]{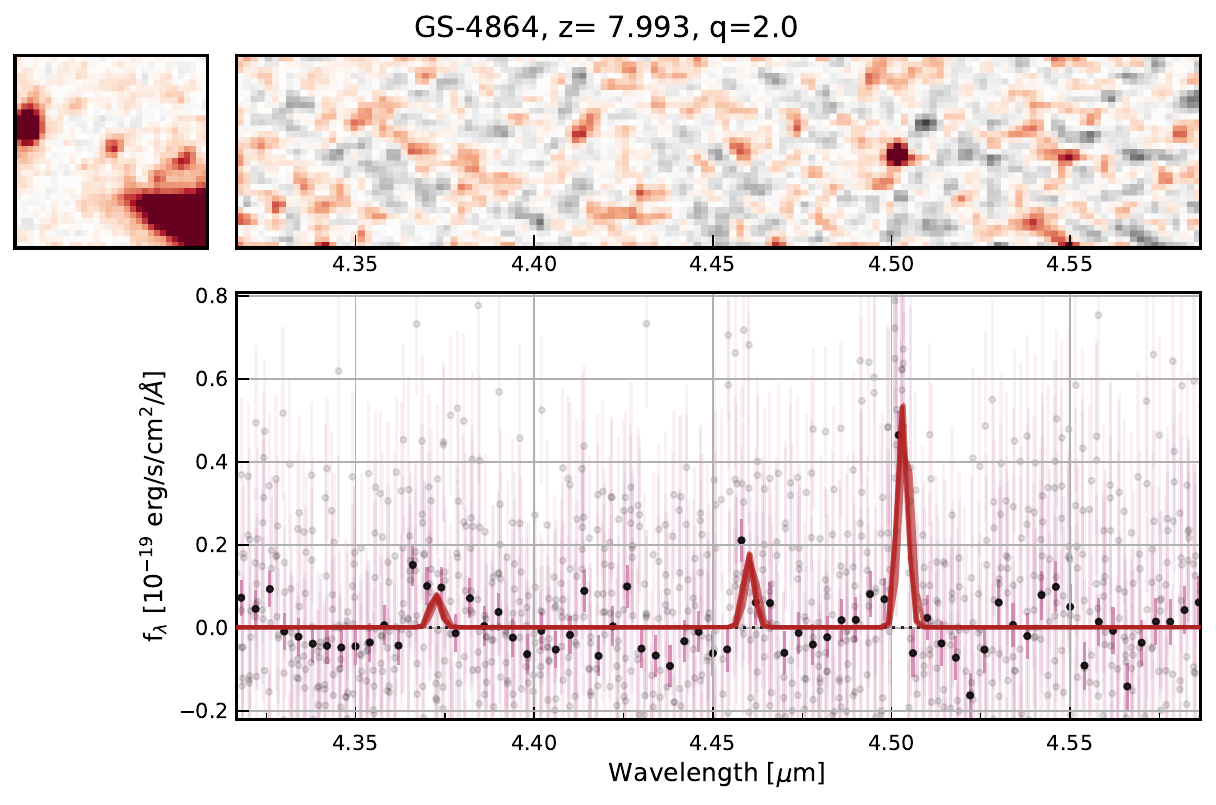}
    \caption{First 8 GOODS-South [\ion{O}{iii}] emitters in the catalogue (omitting 1319 which is merging with 1318). The plots for the full sample are available as supplementary material. }
    \label{fig:all_emitters_GS}
\end{figure*}

\begin{figure*}
\includegraphics[width=0.49\textwidth]{plots_emitters/plot_direct1d2d_GN849.pdf}
\includegraphics[width=0.49\textwidth]{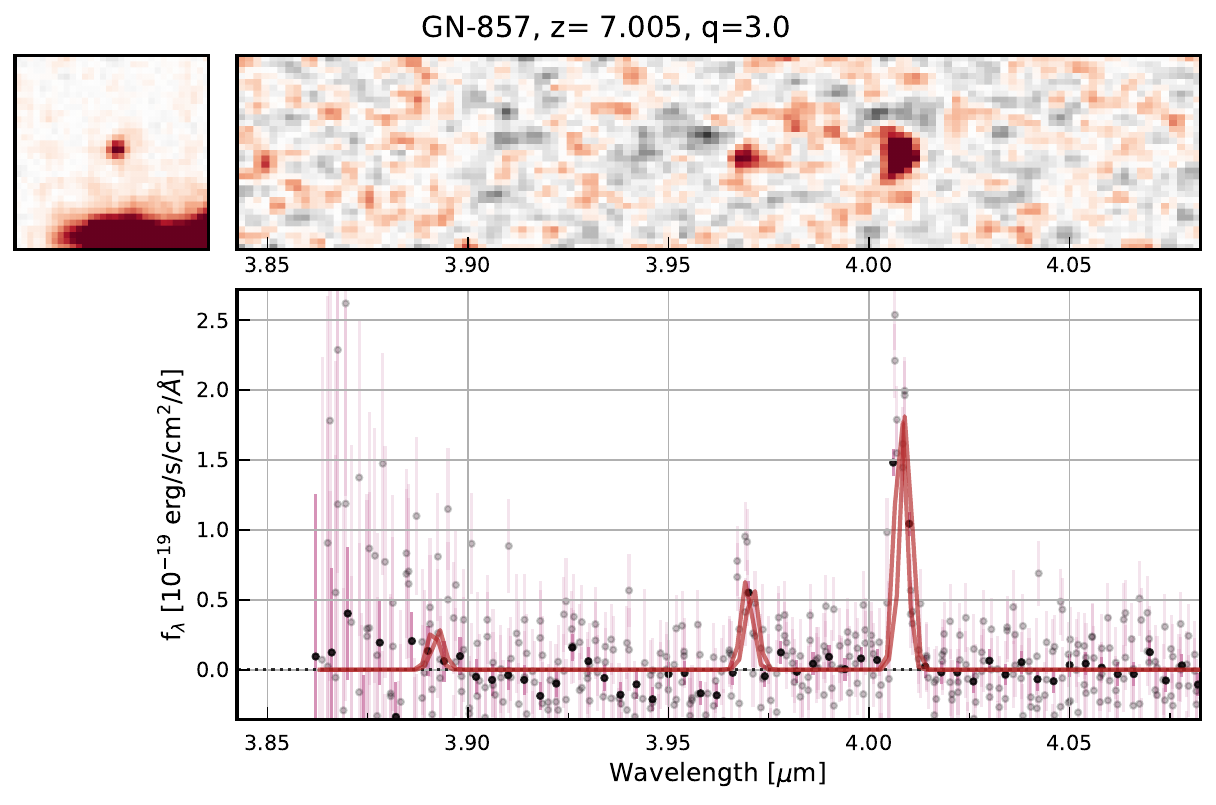} \\
\includegraphics[width=0.49\textwidth]{plots_emitters/plot_direct1d2d_GN1129.pdf} 
\includegraphics[width=0.49\textwidth]{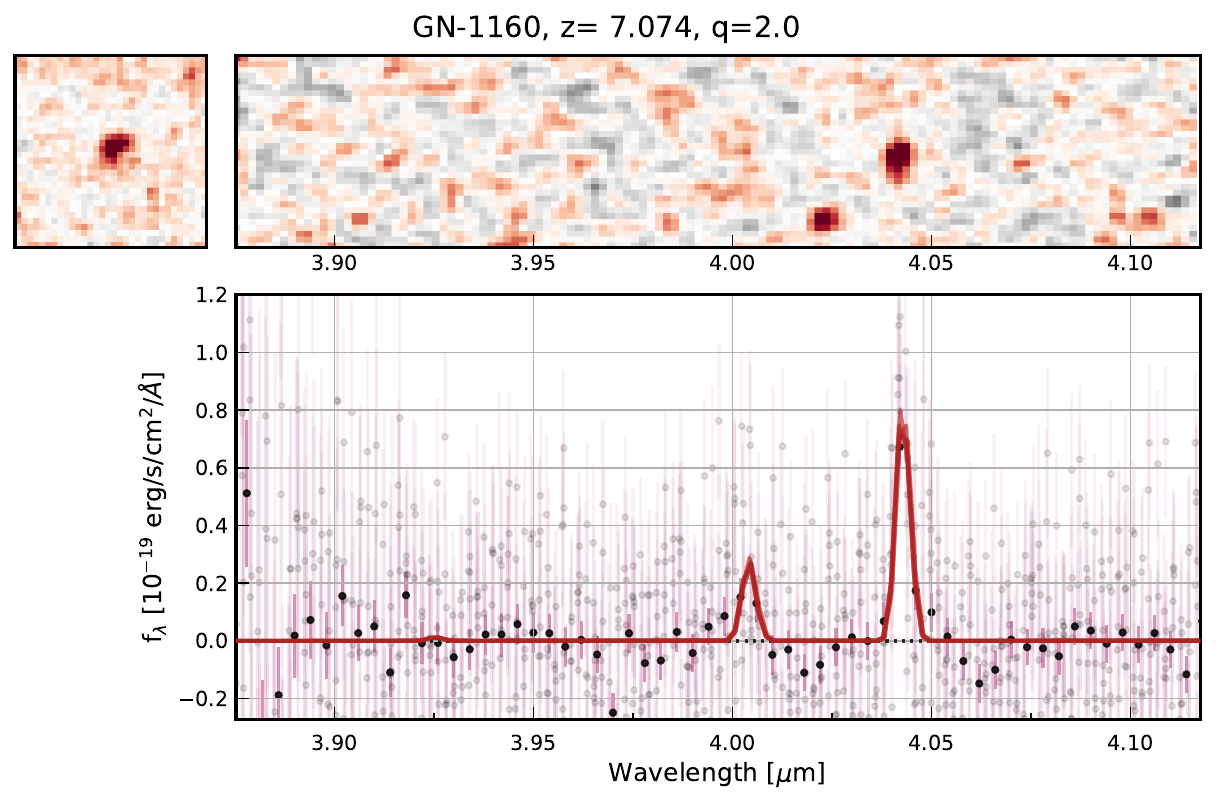} \\
\includegraphics[width=0.49\textwidth]{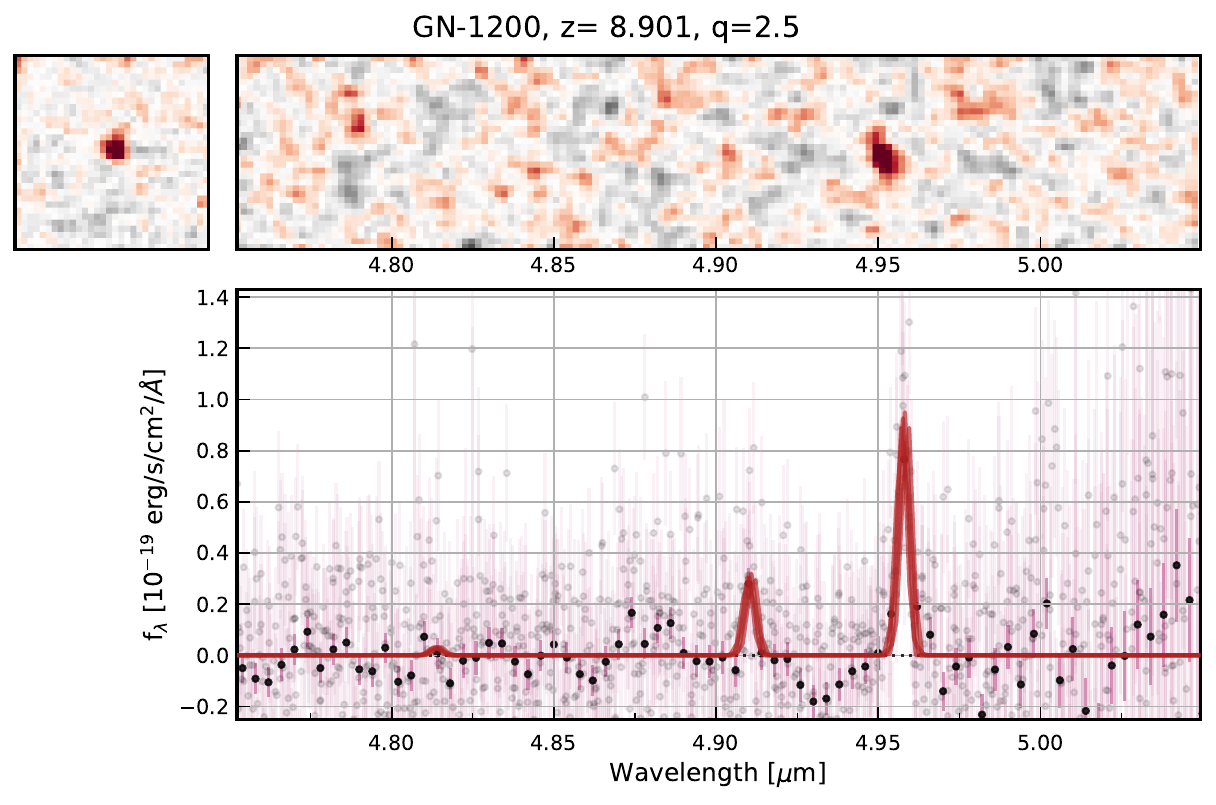} 
\includegraphics[width=0.49\textwidth]{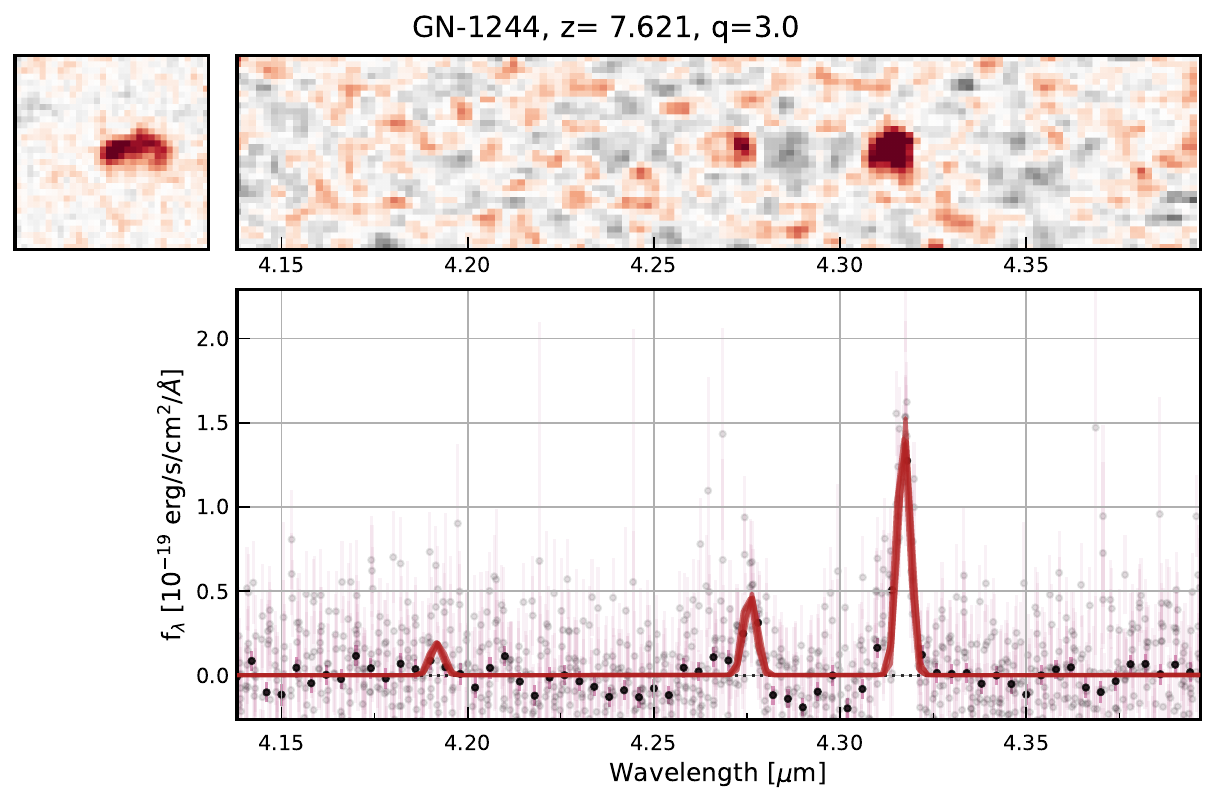}  \\
\includegraphics[width=0.49\textwidth]{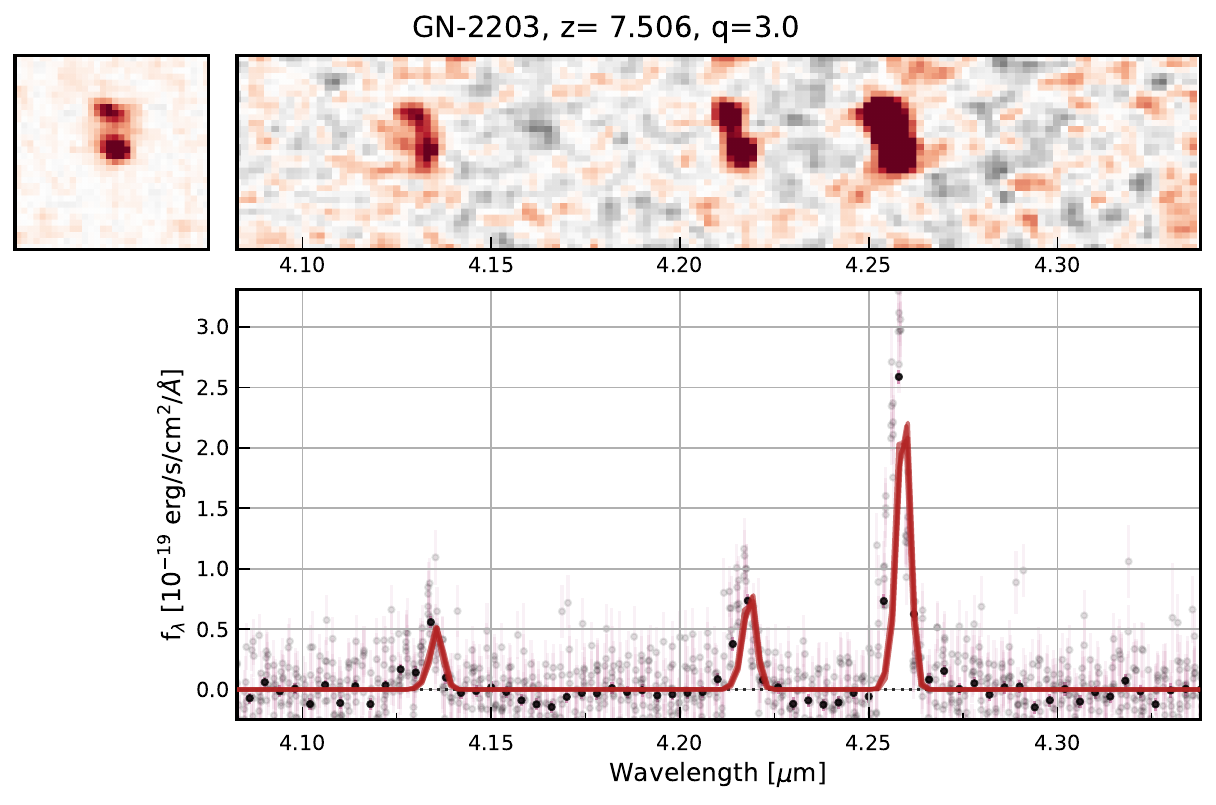} 
\includegraphics[width=0.49\textwidth]{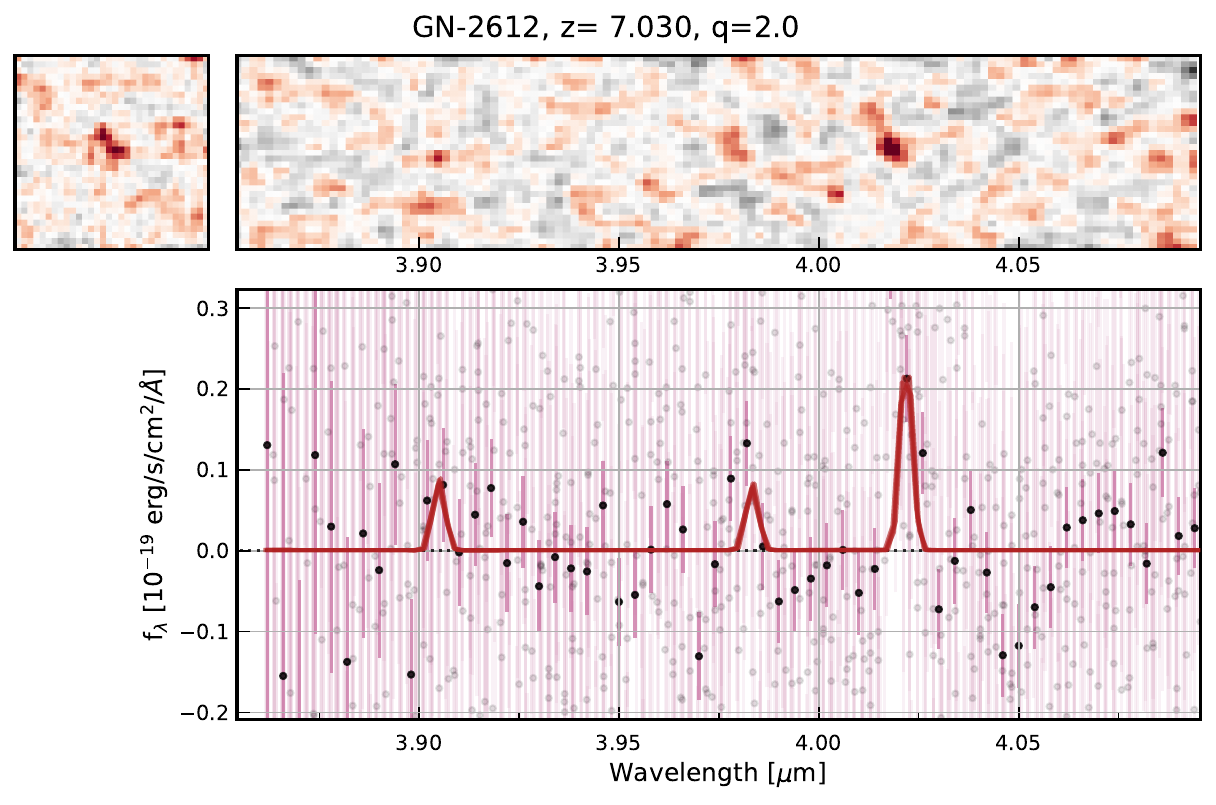}  \\
    \caption{First 8 GOODS-North [\ion{O}{iii}] emitters in the catalogue (we have omitted 1245 and 2204 as they are mergers with 1244 and 2203, respectively, and already visible in their plots). The plots for the full sample are available as supplementary material. }
     \label{fig:all_emitters_GN}
\end{figure*}

In this appendix we present an excerpt of the full [\ion{O}{iii}] emitter catalogue. For the sake of brevity, only the first eight emitters in each field are presented. The identifier numbers, coordinates, redshift, quality flags, magnitudes and line fluxes are given in Table \ref{table:full_cat}. We show the direct imaging, 1D and 2D spectra of the emitters in Fig. \ref{fig:all_emitters_GS} and  \ref{fig:all_emitters_GN} for GOODS-North and GOODS-South, respectively. The full machine-readable catalogue and the plots for each emitter are available as supplementary material, and can also be found at \url{https://github.com/rameyer/fresco/}.

\begin{table*}
    \centering
    \scriptsize
    \caption{Full catalogue of [\ion{O}{iii}] emitters in FRESCO GN/GS. Only the first 8 systems (ordered by detection ID) per field are shown for brevity, matching those shown in Figure \ref{fig:all_emitters_GS} and Figure \ref{fig:all_emitters_GN}. The full machine-readable catalogue is accessible as supplementary material or at \url{https://github.com/rameyer/fresco/}. $^{a,c}$ IDs and fluxes refer to the internal FRESCO data release v7.3. $^{b}$ The rest-frame UV magnitude $M_{\rm{UV}}$ is derived from the \texttt{Prospector} best-fits to the data using the spectroscopic redshift in column 4 (Naidu et al., in prep). }
    \setlength{\tabcolsep}{6pt} 
\renewcommand{\arraystretch}{1.3} 
    \begin{tabular}{c|c|c|c|c|c|c|c|c|c}
        ID$^{a}$ & RA & DEC & $z_{\rm{spec}}$ & q & $M_{\rm{UV}}^{b}$&$f_{\rm{F444W}}^{c}$ &  $f_{\rm{H}\beta}$ & $f_{\rm{[\ion{O}{iii}] 4960}}$ & $f_{\rm{[\ion{O}{iii}] 5008}}$ \\ 
        & [deg] & [deg] & & & [AB] & [nJy] & $[10^{-18} \rm{erg\ s}^{-1} \rm{cm}^{-2}]$ & $[10^{-18} \rm{erg s}^{-1} \rm{cm}^{-2}]$  & $[10^{-18} \rm{erg s}^{-1} \rm{cm}^{-2}]$ \\ 
        \hline
FRESCO-GS-00284 & 53.068819 & -27.731392 & 7.965 & 2.5 & $-21.30^{\tabularrel+0.10}_{\tabularrel-0.09}$ & $199\pm14$ & $0.99 \pm 0.48$ & $2.64 \pm 2.64$ & $5.63 \pm 0.54$ \\ 
FRESCO-GS-01318 & 53.167813 & -27.736141 & 7.559 & 2.5 & $-19.72^{\tabularrel+0.08}_{\tabularrel-0.05}$ & $71\pm4$ & $0.03 \pm 0.26$ & $0.83 \pm 0.83$ & $2.07 \pm 0.26$ \\ 
FRESCO-GS-01319 & 53.167772 & -27.736194 & 7.555 & 2.5 & $-19.86^{\tabularrel+0.09}_{\tabularrel-0.07}$ & $50\pm2$ & $0.09 \pm 0.26$ & $0.65 \pm 0.65$ & $1.14 \pm 0.25$ \\ 
FRESCO-GS-01414 & 53.193895 & -27.736557 & 7.532 & 3.0 & $-18.57^{\tabularrel+0.53}_{\tabularrel-0.43}$ & $62\pm9$ & $0.67 \pm 0.15$ & $0.80 \pm 0.80$ & $1.88 \pm 0.16$ \\ 
FRESCO-GS-01451 & 53.109952 & -27.736655 & 8.381 & 1.5 & $-18.86^{\tabularrel+0.36}_{\tabularrel-0.40}$ & $27\pm7$ & $0.09 \pm 0.25$ & $0.27 \pm 0.27$ & $1.87 \pm 0.27$ \\ 
FRESCO-GS-01744 & 53.169576 & -27.738058 & 7.246 & 2.0 & $-20.29^{\tabularrel+0.08}_{\tabularrel-0.06}$ & $111\pm6$ & $0.08 \pm 0.26$ & $0.81 \pm 0.81$ & $1.84 \pm 0.25$ \\ 
FRESCO-GS-03081 & 53.172590 & -27.743935 & 7.571 & 3.0 & $-21.35^{\tabularrel+0.07}_{\tabularrel-0.04}$ & $331\pm17$ & $1.77 \pm 0.34$ & $3.15 \pm 3.15$ & $9.50 \pm 0.35$ \\ 
FRESCO-GS-04357 & 53.100195 & -27.750296 & 7.224 & 2.5 & $-18.93^{\tabularrel+0.21}_{\tabularrel-0.21}$ & $46\pm6$ & $0.63 \pm 0.16$ & $0.74 \pm 0.74$ & $1.78 \pm 0.18$ \\ 
FRESCO-GS-04864 & 53.192112 & -27.752517 & 7.993 & 2.0 & $-18.92^{\tabularrel+0.08}_{\tabularrel-0.10}$ & $82\pm4$ & $0.27 \pm 0.20$ & $0.50 \pm 0.50$ & $1.99 \pm 0.24$ \\ 
FRESCO-GS-05175 & 53.077924 & -27.753943 & 8.314 & 1.5 & $-20.40^{\tabularrel+0.19}_{\tabularrel-0.18}$ & $92\pm15$ & $0.09 \pm 0.39$ & $0.67 \pm 0.67$ & $1.29 \pm 0.41$ \\ 
\vdots & \vdots & \vdots & \vdots & \vdots & \vdots & \vdots & \vdots  & \vdots & \vdots   \\
FRESCO-GN-00169 & 189.235731 & 62.328352 & 7.001 & 1.5 & $-19.44^{\tabularrel+0.15}_{\tabularrel-0.13}$ & $40\pm7$ & $1.06 \pm 0.36$ & $0.37 \pm 0.37$ & $1.41 \pm 0.18$ \\ 
FRESCO-GN-00617 & 189.187361 & 62.318837 & 7.001 & 2.0 & $-19.89^{\tabularrel+0.11}_{\tabularrel-0.07}$ & $95\pm9$ & -- & $0.76 \pm 0.76$ & $2.07 \pm 0.13$ \\ 
FRESCO-GN-00620 & 189.187537 & 62.318765 & 6.987 & 2.0 & $-19.09^{\tabularrel+0.21}_{\tabularrel-0.18}$ & $46\pm10$ & -- & $0.94 \pm 0.94$ & $2.94 \pm 0.12$ \\ 
FRESCO-GN-00849 & 189.222006 & 62.315764 & 7.602 & 1.5 & $-20.07^{\tabularrel+0.07}_{\tabularrel-0.07}$ & $69\pm7$ & $0.02 \pm 0.23$ & $0.21 \pm 0.21$ & $1.80 \pm 0.23$ \\ 
FRESCO-GN-00857 & 189.262173 & 62.315655 & 7.005 & 3.0 & $-20.02^{\tabularrel+0.06}_{\tabularrel-0.07}$ & $156\pm8$ & $1.02 \pm 0.67$ & $2.08 \pm 2.08$ & $6.86 \pm 0.36$ \\ 
FRESCO-GN-01129 & 189.212314 & 62.312400 & 7.037 & 2.5 & $-19.78^{\tabularrel+0.10}_{\tabularrel-0.07}$ & $84\pm7$ & $0.02 \pm 0.31$ & $1.40 \pm 1.40$ & $2.33 \pm 0.23$ \\ 
FRESCO-GN-01160 & 189.243926 & 62.312127 & 7.074 & 2.0 & $-19.83^{\tabularrel+0.09}_{\tabularrel-0.11}$ & $103\pm8$ & $0.08 \pm 0.33$ & $1.13 \pm 1.13$ & $3.59 \pm 0.31$ \\ 
FRESCO-GN-01200 & 189.217725 & 62.311782 & 8.901 & 2.5 & $-20.04^{\tabularrel+0.13}_{\tabularrel-0.11}$ & -- & $0.15 \pm 0.26$ & $1.28 \pm 1.28$ & $3.56 \pm 0.30$ \\ 
FRESCO-GN-01244 & 189.224394 & 62.311365 & 7.621 & 3.0 & $-20.36^{\tabularrel+0.11}_{\tabularrel-0.07}$ & $112\pm8$ & $0.82 \pm 0.28$ & $1.83 \pm 1.83$ & $6.16 \pm 0.30$ \\ 
FRESCO-GN-01245 & 189.224488 & 62.311281 & 7.609 & 2.5 & $-20.49^{\tabularrel+0.13}_{\tabularrel-0.11}$ & $205\pm15$ & $1.08 \pm 0.40$ & $2.55 \pm 2.55$ & $8.57 \pm 0.42$ \\ 
FRESCO-GN-02203 & 189.157976 & 62.302398 & 7.506 & 3.0 & $-21.15^{\tabularrel+0.05}_{\tabularrel-0.06}$ & $311\pm16$ & $2.14 \pm 0.22$ & $3.11 \pm 3.11$ & $10.33 \pm 0.26$ \\ 
FRESCO-GN-02204 & 189.157791 & 62.302343 & 7.502 & 3.0 & $-20.76^{\tabularrel+0.09}_{\tabularrel-0.06}$ & $287\pm14$ & $1.80 \pm 0.28$ & $2.79 \pm 2.79$ & $8.88 \pm 0.33$ \\ 
FRESCO-GN-02612 & 189.209600 & 62.299283 & 7.030 & 2.0 & $-19.14^{\tabularrel+0.14}_{\tabularrel-0.10}$ & $47\pm8$ & $0.31 \pm 0.30$ & $0.35 \pm 0.35$ & $0.85 \pm 0.22$ \\ 
\vdots & \vdots & \vdots & \vdots & \vdots & \vdots & \vdots & \vdots & \vdots & \vdots  
    \end{tabular}
    \label{table:full_cat}
\end{table*}

\section{Completeness functions of the Gaussian-Matched filtering and Visual Inspection}
\label{app:completeness}
\begin{figure}
    \centering
    \includegraphics[width=0.5\textwidth]{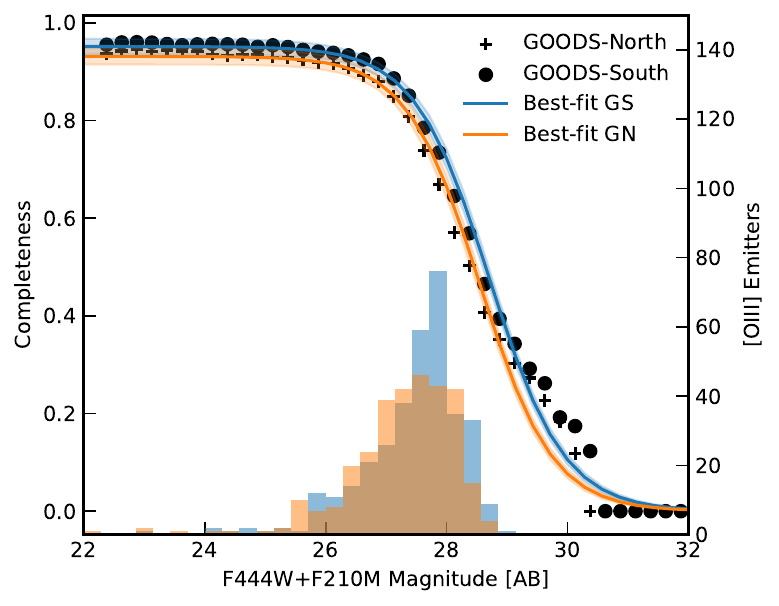}
    \caption{Detection completeness for FRESCO F444W+F210M (black points) and best-fit sigmoid functions (coloured lines and shaded areas). The detection completeness plateaus at $\sim 90-95\%$ depending on the field above a magnitude $<27$, and then slowly declines. We show the distribution of the selected [\ion{O}{iii}] emitters presented in this work in each field as coloured histograms.}
    \label{fig:detection_completeness}
\end{figure}

In this Appendix, we detail the separate completeness functions of the imaging detection, the Gaussian-matched filtering approach and the visual inspection. We first show the detection completeness on Fig. \ref{fig:detection_completeness}. As described in Section \ref{sec:obs_analysis}, the imaging detection completeness is computed using \texttt{GLACIAR2} to iteratively inject and retrieve sources in the FRESCO footprint. The completeness plateaus at $\sim 95\%$ above a magnitude $<27$ AB, and declines slowly down to mag$\sim 30-30.5$. The best-fit sigmoid functions parameters for the GN and GS detection images are presented in Table \ref{table:completeness_params}.

\begin{figure}
    \centering
    \includegraphics[width=0.49\textwidth]{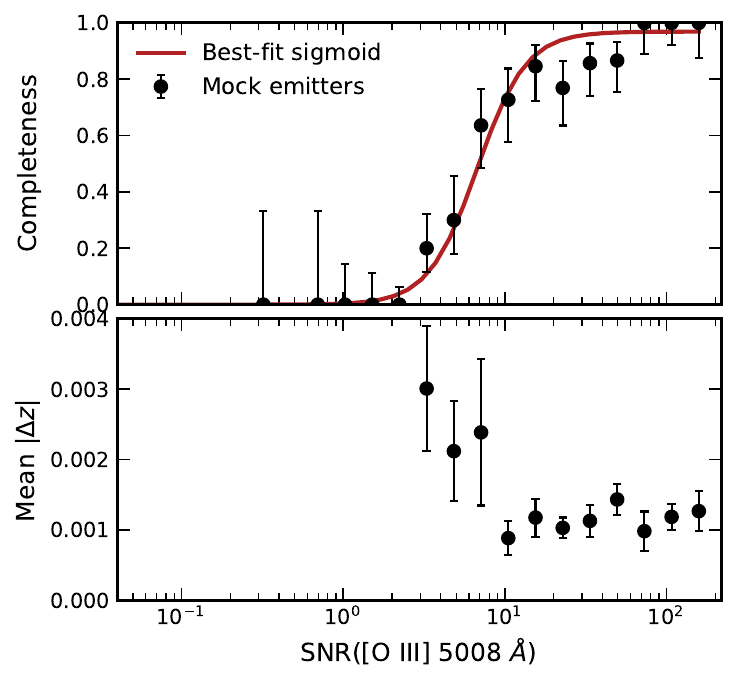}
    \caption{\textit{Upper panel: } Completeness of the Gaussian-matched filtering approach on the mock [\ion{O}{iii}] emitters as a function of the input SNR of the integrated 5008 \AA\ line. The completeness picks up rapidly around input SNR$=6-10$ and reaches a plateau at $(96.9\pm2.2)\%$. The best-fit sigmoid function to the measured completeness is shown in dark red. \textit{Lower panel: } Mean redshift error of the recovered [\ion{O}{iii}] emitters. The redshift error is constant with $|\Delta z| \simeq 0.001$, which corresponds to $\sim 35\ \kms$ at $6.8<z<9.0$. }
    \label{fig:completeness_GM}
\end{figure}

We then show the completeness of the Gaussian-matched filter step on the mock emitters in Fig. \ref{fig:completeness_GM}. In assessing the completeness we run the same Gaussian-matched filter and SNR cuts as for the real sources on the mock emitters. We consider an object to be recovered if the redshift error is $\Delta z<0.05$ or $\leq 2000\ \kms$ at the redshifts considered. This cutoff is chosen to be less than the velocity offset between the 5008 and 4960 \AA \ lines ($2875\ \kms$). Unsurprisingly, the completeness is sharply defined around $\rm{SNR}([\ion{O}{iii}] 5008)\sim 6-12$, which is where the second line of the doublet becomes significant ($\rm{SNR}\simeq2-4$, respectively). We fit the completeness measurement with a sigmoid function, and present the best-fit parameters in Table \ref{table:completeness_params}. The Gaussian-matched filter selection is already $17.6 \%, 60.1\%,81.4\%$ complete at  $\rm{SNR([\ion{O}{iii}] 5008)}=4,8,12$, respectively, and reaches a plateau at $(96.9\pm2.2)\%$ at SNR$\gtrsim20$. The Gaussian-matched filter recovers the redshift of the $\rm{SNR}\geq10$ sources with a mean error $|\Delta z| \simeq 0.001$, which corresponds to $\sim 35\ \kms$ at $6.8<z<9.0$, i.e. a quarter of the nominal spectral resolution.

\begin{table}
\caption{Parameters of the best-fit sigmoid functions to the different completeness functions used in this work. The sigmoid are of the form $c(x) = c_0/ (1+\exp[{a*(x-x_0)]})$. For the detection completeness, $x_0$ and $x$ are in units of mag$_{AB}(\rm{F444W+F210M})$, whereas for the spectroscopic completenesses, they are in $\log_{10}(\rm{SNR([\ion{O}{iii}] 5008)})$ (see further Section \ref{sec:obs_analysis}). \label{table:completeness_params}}
\centering
\begin{tabular}{lrrr}
Type & $c_0$ & $x_0$ & $a$ \\ \hline
Detection - GS & 0.93 & 28.55 & -1.66 \\
Detection - GN & 0.95 & 28.71 & -1.61 \\
Spectroscopy - GM filter & 0.97 & 0.83 & 6.63 \\
Spectroscopy - VI (Visual Inspection) & 0.95 & 1.05 & 7.91 \\
Spectroscopy - GM$\odot$VI (Eq. \ref{eq:completeness}) & 0.96 & 1.08 & 7.20 \\
\end{tabular}
\end{table}

\begin{figure}
    \centering
    \includegraphics[width=0.46\textwidth]{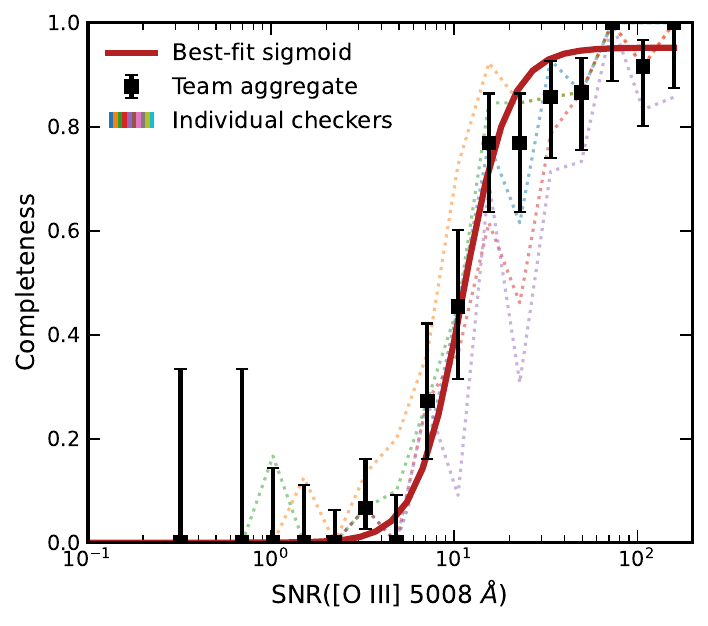}
    \caption{Completeness of the visual inspection for the aggregate team votes (black data points) and individual checkers (coloured dotted lines). The best-fit completeness (with a fixed upper end point $C=1$) is shown in dark red.}
    \label{fig:completeness_vis}
\end{figure}

We show the completeness for each anonymous checker, as well as the median completeness in Fig. \ref{fig:completeness_vis}. We find that, despite no prior instructions on what a real $z>6.8$  [\ion{O}{iii}] emitter should look like in \textit{JWST}/NIRCam data, the scatter between the different individuals is rather limited ($\sigma_C = 0.053$) and much lower than the formal binomial error on the number of sources inspected and recovered in the total sample. As for the visual inspection completeness, we fit the team aggregate completeness with a sigmoid function whose best-fit parameters are displayed in Table \ref{table:completeness_params}. Overall, the visual inspection is rather conservative and less sharply defined, with $39.6\%, 46.6\%, 50.7\%,53.7  \%$ completeness achieved at SNR$([\ion{O}{iii}] 5008)=4,8,12,16$.

\section{RMS pseudo-cube of the FRESCO WFSS data}
\label{app:rms}
In this Appendix, we detail the construction of the empirical rms 3D cube in the FRESCO data. We follow the approach of \citet[][]{Matthee2023_EIGER}{}{} and create a square uniform grid aligned with the FRESCO mosaics and extract spectra in $\Delta \theta = 6\ \rm{arcsec}$ intervals in both x- and y- directions (in the mosaic frame of reference). For each position, we then use the median segmentation profile of [\ion{O}{iii}] emitters to extract one-dimensional spectra from the grism images. We compute the median error and rms noise directly from the spectra in $100$ \AA\ ranges. The rms noise is computed from the median-filtered spectra using three iterations of $3-\sigma$ clipping and taking the standard deviation. We then use the error array values rescaled to the median rms level for the entire FRESCO volume.
\begin{figure*}
    \centering
    \includegraphics[width=0.99\textwidth]{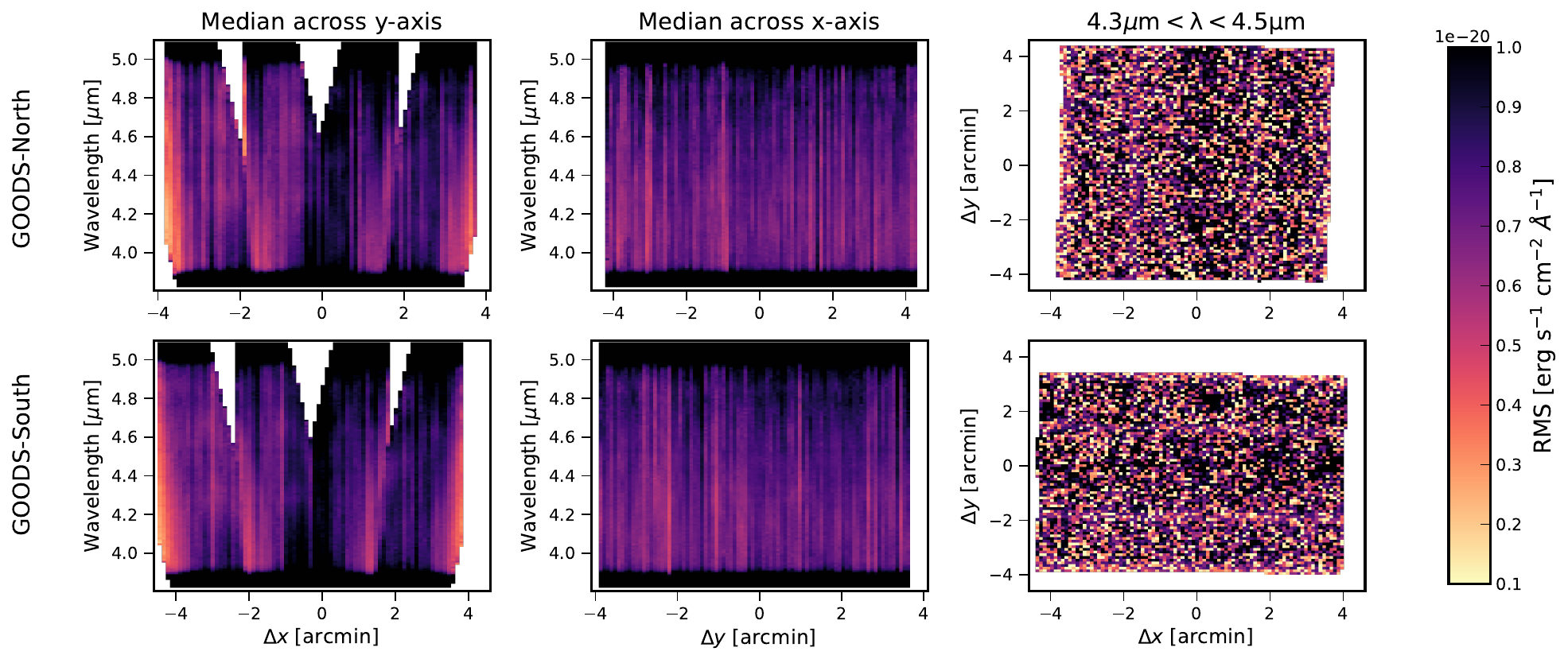}
    \caption{RMS as a function of mosaic position and wavelength in the FRESCO GOODS-North (\textit{upper row}) and GOODS-South (\textit{lower row}). The noise pattern in the x-direction (e.g. along the dispersion direction)  is dominated by the overlaps of the multiple visits and the NIRCam A and B modules in the mosaics (\textit{first column}). The noise in the y-direction (e.g. perpendicular to the dispersion direction) shows a succession of faint and bright lines depending on the presence or absence of bright objects (\textit{second column}). The rms noise evolution with wavelength follows the sensitivity of the NIRCam F444W grism, strongly increased at the edges of the wavelength range.}
    \label{fig:rms}
\end{figure*}
We show projections of the rms noise as a function of position and wavelength in Figure \ref{fig:rms}. Similar to \citet[][]{Matthee2023_EIGER}{}{}, we find that the rms is mostly wavelength-dependent due to the mosaicing pattern of FRESCO, except in select locations around brighter objects. Additionally, we find a pattern along the x-direction due to the overlaps of the module A and B observations and the different visits used to create the mosaic.

\bsp	
\label{lastpage}
\end{document}